\documentclass[fleqn,usenatbib]{mnras}

\usepackage{newtxtext,newtxmath}

\usepackage[T1]{fontenc}
\usepackage{ae,aecompl}

\usepackage{graphicx}	
\usepackage{amsmath}	
\usepackage{float}
\floatstyle{plaintop}
\usepackage{nccmath}
\usepackage{enumitem}

\title[The mid-infrared Leavitt Law for the MCs]{The mid-infrared Leavitt Law for Classical Cepheids in the Magellanic Clouds}

\author[A. H. Chown et al.]{
Abigail H. Chown\thanks{E-mail: A.H.Chown@bath.ac.uk},
Victoria Scowcroft, and Stijn Wuyts
\\
Department of Physics, University of Bath, Claverton Down, Bath, BA2 7AY, UK\\
}

\date{Accepted XXX. Received YYY; in original form ZZZ}

\pubyear{2020}

\begin{document}
\label{firstpage}
\pagerange{\pageref{firstpage}--\pageref{lastpage}}
\maketitle

\begin{abstract}
The Cepheid Leavitt Law (LL), also known as the Period-Luminosity relation, is a crucial tool for assembling the cosmic distance ladder. By combining data from the OGLE-IV catalogue with mid-infrared photometry from the \textit{Spitzer Space Telescope}, we have determined the $3.6$ $\mu$m and $4.5$ $\mu$m LLs for the Magellanic Clouds using $\sim 5000$ fundamental-mode Classical Cepheids. Mean magnitudes were determined using a Monte Carlo Markov Chain (MCMC) template fitting procedure, with template light curves constructed from a subsample of these Cepheids with fully-phased, well-sampled light curves.  
The dependence of the Large Magellanic Cloud LL coefficients on various period cuts was tested, in addition to the linearity of the relationship. The zero point of the LL was calibrated using the parallaxes of Milky Way Cepheids from the \textit{Hubble Space Telescope} and \textit{Gaia} Data Release 2. Our final calibrated relations are $M_{[3.6]} = -3.246(\pm 0.008)(\log(P)-1.0)-5.784(\pm 0.030)$ and $M_{[4.5]} = -3.162(\pm 0.008)(\log(P)-1.0)-5.751(\pm 0.030)$.
\end{abstract}

\begin{keywords}
stars: variables: Cepheids -- Magellanic Clouds -- infrared: stars
\end{keywords}



\section{Introduction}

Since its initial discovery around a century ago, the Cepheid Leavitt Law (\citealt{Leavitt1908,Leavitt1912}), also known as the Period-Luminosity relation, has proven to be an imperative tool for measuring astronomical distances. The Leavitt Law (LL) relates the luminosity of a Cepheid to its pulsation period, revealing that Cepheids with longer periods are intrinsically brighter than their shorter period counterparts. Thus, Cepheid variables play a vital role as distance indicators; in particular, their LL is often used as the first rung of the traditional cosmic distance ladder, which in turn can be used to measure the Hubble constant, $H_0$. In recent years, a growing discrepancy between local- and early-Universe measurements of $H_0$ has arisen (see \citealt{Riess2019} for a review). Therefore, reducing the uncertainties associated with the Cepheid LL will prove essential for understanding the reason behind this discrepancy.

In her pioneering work, Henrietta Leavitt determined the LL for 25 Cepheids in the Small Magellanic Cloud \citep{Leavitt1912}. Since then, a large and growing body of literature has emerged, with many Local Group galaxies now possessing their own LL (e.g.: MW, \citealt{Tammann2003}; NGC 6822, \citealt{Madore2009}; M33, \citealt{Scowcroft2009}; M31, \citealt{Kodric2015}). Of particular importance are the Large and Small Magellanic Clouds (LMC and SMC, respectively). These galaxies are relatively nearby, making them excellent targets for stellar population studies as their individual stars can be resolved. Notably, they each have a substantial Cepheid population, with the Optical Gravitational Lensing Experiment (OGLE) \citep{Udalski1999} cataloguing $\sim 5000$ fundamental-mode Classical Cepheids across both galaxies. Due to the LMC being used as the current anchor galaxy for the extragalactic distance scale, its Cepheid population has been studied extensively. Although the SMC is not used in anchoring the distance scale, it serves as a key galaxy for examining the effect of metallicity on the LL, as it is more metal-poor than both the LMC and Milky Way (MW). 

The LL for the Magellanic Clouds has been studied at many wavelengths, from the optical through to the mid-infrared. Numerous studies have been performed at optical wavelengths (e.g. \citealt{Gieren2005, Sandage2009, Ngeow2015, Groenewegen2018}). \cite{Fouque2007} suggested that the LL is universal, as they found no significant difference in the slopes of the MW and LMC LL for various bands (from $B$ to $K$). In the near-infrared, \cite{Ripepi2016a} used data from the VISTA survey of the Magellanic Clouds (VMC) \citep{Cioni2011} to determine the $YJK_{s}$ LLs for 4172 SMC Classical Cepheids. For the LMC, a near-infrared synoptic survey of its central region was performed by \cite{Macri2015} to obtain the $JHK_{s}$ LLs for 1417 LMC Classical Cepheids. At mid-infrared (mid-IR) wavelengths, the Carnegie Hubble Program (CHP) performed observations of fundamental-mode Classical Cepheids using the \textit{Spitzer Space Telescope} to determine the $3.6$ $\mu$m and $4.5$ $\mu$m LLs for 37 MW (\citealt[hereafter M12]{Monson2012}), 85 LMC (\citealt[hereafter S11]{Scowcroft2011}) and 90 SMC (\citealt[hereafter S16]{Scowcroft2016a}) Cepheids. The observations for these Cepheids were designed to uniformly sample the pulsation cycle, resulting in exquisite light curves and highly-precise mean magnitudes. Complementary analyses of the \textit{Spitzer} SAGE (Surveying the Agents of a Galaxy's Evolution) \citep{Meixner2006} catalogue used much larger samples of Cepheids than the CHP to determine the mid-IR LL for the LMC and SMC, at the expense of a much coarser sampling of the light curves (one or two epochs only), thus leaving more significant uncertainties in the mean magnitudes (\citealt{Ngeow2008, Ngeow2009, Ngeow2010}).

In this study, we use mid-IR data of Cepheids, as observations at these longer wavelengths have several fundamental advantages over optical wavelengths. First, the effects of interstellar extinction are substantially reduced in the infrared, with the \textit{Spitzer} $3.6~\mu$m band extinction being around 20 times smaller than the extinction in the $V$ band \citep{Indebetouw2005}. Second, infrared Cepheid light curves have smaller amplitudes than their optical analogues. This phenomenon occurs because the mid-IR is situated in the Rayleigh–Jeans tail of the stellar spectrum, resulting in minimal temperature effects on the light curve and a more sinusoidal shape. Finally, the dispersion of the mid-IR LL is smaller and its slope is steeper than its optical counterparts, also arising from these wavelengths being dominated by radius variations rather than temperature variations. Consequently, magnitudes inferred from a mid-IR LL will carry smaller systematic uncertainties than those inferred from optical relations (S16).

The work presented in this paper combines the CHP sample of Cepheids from S11 and S16 with the \textit{Spitzer} SAGE survey to obtain $3.6~\mu$m and $4.5$ $\mu$m magnitudes for all known fundamental-mode Classical Cepheids in the Magellanic Clouds, as provided in the OGLE-IV catalogue \citep{Soszynski2017}. The previously mentioned studies that used the SAGE catalogue (\citealt{Ngeow2008, Ngeow2009, Ngeow2010}) only used the OGLE-III catalogue of Cepheids. Since then, OGLE have discovered more Cepheids in the Magellanic Clouds, resulting in the OGLE-IV catalogue. Thus, our work is the most comprehensive mid-IR Cepheid study in the Magellanic Clouds to date. In a separate companion paper (Chown et al. 2020, in prep), we exploit this highly complete and refined catalogue of Cepheid distances to map the 3D distribution of the Magellanic System.

The paper is organised as follows: Section 2 describes the datasets used in this work. In Section 3 we outline the fully-automated photometry pipeline that was developed to process the data. Section 4 describes our template light curve construction and fitting methods. In Section 5 we present our mid-IR catalogue of Magellanic Cloud Cepheids. In Section 6 we assess the dependency of the LL on sample characteristics, and also compare our relations to the literature. Section \ref{sec:finalLL} gives our final adopted LL and its zero point calibration. We present our conclusions in Section 7. 

Throughout this paper, the term ``Cepheid'' will refer to Type-I Classical Cepheids pulsating in the fundamental mode.

\section{Data \& Observations}

\begin{table}
\centering
\caption{\label{tab:sampleStats} Sample statistics for the calibrating and complete samples.}
\begin{tabular}{ccccc}
\hline
                        & \multicolumn{2}{c}{LMC} & \multicolumn{2}{c}{SMC} \\
                        & Cal  & Complete & Cal  & Complete \\ \hline
$N_{\text{Cepheids}}$      & 85           &   2477       & 90           &     2754     \\
Median $N_{\text{obs}}$ & 24           &    3      & 12           &    4      \\
Period range (d)            & $6 - 134$    &   $1 - 134$       & $6 - 209$    &     $1 - 209$     \\
Mean period (d)             &     25.62         &   5.22       &       22.08       &   3.49       \\ \hline \hline
\end{tabular}
\end{table}

The data used in this work are from \textit{Spitzer}, which launched in 2003 and recently ceased operations in January 2020. Onboard \textit{Spitzer} was the Infrared Array Camera \citep[IRAC,][]{Fazio2004}, which initially operated simultaneously at four mid-infrared wavelengths (3.6 $\mu$m, 4.5 $\mu$m, 5.8 $\mu$m and 8.0 $\mu$m) during its ``cold'' mission. Once its cryogenic fuel was exhausted in 2009, \textit{Spitzer} entered its ``warm mission'' phase, operating only at the two shorter wavelengths, 3.6 $\mu$m and 4.5 $\mu$m. We combine data from both the cold and warm missions, using only the data at 3.6 $\mu$m and 4.5 $\mu$m (henceforth denoted as $[3.6]$ and $[4.5]$, respectively).

We use data from both the CHP sample (S11; M12; S16) of Cepheids and the \textit{Spitzer} SAGE survey (\citealt{Meixner2006, Gordon2011, Riebel2015}). While the CHP sample has fully-phased well-sampled observations for every Cepheid (number of observations $N_{\text{obs}} = 24$ for the LMC and 12 for the SMC), the SAGE data only provides a few observations (median number of observations are 3 and 4 for the LMC and SMC, respectively) that do not uniformly sample the light curve. Therefore, in order to obtain precise mean magnitudes of the Cepheids with sparsely sampled light curves, we have developed and applied a template light curve fitting procedure. The templates are constructed from the CHP Cepheids, hereafter referred to as the ``calibrating sample''. These templates are then fit to the sample of all known fundamental mode Cepheids in the Magellanic Clouds, hereafter referred to as the ``complete sample''. Sample statistics for the calibrating and complete samples are provided in Table \ref{tab:sampleStats}. 

\subsection{Calibrating sample}

The calibrating sample have photometric data taken as part of the CHP \citep{Freedman2011a}. Our sample consists of 85 LMC Cepheids (PI Freedman, Program IDs: 61000, 04, 05, 06, 07; S11) with periods ($P$) in the range $6 \leq P \leq 134$ days, and 90 SMC Cepheids (PI Madore, Program ID 70010; S16) with $6 \leq P \leq 209$ days. All observations were taken during \textit{Spitzer's} warm mission. Each observation consists of ten frames (five in each of the two wavelengths) where the Cepheid is in the field of view. Each frame has a frame time of 2.0 s, resulting in
an effective per-pixel integration time of 1.2 s. A medium-scale, five-point Gaussian dither pattern was used to reduce the effects of cosmic rays hitting the detector and also improves the S/N for the Cepheid.

Cepheids with $P \geq 12$ days have full-phase coverage, with approximately equally spaced observations over one pulsation cycle. These observations were taken using deterministic sampling in approximately $P/24$ steps for the LMC and $P/12$ steps for the SMC. A full discussion and justification of this observing strategy is given in the appendix of S11. To summarise, deterministic sampling requires fewer observations than random sampling to achieve the same precision, as the error on the mean follows $\sigma \sim 1/N_{\text{obs}}$ compared to random sampling where $\sigma \sim 1/\sqrt N_{\text{obs}}$, for $N_{\text{obs}}$ observations. For the deterministic sampling strategy and $N_{\text{obs}}$ described above, the error on the mean magnitude for each Cepheid reduces by a factor of $\sim5$ and $\sim3$ for the LMC and SMC, respectively, compared to random sampling. 

The shorter period Cepheids with $P \leq 12$ days could not be observed in such a tightly phased way, as this would overburden the \textit{Spitzer} schedule. Instead, each observation epoch was separated by $12 \pm 4$ days to ensure minimal overlap of phase points, thus reducing the likelihood of redundant data. The uncertainties on the magnitudes for these Cepheids follow $\sigma \sim 1/\sqrt N_{\text{obs}}$, as their light curve is randomly sampled.

\subsection{Complete sample}

\subsubsection{Sample selection}
\begin{figure}
    \centering
    \includegraphics[width=0.4\textwidth]{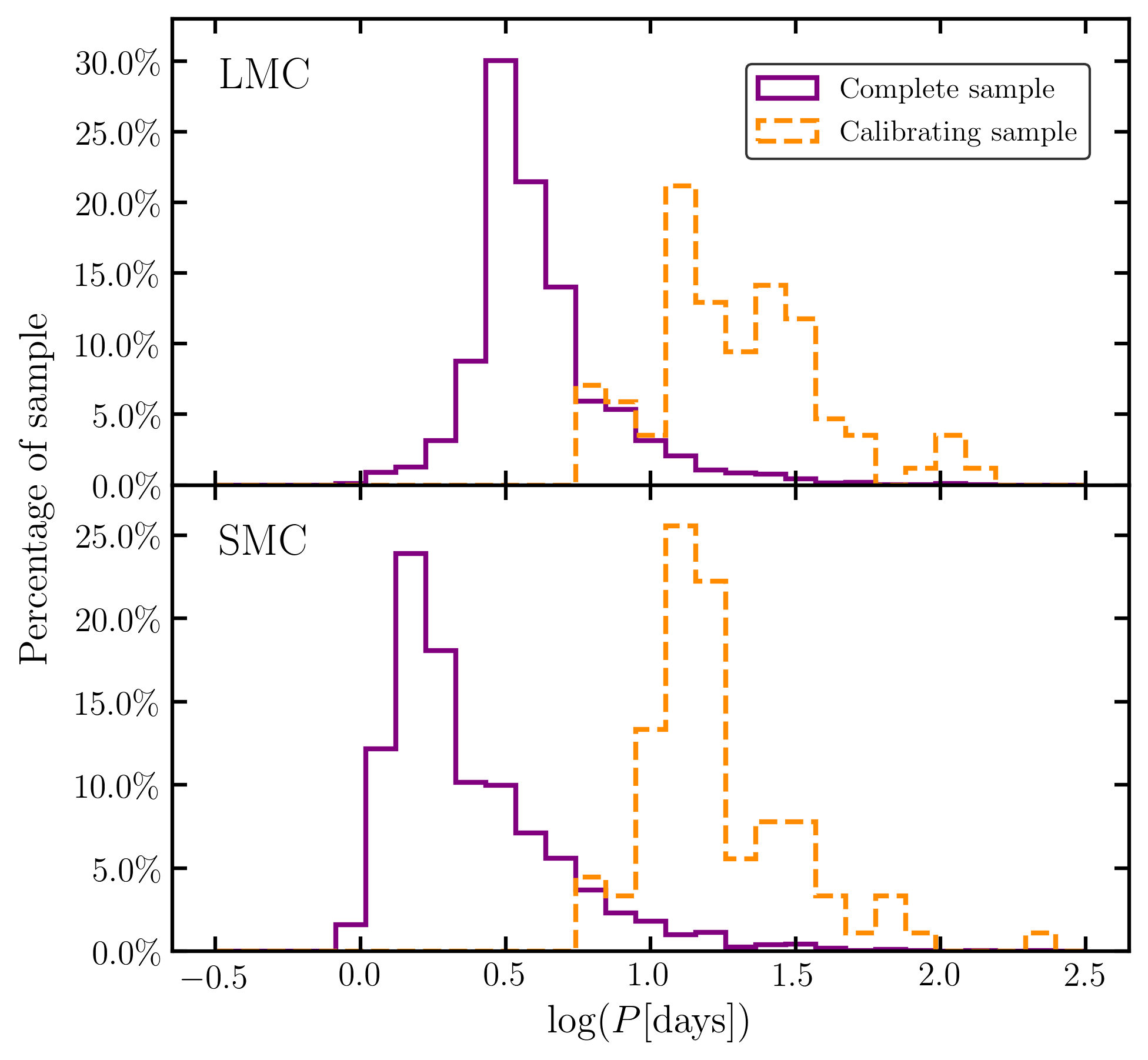}
    \caption{\textit{Top:} The period distribution of the complete (purple solid line) and calibrating (orange dashed line) samples for the LMC. \textit{Bottom:} Same as top panel, but for the SMC. In both panels, the periods are taken from the OGLE-IV catalogue \citep{Soszynski2017}.}
    \label{fig:period_distribution}
\end{figure}

Our complete sample contains 2477 LMC and 2754 SMC fundamental-mode Classical Cepheids. These are all the known Cepheids of this type found in the Magellanic Clouds, as given in the OGLE-IV catalogue of variable stars \citep{Soszynski2017}. Figure \ref{fig:period_distribution} shows the period distribution for both the complete and calibrating samples, where the periods have been taken from the OGLE-IV catalogue \citep{Soszynski2017}. For our work, we adopt these periods as they are extremely precise, with uncertainties on the order of $10^{-5}$s. For the complete sample, shown by the purple solid lines in Figure \ref{fig:period_distribution}, both galaxies have distributions that are skewed towards short periods, with average periods of 5.22 and 3.49 days for the LMC and SMC populations, respectively. As a result of the CHP selecting Cepheids that were bright, we see that the calibrating sample, shown by the orange dashed lines, occupies the long period end of the complete sample distribution. Along with the period, the OGLE catalogue also provides the RA $(\alpha)$ and Dec $(\delta)$ positions for each Cepheid. These coordinates are adopted in this work and used to determine the positions of the Cepheids in the SAGE images.
 
\subsubsection{Photometric data}

The mid-IR observations for the complete sample of Cepheids were taken as part of the \textit{Spitzer} SAGE legacy program. SAGE performed a uniform and unbiased imaging survey of the LMC \citep{Meixner2006}, SMC and Magellanic Bridge (MB) \citep{Gordon2011}. The LMC footprint is shown in figure 3 of \citet{Meixner2006}, while the SMC footprint is shown in figure 2 of \citet{Gordon2011}. The SAGE observations were carried out in 2005 during \textit{Spitzer's} cold mission at two epochs separated by approximately three months. A follow-up program to SAGE, SAGE-Var \citep{Riebel2015}, was conducted between August 2010 and June 2011 during the warm mission. These observations focused on the bar of the LMC and the central region of the SMC, providing up to an additional four epochs for these regions. Figures 1 and 2 of \citet{Riebel2015} show the SAGE-Var footprint for the LMC and SMC, respectively.

As a result of the extensive spatial coverage of the SAGE survey, each SAGE Cepheid has between two and six epochs of observation, depending on its location within its host galaxy, providing several phase points of each mid-IR light curve. Although SAGE performed its observations at all four \textit{Spitzer} wavelengths, we only use the $[3.6]$ and $[4.5]$ data so that the CHP and SAGE observations can be combined. In addition to providing observations for Cepheids in the complete sample, SAGE also yields additional observations for the calibrating Cepheids, since these Cepheids are a subset of the complete sample. 

\section{Photometry pipeline}
\label{sec:photometry}

We developed a fully-automated photometry pipeline to obtain intensity-averaged mean magnitudes and light curves for each Cepheid in our sample. While the photometry for the calibrating and complete samples was obtained using an identical pipeline to ensure consistent analysis, the determination of the intensity-averaged mean magnitudes and light curves differs between the two samples, due to the calibrating sample having vastly more observations than the complete sample. For the calibrating sample, we used GLOESS fitting (further details in Section \ref{sec:gloess}) to obtain the mid-IR mean magnitudes and light curves. For the complete sample, we developed and applied a template fitting procedure, as described in Section \ref{sec:temp_creation}, where the templates were constructed from the fully-phased well-sampled light curves of the calibrating sample. 

The input files are individual Basic Calibrated Data (CBCD) FITS images downloaded from the \textit{Spitzer} Heritage Archive (SHA). These images were initially processed using the \textit{Spitzer} processing pipeline, Version S19.2. To perform the photometry, we used the \verb|DAOPHOT/ALLSTAR/ALLFRAME| (\citealt{Stetson1987, Stetson1988, Stetson1994}) packages, kindly provided by Peter Stetson. Our photometry pipeline is written entirely in Python, using the \verb|pexpect| package to control the photometry packages. The code used for this work is publicly available\footnote{\url{https://github.com/abichown/DAOPHOT-Scripts}}. The remainder of this section outlines the main steps of our fully-automated photometry pipeline. 

\subsection{PSF photometry}
\label{sec:phot}

Initial star lists for each CBCD frame were created using the \verb|DAOPHOT| \verb|FIND| routine, with a $6\sigma$ and $4\sigma$ detection threshold for the LMC and SMC, respectively. A lower threshold value for the SMC was required as it is the more distant galaxy, resulting in fainter magnitudes which are harder to detect. Aperture photometry was then performed using the \verb|PHOT| routine to obtain rough instrumental magnitudes. A 3 pixel aperture radius and sky annulus from 12 to 20 pixels was used. This aperture radius is smaller than the standard IRAC Vega system value, as given in \citet[hereafter R05]{Reach2005}, but this difference is corrected for (see Section \ref{sec:cal}). A smaller aperture radius was preferred as some Cepheids have neighbours within a 10 pixel radius; if the larger aperture were used instead, these neighbours would contaminate the magnitude measurement of our Cepheid. The instrumental magnitudes computed at this stage are only initial estimates as some fields are fairly crowded, which causes the sky annulus to be contaminated. 

Next, we created a median image by combining all available CBCD images that contained the target Cepheid. Due to the scheduling and design of the observations, there are two fields for each Cepheid: one field contains the target Cepheid, referred to here as the ``on-target'' field, and the other does not. For the LMC Cepheids, there are 240 (24 epochs $\times$ 5 dithers $\times$ 2 wavelengths) on-target frames, while there are 120 (the same as for the LMC but with 12 epochs) on-target frames for the SMC Cepheids. Initial coordinate transformations between each of the on-target frames were determined using \verb|DAOMATCH|, which uses the 30 brightest stars in each image to compute these transformations. These initial transformations are then refined using \verb|DAOMASTER|, which uses all of the stars in the image to improve the \verb|DAOMATCH| transformations. The median image was created by passing these coordinate transformations through the \verb|MONTAGE2| package. 

The creation of the median image serves two purposes: it is used to create a master star list and is also used to create a Point Spread Function (PSF) model. Due to the stacking of a large number of frames, the signal-to-noise (S/N) ratio is much greater for the median image than for the individual CBCD frames. This high S/N ratio is beneficial for both the creation of a PSF model and obtaining a master star list. We obtained the master star list by performing \verb|FIND| and \verb|PHOT| on the median image. The detection threshold was increased to 20$\sigma$, as this was determined to be the optimal threshold which ensured the majority of stars were detected without a significant number of false detections. The master star list contains coordinates for all stars in the on-target field and is used in subsequent steps of the pipeline.

In addition to the master star list, a suitable PSF model was created from the median image. When the stellar field is crowded, PSF photometry is vastly superior to aperture photometry, as multiple PSF models can be fit simultaneously to groups of overlapping stars. The creation of the PSF model is the most crucial step of the photometry pipeline, as the model must be representative of all the stars in the image so that accurate magnitudes can be obtained. 

To create the PSF model, a sample of ``good'' stars are selected as input; these stars should be bright and relatively free from crowding. The 20 brightest stars in the master star list that were not saturated were initially selected as PSF stars. These stars are then accepted or rejected depending on their position in the median image; stars are rejected if they are too close to the edge of the frame, as there are less individual CBCD images at the edges of the median image, resulting in a lower S/N. A star was also rejected if it had any neighbours that could affect its use as a PSF star. After these tests, the resulting list of stars was passed to the \verb|PSF| routine to create the PSF model, with the smallest ``Chi'' value\footnote{DAOPHOT II User's Manual, Peter Stetson} being chosen as the final model. 

Using the \verb|ALLSTAR| package and providing it with the PSF model previously created, we obtain instrumental magnitudes via PSF photometry for each individual CBCD image. While the magnitudes from \verb|ALLSTAR| are reasonable, the uncertainties can be reduced using \verb|ALLFRAME|. \verb|ALLFRAME| works in a similar way to \verb|ALLSTAR|, except that the $(x,y)$ pixel positions of the stars are fixed. From \verb|ALLFRAME|, the pipeline outputs instrumental PSF magnitudes for all stars in every individual CBCD image, which includes the magnitude of the target Cepheid.

\subsection{Photometric calibration}
\label{sec:cal}

The instrumental PSF magnitudes from our photometry pipeline were calibrated to the standard IRAC Vega system (R05). In addition, we applied corrections based on a star's position in the image, as recommended in the IRAC handbook (Version 2.1.2). These steps are essential to ensure we obtain high precision photometry. To this end, four calibration steps were applied to our photometry results: an aperture correction, a zero point correction, a location correction, and a pixel-phase response correction.

We first calibrated our magnitudes to the standard IRAC Vega system (R05). We corrected the standard zero points used by \verb|DAOPHOT| to IRAC zero points of 17.30 and 16.81 for the $[3.6]$ and $[4.5]$ band, respectively. An aperture correction of 1.1132 for the $[3.6]$ band and 1.1126 for the $[4.5]$ band (S. Carey, private communication) were applied to convert to the same aperture system as R05.

To ensure we have high-precision photometry, we also corrected for effects due to the different responses of the pixels across the detector. These corrections are small but it is important to correct for all possible sources of photometric error. The two corrections that were applied account for both the star's $(x,y)$ pixel location in the image and also its precise location \textit{within} that pixel. The need to correct the magnitude based on the star's location in the array is due to two effects that are not corrected for by flat-fielding: the first being significant scattering and distortion in the images, and the second being a variation in the effective filter bandpass as a function of the angle of incidence\footnote{\url{https://irsa.ipac.caltech.edu/data/SPITZER/docs/irac/iracinstrumenthandbook/19/\#_Toc410728307}}.  

The location correction accounts for the star's $(x,y)$ position in the image. To apply the location correction, wavelength-specific correction images were downloaded from the \textit{Spitzer} website\footnote{\url{https://irsa.ipac.caltech.edu/data/SPITZER/docs/irac/calibrationfiles/locationcolor/}}. For each star, its $(x,y)$ position was obtained and the corresponding correction value from the correction image was also found. The measured flux of the star is then multiplied by this correction factor. These corrections are small, with the location-corrected magnitude being between 0.022 mag brighter and 0.067 mag dimmer than the uncorrected magnitude.

The pixel-phase correction accounts for the fact that a star's precise location within a pixel affects the response of that pixel to the star. This effect is believed to be caused primarily by intra-pixel quantum efficiency variations \citep{Mighell2008}. To correct for this effect, \verb|IDL| code is available\footnote{\url{https://irsa.ipac.caltech.edu/data/SPITZER/docs/irac/calibrationfiles/pixelphase/}}, which we converted to Python for consistency with the rest of our pipeline. A correction value is obtained and applied for each individual star. The pixel-phase corrected magnitude can be between 0.052 mag brighter and 0.042 mag dimmer than the uncorrected magnitude. Completion of this final step of the calibration procedure yields calibrated magnitudes on the standard IRAC Vega system for all calibrating Cepheids. To determine the average uncertainty on the magnitudes obtained, we took a randomly chosen frame and computed the average uncertainty of all its stars, not just the Cepheids. We found that the typical uncertainty on a single photometric point was $\sim 0.06$ mag.

\subsection{GLOESS fitting}
\label{sec:gloess}

The LMC and SMC Cepheids in the calibrating sample have 24 and 12 epochs of observation, respectively. Due to the five-dither pattern implemented in the scheduling of observations, each epoch has five magnitude measurements of the target Cepheid. We computed the intensity-averaged magnitude for each epoch as the uncertainty weighted mean of the fluxes, with weightings given by the inverse variance. The typical uncertainty on the magnitude at each epoch of observation is $\sim 0.03$ mag. These magnitudes were then phased using the pulsation periods from OGLE and the resulting light curves were fit with GLOESS, a Gaussian local estimation algorithm \citep{Persson2004}. GLOESS fits the data via a local regression method using a second-order polynomial. The algorithm produces a smooth curve through the $[3.6]$ and $[4.5]$ observations independently by interpolating between each two consecutive data points using a Gaussian window. The colour curve is computed as the difference between the $[3.6]$ and $[4.5]$ curves.

The mean magnitude of the Cepheid in each photometric band is determined as the intensity-averaged value of its GLOESS fitted curve across one entire pulsation cycle. Thus, the mean magnitude does not necessarily lie exactly halfway between the minimum and maximum value of the curve. Instead, its value depends on the fraction of one entire pulsation cycle that is spent at brighter magnitudes compared to the amount spent at fainter magnitudes. Similarly, the mean $[3.6] - [4.5]$ colour is computed as the average value of the colour curve across one complete cycle. For both the light curves and colour curve, the amplitude is computed as the maximum minus the minimum magnitude for the GLOESS curve, not the difference between the maximum and minimum phase points.

\begin{figure}
    \centering
    \includegraphics[width=0.48\textwidth]{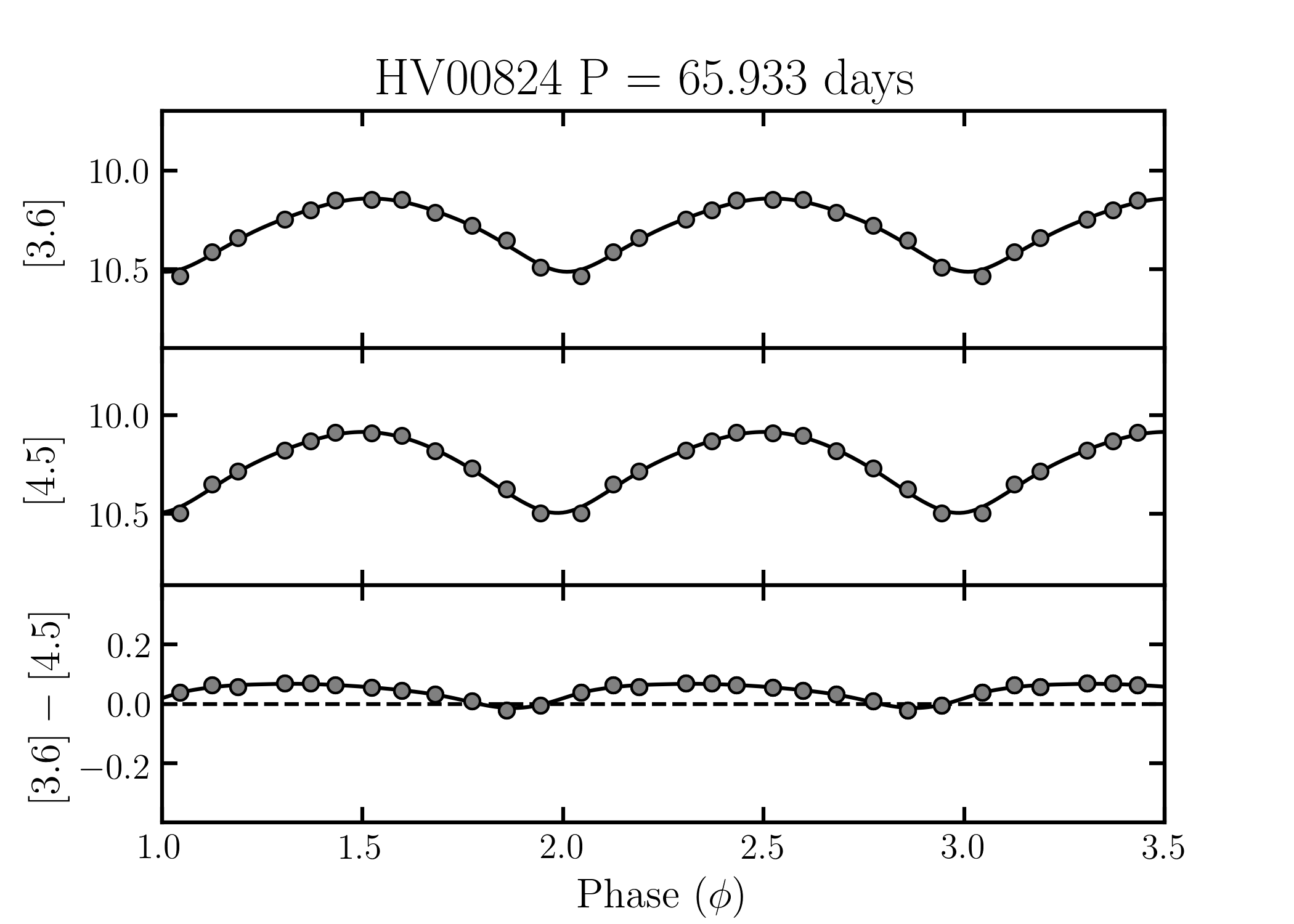}
    \caption{Example mid-IR light and colour curves produced with our photometry pipeline. This SMC Cepheid HV00824 has 12 equally spaced observations. The top two panels show the infrared light curves, with uncertainties comparable to the point size. The bottom panel shows the infrared colour curve, which illustrates the destruction and formation of CO in the Cepheid's atmosphere \citep{Scowcroft2016b}. The full set of light curves are available in Appendix \ref{sec:ap1}.}
    \label{fig:lightcurves}
\end{figure}

Example light and colour curves obtained are shown in Figure \ref{fig:lightcurves} for the SMC Cepheid HV00824. The top two panels of Figure \ref{fig:lightcurves} show the $[3.6]$ and $[4.5]$ light curves, while the bottom panel shows the $[3.6] - [4.5]$ colour curve. The uncertainties on the individual epoch measurements for the light curves are comparable to the size of the points. We confirm that the 12 equally-spaced observations allow for excellent determination of the Cepheid's light curve and hence its mean magnitude. In the bottom panel, we observe how the infrared colour curve illustrates the destruction and formation of CO in the Cepheid's atmosphere \citep{Scowcroft2016b}. 

We compared our mean magnitudes for these calibrating Cepheids to those from S11 and S16 for an identical sample of Cepheids. Combining the LMC and SMC samples, we computed the mean difference in magnitude as $0.015 \pm 0.030$ mag and $0.016 \pm 0.035$ mag for the $[3.6]$ and $[4.5]$ band, respectively. Differences between these results could be due to differences between our pipeline and the one used by S11 and S16, such as the use of different versions of the \textit{Spitzer} processing pipelines or that we used PSF photometry rather than PRF photometry. 

\subsection{SAGE photometry pipeline}

In order to process the SAGE data, we made small modifications to the photometry pipeline described in Sections \ref{sec:phot} and \ref{sec:cal} to account for the differences between the SAGE data and the CHP calibrating data. As much as possible, we kept the pipeline the same to ensure that the photometry was obtained consistently for both datasets. As with the calibrating sample, the SAGE photometry pipeline puts magnitudes on the standard IRAC Vega photometric system. The main modifications that were made are the following:

\begin{enumerate}
    \item Instead of working with the individual CBCD images and using \verb|MONTAGE2| to create a medianed image, we used \verb|MOPEX| \citep{Makovoz2005} to create mosaics of an entire AOR. This process conserves flux, hence we can use these mosaics to measure magnitudes.
    \item The \verb|DAOPHOT| parameters for the \verb|FIND| routine were updated to account for the SAGE images being mosaics rather than single CBCD images.
    \item The \verb|DAOPHOT| parameters for the \verb|PHOT| routine were updated to account for the change in pixel size. Previously, the pixel size of the individual CBCDs were $1\farcs2 \times 1\farcs2$, while the SAGE mosaic images have a pixel size of $0\farcs6 \times 0\farcs6$.  Rather than using a 3 px aperture with a $12-20$ px sky annulus, a 10 px aperture with $24 - 40$ px sky annulus was used.
    \item Many of the mosaics - particularly in the central regions of each galaxy - are extremely crowded and it is therefore not appropriate to obtain an individual PSF model for each mosaic. Therefore, we used the mosaic image of an uncrowded region in each galaxy to produce a PSF model for each wavelength. We also used the PSF stars that created each PSF model to compute the corresponding aperture correction.
    \item The calibrating sample were observed entirely during \textit{Spitzer's} warm mission. However, as the SAGE observations were carried out during both cold and warm missions, we updated the photometry pipeline to apply the appropriate calibration procedure depending on when the observations were taken. 
\end{enumerate}

\section{Template light curve fitting methods}

The calibrating Cepheids have well-sampled light curves in both the optical and mid-IR bands, making them excellent for use in creating template light curves. These templates can then be applied to the complete sample of Cepheids in the OGLE-IV catalogue in combination with our photometric measurements from the SAGE images, to obtain mean-light magnitudes in the mid-infrared. The template fitting procedure is a necessary step as the data available from SAGE are sparse and do not uniformly sample the Cepheid light curve. Applying a template to the available phase points produces significantly more precise mean magnitudes and colours compared to just averaging the available magnitude measurements. We now describe our template light curve construction process as well as the method for fitting the templates. 

\subsection{Template light curve construction}
\label{sec:temp_creation}

\begin{table}
\caption{\label{tab:MAR} The median absolute deviation (MAD) for various groupings of the calibrating sample.}
\centering
\begin{tabular}{ccc}
\hline 
Grouping method & Subsample characteristic             & MAD  \\ \hline
Host galaxy          & LMC                                  & 0.065      \\
                & SMC                                  & 0.079  \\ \hline
Wavelength      & $[3.6]$                              & 0.064     \\
                & $[4.5]$                              & 0.071  \\ \hline
Period        & $P \leq 10.88$                      & 0.133     \\
                & $10.88 < P \leq 13.63$             & 0.094   \\
                & $13.63 < P \leq 15.83$             & 0.071   \\
                & $15.83 < P \leq 22.65$             & 0.055  \\
                & $22.65 < P \leq 32.95$             & 0.052  \\
                & $P > 32.95$                         & 0.047     \\ \hline
Mid-IR colour   &  $ [3.6] - [4.5] \leq -0.03$          & 0.049     \\
               & $-0.03 < [3.6]-[4.5]  \leq -0.01$ & 0.059  \\
                & $-0.01 < [3.6]-[4.5]  \leq 0.00$  & 0.078  \\
                & $0.00 < [3.6]-[4.5]  \leq 0.02$   & 0.076  \\
                & $0.02< [3.6]-[4.5]  \leq 0.04$    & 0.077   \\
                & $[3.6]-[4.5] > 0.04$              & 0.093    \\ \hline \hline
\end{tabular}
\end{table}

Template light curves that are representative of a population of Cepheids can be obtained by compiling the observations from a large number of these Cepheids into a single light curve. In order to compile data in this way, a necessary step was to modify the way in which our data were phased. Thus far, the observations were phased using the periods from OGLE-IV \citep{Soszynski2015} using 

\begin{equation}
    \hspace{3cm}
    \phi = \text{mod}\Big(\frac{JD}{P}\Big),
\label{eq:phase}
\end{equation}

\noindent where $\phi$ is the phase, $JD$ is the Julian Date of the observation and $P$ is the period. However, phasing in this way does not allow for data from multiple Cepheids to be compiled together as there is no consistent zero-phase point. To set the zero point of their phasing, \cite{Soszynski2005} use the time of maximum brightness. However, due to the Hertzsprung progression \citep{Hertzsprung1926}, Cepheids with periods between 6 and 16 days exhibit a secondary bump in their optical light curves near maximum brightness. Therefore, \cite{Inno2015} argue that a more robust zero-phase point is the phase of the mean magnitude along the rising branch of the light curve, as it is not affected by this phenomena. Since approximately $11\%$ of our complete sample are within this period range, we choose to follow the approach of \cite{Inno2015}, anchoring our observations to the time of mean magnitude along the rising branch, $JD_{\text{mean}}$. Equation \ref{eq:phase} with this new zero-phase point is given by

\begin{align}
    \hspace{2.3cm}
    \phi = \text{mod}\Big(\frac{JD - JD_{\text{mean}}}{P}\Big).
\end{align}

\noindent With a consistent zero-phase point established, we followed the approach of \cite{Soszynski2005}, who fit templates to single-epoch $JHK$ observations, to create template light curves from the calibrating Cepheids. We first normalised the light curves using equation 2 of \cite{Soszynski2005},

\begin{equation}
    \hspace{2.6cm}
    T(\phi) = \frac{m_{\lambda}(\phi) - \langle m_{\lambda}\rangle} {A_{\lambda}},
\end{equation}

\noindent where $T(\phi)$ and $m_{\lambda}(\phi)$ are the normalised and observed magnitudes at phase $\phi$, respectively, $\langle m_{\lambda}\rangle$ is the intensity-averaged mean magnitude, and $A_{\lambda}$ is the amplitude of the light curve, all observed at wavelength $\lambda$ of either $[3.6]$ or $[4.5]$. Each normalised light curve has an amplitude of one and a mean magnitude of zero.

\begin{figure}
\centering
\includegraphics[width=0.48\textwidth]{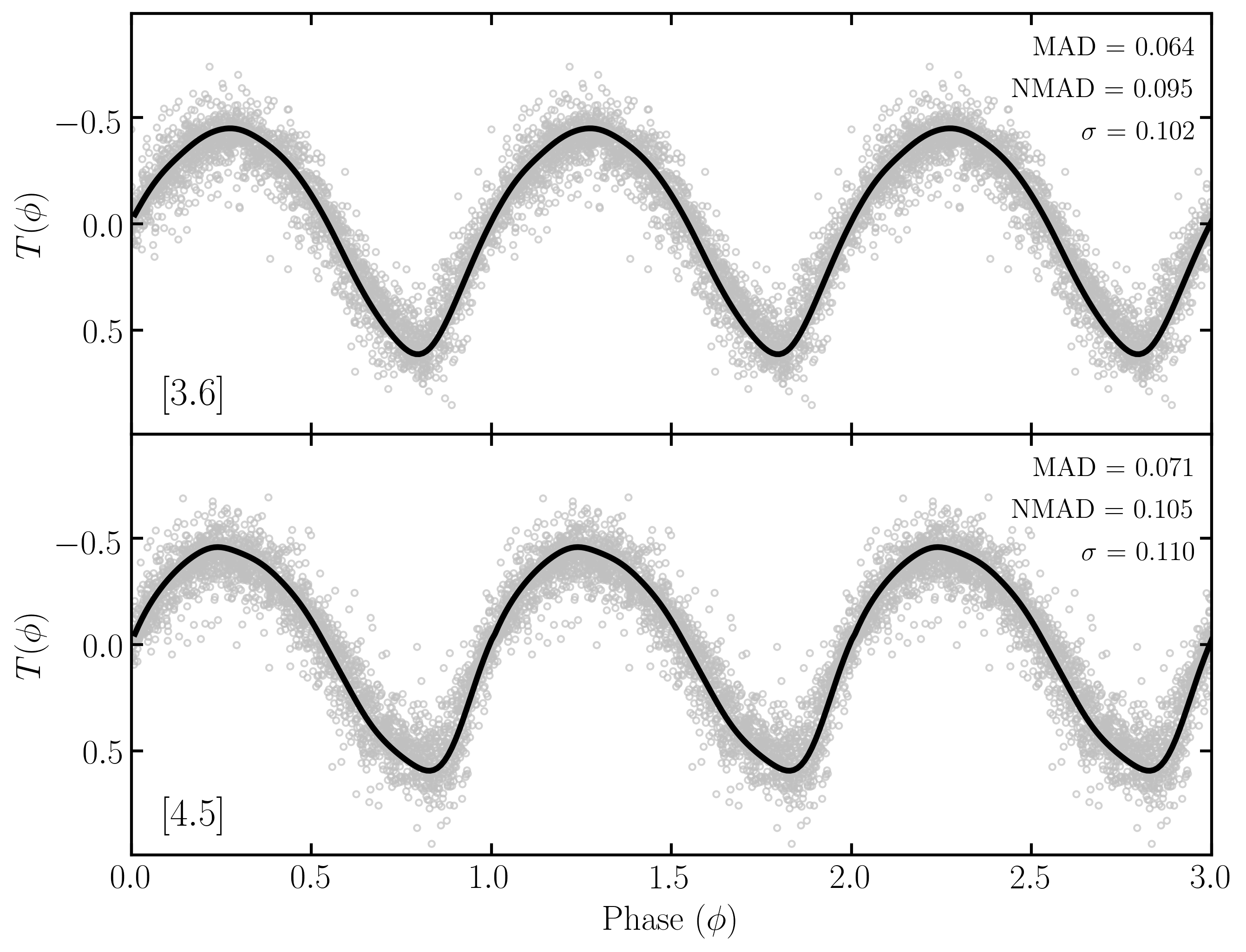}
\caption{\label{fig:templates} Template light curves produced following the normalisation procedure of
\protect\cite{Soszynski2005}
with a zero-phase point given by the phase of mean magnitude along the rising branch. Each normalised light curve has a mean magnitude of zero and an amplitude equal to one. Various binning methods were trialled, with splitting the observations by wavelength resulting in the smallest median absolute deviation (MAD). For a normal distribution, the normalised median absolute deviation (NMAD) should equal the standard deviation, $\sigma$. The top panel shows the $[3.6]$ normalised observations, while the bottom panel shows the $[4.5]$ normalised observations. In each panel, the black solid line  shows the fitted GLOESS curve to the normalised phase points, given by the grey circles.}
\end{figure}

To allow for the creation of multiple template light curves, the normalised light curves of the calibrating sample were grouped into subsamples based on various physical and observed properties of the Cepheids. Various grouping techniques have been used in the literature, including separating the sample by period \citep{Inno2015}, host galaxy \citep{Soszynski2005}, variable type \citep{Inno2015}, light curve shape \citep{Ripepi2016a}, and wavelength (\citealt{Soszynski2005,Inno2015,Ripepi2016a}). We tested grouping our normalised light curves by galaxy (LMC and SMC), wavelength ($[3.6]$ and $[4.5]$), period (6 approximately equally-populated bins) and by mid-IR colour (6 approximately equally-populated bins), fitting each group of normalised points with GLOESS. 

\begin{table*}
\caption{\label{tab:ogle_numbers} Percentage of Cepheids in the complete sample with $V$ and $I$ photometry from either OGLE-IV or OGLE-III.}
\begin{tabular}{ccccccc}
\hline
    & \multicolumn{3}{c}{$V$}       & \multicolumn{3}{c}{$I$}       \\
    & OGLE-IV  & OGLE-III & Neither & OGLE-IV  & OGLE-III & Neither \\ \hline
LMC & $92.4\%$ & $5.8\%$  & $1.8\%$ & $98.1\%$ & $0.6\%$  & $1.3\%$ \\
SMC & $95.0\%$ & $4.4\%$  & $0.6\%$ & $99.5\%$ & $0.5\%$  & $0.0\%$   \\ \hline \hline
\end{tabular}
\end{table*}

To quantify the optimal grouping method, we computed the median absolute deviation (MAD) between the observed $T(\phi)$ value and the template value at $\phi$. The MAD for each of the grouping methods is provided in Table \ref{tab:MAR}. While the long period Cepheids with $P > 32.95$ days have the smallest MAD value of 0.047 mag, we find that the MAD value monotonically increases as we move to shorter periods, with the shortest period group having a MAD of 0.133 mag. As observed in Figure \ref{fig:period_distribution}, the majority of our Cepheids in the complete sample would be in this shortest period group. Thus, on average we find that there is not an overall improvement when grouping by period compared to host galaxy or wavelength. For our templates we decided to group our Cepheids by wavelength only as it had the smallest average MAD value across the four grouping methods. The addition of separating further by galaxy, period or colour (or a combination of these) did not improve the results. Our final template light curves are shown in Figure \ref{fig:templates}.

\subsection{Cepheid optical-infrared parameter relations}

Applying a template light curve to a sparse set of observations requires knowledge of the appropriate scale and shift factors for amplitude and phase, respectively, to best fit the observations. However, due to the limited number of SAGE observations, we cannot compute the amplitude and phase directly from the infrared data alone. Instead, we used our calibrating sample to establish relationships between the optical and mid-IR Cepheid parameters, which are then used to infer the mid-IR amplitude and phase of mean magnitude along the rising branch for the complete sample.

\begin{figure*}
    \centering
    \includegraphics[width=\textwidth]{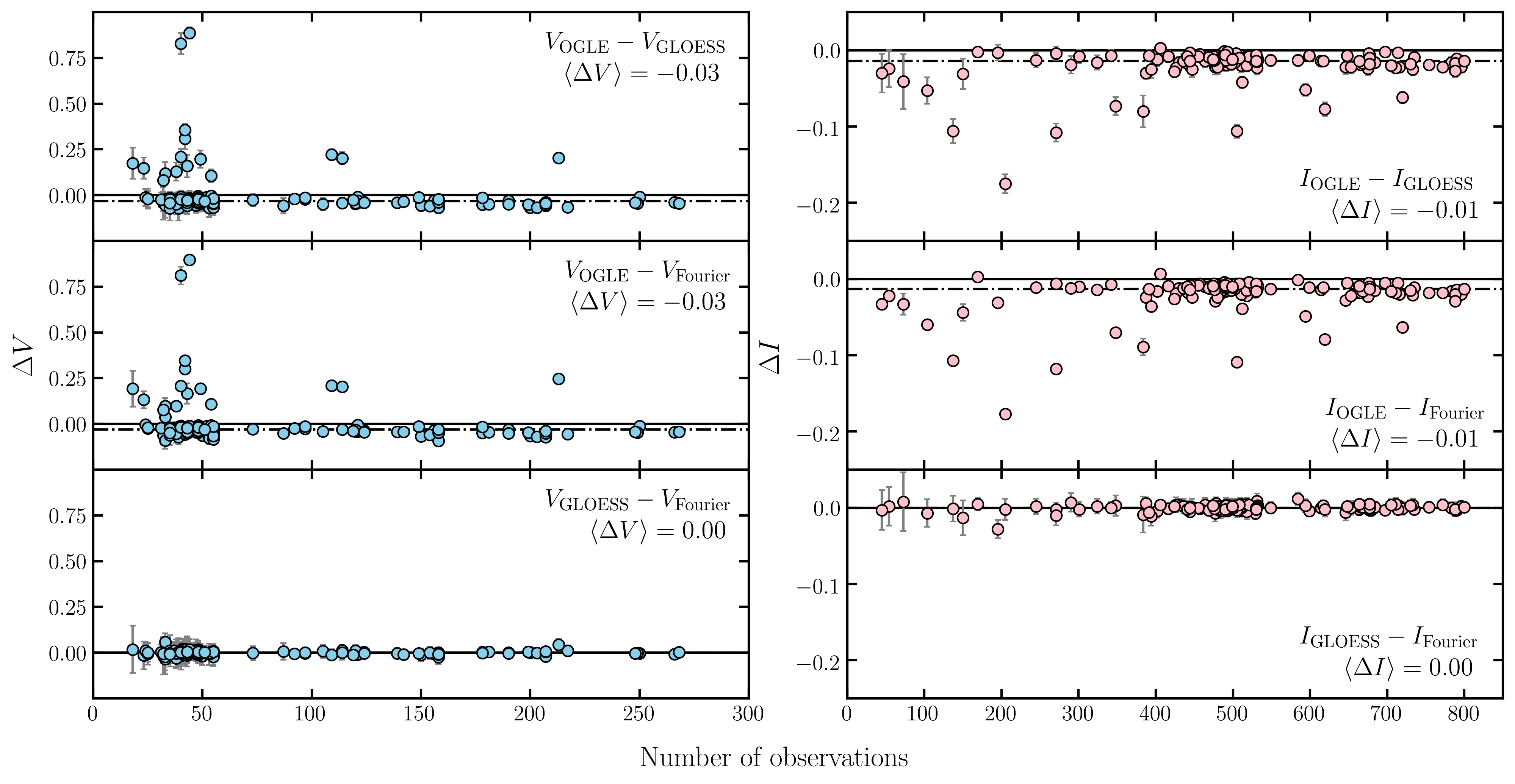}
    \caption{Comparison between the mean magnitudes obtained for the calibrating sample of Cepheids using GLOESS fitting, those provided in the OGLE-IV catalogue (using Fourier decomposition), and those obtained through our own Fourier decomposition. \textit{Left:} Mean magnitude comparison between GLOESS and the OGLE-IV catalogue (top), between the OGLE-IV catalogue and our own Fourier decomposition (middle), and between GLOESS and our own Fourier decomposition (bottom) for the $V$ band. Uncertainties are shown by the grey bars and were added in quadrature for the GLOESS and Fourier mean magnitudes. No uncertainties on mean magnitude are provided in the OGLE catalogue. \textit{Right:} Same as left panels but for the $I$ band. The solid line in each panel shows $\Delta \text{mag} = 0$ and the dash-dot line in each panel shows the median absolute magnitude difference $\langle \Delta \text{mag} \rangle$. } 
    \label{fig:mags_compare_gloess_fourier}
\end{figure*}

\begin{figure*}
    \centering
    \includegraphics[width=\textwidth]{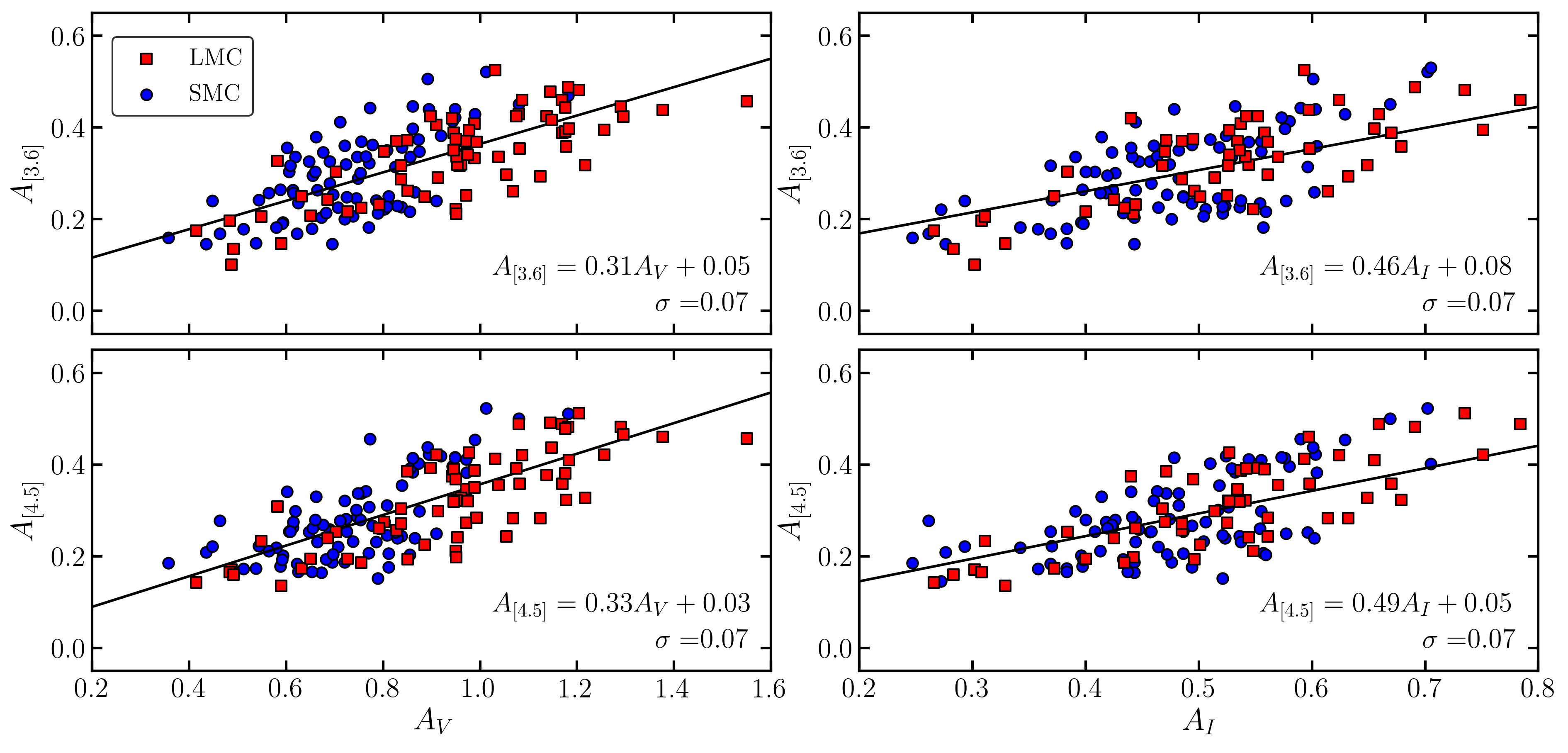}
    \caption{The amplitude relations used to infer the infrared light curve amplitude using the corresponding optical light curve amplitude. The left panels show the $[3.6]$ (top) and $[4.5]$ (bottom) amplitude relations using $V$ band data. The right panels show the $[3.6]$ (top) and $[4.5]$ (bottom) amplitude relations using $I$ band data. In every panel, the SMC Cepheids are shown by blue circles and the LMC Cepheids by red squares. The best fitting linear relation is shown by the solid line, with the derived parameters given in Table \ref{tab:amp_ratios}.}
    \label{fig:amp_relations}
\end{figure*}

\begin{figure}
    \centering
    \includegraphics[width=0.45\textwidth]{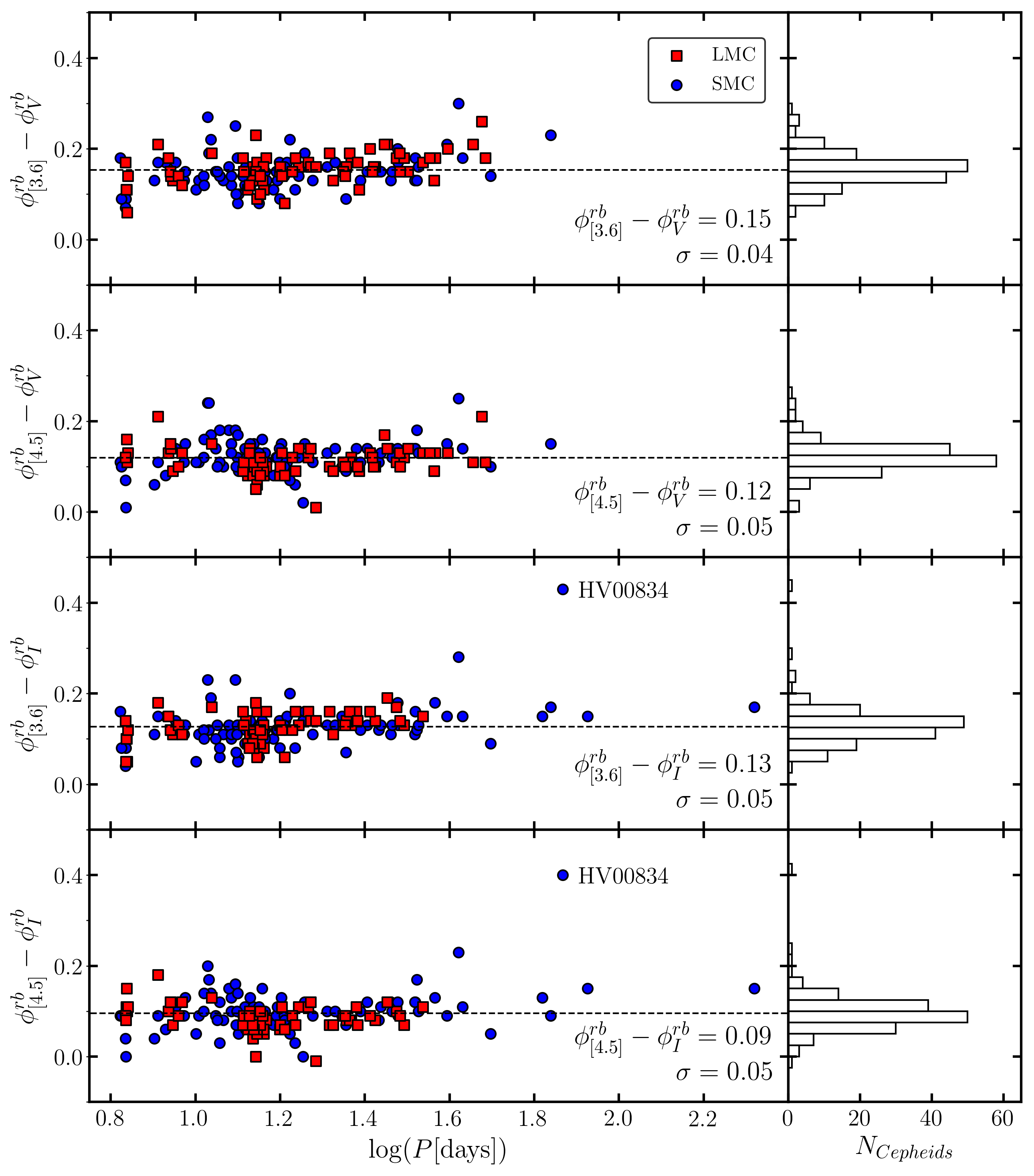}
    \caption{The phase relations used to infer a value of the mid-IR phase of mean magnitude along the rising branch, $\phi^{\text{rb}}$, using the equivalent optical phase. The top two panels show the $[3.6]$ (top) and $[4.5]$ (bottom) phase relations using $V$ band data. The bottom two panels show the $[3.6]$ (top) and $[4.5]$ (bottom) phase relations using $I$ band data. In each panel, the SMC Cepheids are shown by blue circles and the LMC Cepheids by red squares. The mean phase difference is shown by the dashed line.}
    \label{fig:phase_relations}
\end{figure}

\subsubsection{Optical data}
To establish optical-infrared parameter relations, we obtained all available photometric data in the $V$ and $I$ bands from the OGLE-IV catalogue\footnote{\url{http://ogledb.astrouw.edu.pl/~ogle/OCVS/}} for Magellanic Clouds Cepheids. For some Cepheids, the photometric data from the OGLE-III catalogue \citep{Udalski2008} was used as OGLE-IV measurements were not available. Table \ref{tab:ogle_numbers} shows the proportion of Cepheids with data from OGLE-IV and OGLE-III. The vast majority of Cepheids have OGLE-IV $I$ band photometry available ($98.1\%$ and $99.5\%$ for the LMC and SMC, respectively), with only a small percentage of Cepheids ($0.6\%$ and $0.5\%$ for the LMC and SMC, respectively) requiring the use of OGLE-III data. While all SMC Cepheids had photometric data in at least one of the two bands, no photometric data was available in either band for 13 LMC Cepheids. These Cepheids were identified to be the longest period, and hence brightest, Cepheids, which would cause the detector to saturate. For these Cepheids, the OGLE team used their Difference Image Analysis pipeline, which produces artificial objects in close vicinity to the Cepheid, in order to determine their magnitude (I. Soszy\'{n}ski, private communication).

Mean $V$ and $I$ band magnitudes and $I$ band amplitudes are already given in the OGLE-IV and OGLE-III catalogues, and were obtained by Fourier decomposition of the light curves. However, $V$ band amplitudes are not provided. As this parameter is required to derive our template fitting parameter relations, we decided to fit the $V$ data with GLOESS to determine its amplitude. In addition, since we use GLOESS fitting throughout this work, we chose to recompute the $V$ and $I$ magnitudes and $I$ band amplitudes using GLOESS, in order to maintain consistency within our work. 

In the two top panels of Figure \ref{fig:mags_compare_gloess_fourier}, we compare the mean magnitudes of our calibrating Cepheids computed by GLOESS fitting to those given in the OGLE-IV catalogue. In the $V$ band, we find a significant number of Cepheids with much brighter ($\Delta V \geq 0.10$ mag) GLOESS magnitudes than their OGLE-IV catalogue counterparts. For the most discrepant Cepheid HV00877 (OGLE-LMC-CEP-0461), the GLOESS and OGLE-IV catalogue magnitudes are 13.356 and 14.246 mag, respectively ($\Delta V = 0.890$ mag). Looking at the photometric data, we find that the magnitude given in the OGLE-IV catalogue is unfeasible, as the photometric data only range from $13.0 - 13.7$ mag. However, we find that the $V$ band magnitude given in the OGLE-III catalogue is 13.344 mag, which is consistent with our GLOESS magnitude ($\Delta V = -0.012$ mag). Aside from these Cepheids that have much brighter GLOESS magnitudes, all remaining Cepheids exhibit a systematic offset between the GLOESS and OGLE-IV magnitudes, with the OGLE-IV magnitudes being systematically brighter. This offset is also observed for the $I$ band. 

While a detailed analysis of these OGLE discrepancies is beyond the scope of the paper, we briefly discuss Cepheids with $\Delta V > 0.05$ mag and $\Delta I < -0.05$ mag. $40\%$ of these Cepheids (7 out of 17) were discrepant in both $V$ and $I$, $30\%$ have both $V$ and $I$ data but are only discrepant in one band, and the remaining $30\%$ have only single band data. Several of the discrepant Cepheids have a similar issue as HV00877, where the mean magnitude given in the OGLE-IV catalogue is outside the range of photometric measurements for that star. Some Cepheids have large amplitudes, in excess of one magnitude, such that their small discrepancies (of around 0.08 mag) are not as significant. Finally, several Cepheids have OGLE-IV mean magnitudes that are too bright or too faint, given the range of photometric measurements. For example, HV11211 has an OGLE-IV mean magnitude of 12.965 mag, which occurs right at the top of the light curve. At this time, we are unable to provide explanation for the cause of this discrepancy and further investigation is required.

In light of these magnitude discrepancies, we also fit the photometric data with a fourth order Fourier series of the form

\begin{equation}
\hspace{1.8cm}
   m(t) = m_0 + \sum_{n=1}^{4} a_{n} \cos \left( \frac{2\pi t}{P} + \phi_{n} \right),
\end{equation}

\noindent where $m(t)$ is the magnitude at time $t$, $m_0$ is the mean magnitude, and $a_n$ and $\phi_n$ are the amplitude and phase parameters of the $n^{\text{th}}$ harmonic, respectively. The differences between our Fourier-derived mean magnitudes and those provided in the OGLE-IV catalogue, which also uses Fourier decomposition, are shown in the middle panels of Figure \ref{fig:mags_compare_gloess_fourier}. Again, we find large discrepancies in mean magnitude for a substantial number of Cepheids, as well as a systematic offset for the remainder of the Cepheids. However, as shown in the bottom panels of Figure \ref{fig:mags_compare_gloess_fourier}, we find the GLOESS mean magnitudes and those derived from our own application of Fourier decomposition are consistent. 

\begin{table*}
\caption{\label{tab:amp_ratios} Amplitude relations of the form $A_{\text{IR}} = \alpha A_{\text{opt}} + \beta$, as derived from the LMC and SMC calibrating sample shown in Figure \ref{fig:amp_relations}. These relations are used in the template light curve fitting procedure to provide an estimate of the Cepheid's mid-IR amplitude $(A_{\text{IR}})$ given its optical band amplitude $(A_{\text{opt}})$.}
\begin{center}
\begin{tabular}{ccccccc}
\hline
$A_{\text{IR}}$  & \multicolumn{3}{c}{{[}3.6{]}} & \multicolumn{3}{c}{{[}4.5{]}} \\ 
$A_{\text{opt}}$ & $\alpha$ & $\beta$ & $\sigma$ & $\alpha$ & $\beta$ & $\sigma$ \\ \hline
$V$       & 0.310 $\pm$ 0.026     & 0.054 $\pm$ 0.022    & 0.07     &  0.334 $\pm$ 0.025     & 0.023 $\pm$ 0.022    & 0.07    \\
$I$       &  0.461 $\pm$ 0.049     & 0.077 $\pm$ 0.025    & 0.07    & 0.493 $\pm$ 0.046     & 0.047 $\pm$ 0.024    & 0.07     \\ \hline \hline
\end{tabular}
\end{center}
\end{table*}

As a result of the above analysis, we decided to use the mean magnitudes computed by GLOESS fitting. We chose to do this for a number of reasons. First, there are unexplained discrepancies between OGLE-III and OGLE-IV magnitudes for the same Cepheid. Second, we have verified that our GLOESS magnitudes are consistent with the magnitudes obtained through our own Fourier decomposition. Finally, to avoid introducing any possible systematic, we fit the optical data using GLOESS as our photometry pipeline fits the mid-IR data with GLOESS.

In the remainder of this work, any optical parameters used will be the GLOESS-derived values as opposed to the values provided in the OGLE catalogues. However, we do continue to use the periods from OGLE, as these are extremely accurate, with uncertainties on the order of $10^{-5}$s.

\subsubsection{Optical-infrared amplitude relation}
\label{sec:amp_scale}

In order to fit a template to sparse data, we must be able to scale our template by the appropriate amplitude that best fits the available observations. In Figure \ref{fig:amp_relations}, we show the relationship between Cepheid amplitudes in the various optical and mid-IR photometric bands; we find a positive ($R_{\text{pearson}} \sim 0.7$) correlation in each case. For each combination of bands, we performed an unweighted linear least squares fit, with the derived parameters and their uncertainties provided in Table \ref{tab:amp_ratios}. We find a dispersion of 0.07 mag for all relations. These relations allow us to estimate the mid-IR amplitude of a Cepheid based solely on its optical amplitude. 

\subsubsection{Optical-infrared phase relation}
\label{sec:phase_shift}

In addition to the amplitude, we must also know the phase at which the mean magnitude occurs along the rising branch, denoted $\phi^{\text{rb}}$, for the mid-IR light curve. To this end, we define the phase lag as the difference in $\phi^{\text{rb}}$ between the mid-IR and optical light curves of a Cepheid, given by

\begin{equation}
    \hspace{3.1cm}
       \phi_{\text{lag}} = \phi_{\lambda}^{\text{rb}} - \phi_{\xi}^{\text{rb}},
\end{equation}

\noindent where $\lambda$ is either $[3.6]$ or $[4.5]$ and $\xi$ is either $V$ or $I$.

\begin{table}
\caption{\label{tab:phase_relations} Phase lag relations of the form $\phi_{\text{lag}} = \phi_{\lambda}^{\text{rb}} - \phi_{\xi}^{\text{rb}}$, as derived from the LMC and SMC calibrating sample shown in Figure \ref{fig:phase_relations}. These relations are used to estimate a Cepheid's mid-IR $\phi^{\text{rb}}$ from its corresponding optical value. The standard deviation is given by $\sigma$.}
\centering
\begin{tabular}{ccccc}
\hline
$\xi$     & \multicolumn{2}{c}{$V$}        & \multicolumn{2}{c}{$I$}        \\
$\lambda$ & $\phi_{\text{lag}}$ & $\sigma$ & $\phi_{\text{lag}}$ & $\sigma$ \\ \hline
$[3.6]$   &     0.15                &    0.04      &            0.13         &     0.05     \\
$[4.5]$   &      0.12               &   0.05       &              0.09       &      0.05    \\ \hline \hline
\end{tabular}
\end{table}

In Figure \ref{fig:phase_relations}, we present the phase lags as a function of period for each optical and mid-IR band combination. We find that, with the exception of HV00834 which is highlighted in the bottom two panels of the Figure, the majority of the phase lags fall around a similar value, with the mean phase lag and dispersion for each combination of bands provided in Table \ref{tab:phase_relations}. These mean values are shown by the dashed line in each of the panels. 

We checked whether there was any dependence of the phase lags on pulsation period. We fit a linear relation of the form $\phi_{\text{lag}} = a\log(P) + b$ to the phase lags and found that for all band combinations the slope, $a$, was consistent with zero, verifying that there is no significant dependence on period. Similarly, we found no significant dependence on the host galaxy so the LMC (red squares) and SMC (blue circles) samples were combined. 

While HV00834 $V$ band photometry is not available, the $I$ band phase lags are discrepant from the general trend. Looking at the $I$ band light curve for this Cepheid, we found the observations to be very noisy, making it difficult to determine a reliable determination of $\phi_{I}^{\text{rb}}$. However, $I$ band photometry is also available from \cite{Moffett1998}\footnote{The ID for HV00834 is incorrectly identified as HV00824 in the \cite{Moffett1998} Vizier catalogue.}. When using these data, we find $\phi_{[3.6]}^{\text{rb}} - \phi_{I}^{\text{rb}} = 0.21$ and $\phi_{[3.6]}^{\text{rb}} - \phi_{I}^{\text{rb}} = 0.18$, which makes HV00834 consistent with the rest of the sample.

To summarise, the relations given by Tables \ref{tab:amp_ratios} and \ref{tab:phase_relations} allow us to estimate the mid-IR amplitude and phase of mean magnitude along the rising branch for a Cepheid, based solely on its optical data. Thus, we can use these relations to scale our normalised template light curves to the sparsely-sampled SAGE data to compute the mean magnitude for each Cepheid in our complete sample.

\subsection{Template fitting method}
\label{sec:temp_fitting_method}

Now that we have created our templates and can scale and shift them in amplitude and phase, we outline the process of fitting the templates to data. Rather than assuming the infrared amplitude and phase inferred using the results of the previous section to be undoubtedly true, we instead use these values as initial starting guesses and employ Markov Chain Monte Carlo (MCMC) methods via the \verb|emcee| Python package \citep{Foreman-Mackey2013} to find the optimal value of each parameter. The parameters to be determined for each Cepheid are $\theta = \{[3.6], A_{[3.6]}, \phi^{\text{rb}}_{[3.6]}, [4.5], A_{[4.5]}, \phi^{\text{rb}}_{[4.5]}\}$, resulting in a six-dimensional parameter space. Each Cepheid has a set of observations, $\mathcal{D} = \{\phi_{[3.6]}, m_{[3.6]}, \sigma_{[3.6]}, \phi_{[4.5]}, m_{[4.5]}, \sigma_{[4.5]}\}$.

As the template light curve is scaled by the optimal amplitude, shifted in phase by the optimal $\phi^{\text{rb}}$ and then shifted in magnitude by the optimal magnitude, it is necessarily the case that this optimal magnitude lies exactly halfway between maximum and minimum light. Therefore, instead of taking this optimal magnitude as our final value, we use the final fitted light curve to compute the intensity-averaged mean magnitude.  

We explored the parameter space using 50 walkers, with each walker performing 1000 burn-in steps followed by a further 10,000 steps. The initial starting amplitude and phase of the walkers are given by the relations provided in Tables \ref{tab:amp_ratios} and \ref{tab:phase_relations}, and an initial magnitude in each band is computed as the average of the available observations. If both $V$ and $I$ band data are available, then we prefer to use the $I$ band values to compute the initial parameter estimates, as these were derived from many more observations than the $V$ band (as shown in Figure \ref{fig:mags_compare_gloess_fourier}). If $I$ band data are unavailable, then we use the $V$ band data. For the 13 LMC Cepheids with no available optical data, we set the initial amplitude to be the average $I$ band amplitude of the complete sample ($0.39$ mag) and we set the initial phase to be $0.50$.

We impose a series of Gaussian and uniform priors on our parameters. The mid-IR amplitudes follow a Gaussian prior, centered on the initial estimate from Table \ref{tab:amp_ratios} with $\sigma$ given by the corresponding scatter on the relation. The mid-IR phase $\phi^{\text{rb}}$ follow a Gaussian prior centered on the initial estimate from Table \ref{tab:phase_relations} with the corresponding $\sigma$. The mid-IR mean magnitudes follow a Gaussian prior centered on an initial mean magnitude given by the average of the available observations, with $\sigma = 0.10$ mag chosen as this is the intrinsic width of the instability strip in the mid-IR. We also impose restrictions on the amplitude ratio, phase difference and magnitude ratio between the $[3.6]$ and $[4.5]$ bands, adopting uniform priors such that $A_{[3.6]}/A_{[4.5]} \sim \mathcal{U}[0.50,1.50]$, $\phi_{[3.6]}-\phi_{[4.5]} \sim \mathcal{U}[-0.20,0.20]$ and $m_{[3.6]}/m_{[4.5]} \sim \mathcal{U}[0.99,1.02]$. The justification for these priors is given in Appendix \ref{sec:ap2}.

For each MCMC step, we have a trial combination of our parameters (${A_{\text{trial}}, \phi^{\text{rb}}_{\text{trial}}, m_{\text{trial}}}$) for each mid-IR band, which are used to scale the appropriate template light curve. For the likelihood function, we use the standard $\chi^2$-squared minimisation to determine how well the scaled template and thus the current trial parameters fit to the observed data.

\begin{table}
\caption{\label{tab:mcmc_validation} Mean absolute differences between the mean magnitudes obtained from template-fitting (tmpl) to $N$ observations and those obtained from just computing the arithmetic average (arth) of the $N$ observations. The standard deviation for each value is given by $\sigma.$}
\begin{tabular}{cccccc}
\hline
            & $N$ & $\Delta \text{mag}_{\text{tmpl}}$ & $\sigma_{\text{tmpl}}$ & $\Delta \text{mag}_{\text{arth}}$ & $\sigma_{\text{arth}}$ \\ \hline
LMC \hspace{0.26cm}$[3.6]$ & 1   & 0.048                             & 0.040                  & 0.102                             & 0.064                  \\
            & 2   & 0.023                             & 0.018                  & 0.073                             & 0.060                  \\
            & 3   & 0.020                             & 0.016                  & 0.053                             & 0.042                  \\
            & 4   & 0.018                             & 0.017                  & 0.046                             & 0.035                  \\
            & 5   & 0.014                             & 0.012                  & 0.042                             & 0.030                  \\
            & 6   & 0.011                             & 0.009                  & 0.030                             & 0.026                  \\
\hspace{1cm}$[4.5]$     & 1   & 0.042                             & 0.031                  & 0.106                             & 0.061                  \\
            & 2   & 0.025                             & 0.021                  & 0.068                             & 0.061                  \\
            & 3   & 0.022                             & 0.018                  & 0.051                             & 0.039                  \\
            & 4   & 0.018                             & 0.015                  & 0.042                             & 0.034                  \\
            & 5   & 0.012                             & 0.009                  & 0.041                             & 0.031                  \\
            & 6   & 0.011                             & 0.009                  & 0.033                             & 0.028                  \\ \hline
SMC \hspace{0.26cm}$[3.6]$ & 1   & 0.046                             & 0.034                  & 0.091                             & 0.058                  \\
            & 2   & 0.029                             & 0.021                  & 0.062                             & 0.050                  \\
            & 3   & 0.023                             & 0.021                  & 0.049                             & 0.038                  \\
            & 4   & 0.014                             & 0.012                  & 0.043                             & 0.034                  \\
            & 5   & 0.013                             & 0.011                  & 0.037                             & 0.026                  \\
            & 6   & 0.011                             & 0.009                  & 0.028                             & 0.023                  \\
\hspace{1cm}$[4.5]$     & 1   & 0.043                             & 0.034                  & 0.080                             & 0.055                  \\
            & 2   & 0.026                             & 0.023                  & 0.060                             & 0.048                  \\
            & 3   & 0.020                             & 0.017                  & 0.049                             & 0.034                  \\
            & 4   & 0.016                             & 0.011                  & 0.040                             & 0.035                  \\
            & 5   & 0.013                             & 0.011                  & 0.033                             & 0.024                  \\
            & 6   & 0.011                             & 0.010                  & 0.028                             & 0.022                  \\ \hline \hline
\end{tabular}
\end{table}

\begin{figure*}
    \centering
    \includegraphics[width=0.8\textwidth]{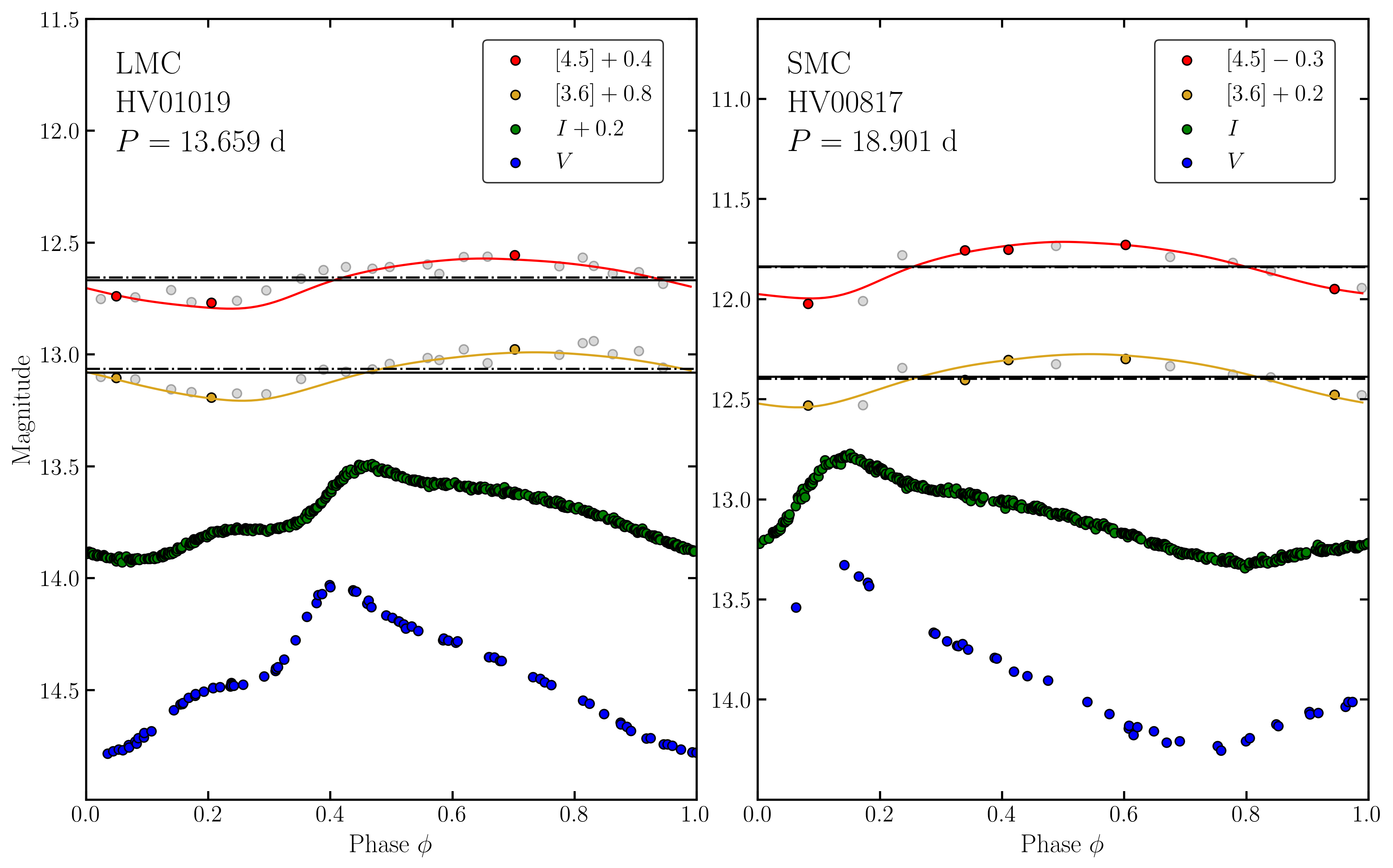}
    \caption{Examples of the mid-IR light curves obtained from our template fitting procedure on the calibrating sample. \textit{Left}: The LMC Cepheid HV01019 with three observations in each band that were used to derive template-fitted mean magnitudes. \textit{Right}: The SMC Cepheid HV00817 with five observations in each band that were used to derive template-fitted mean magnitudes. The observations used in the template fitting are given by the yellow and red points for the $[3.6]$ and $[4.5]$ band, respectively. The grey points are the remaining CHP observations that were not used in the template fitting process. The black solid lines are the ``true'' mean magnitudes obtained from GLOESS fitting to all available observations while the dashed-dot line shows the template-derived mean magnitude using the sparse set of observations. These results clearly demonstrate that our template fitting procedure can reproduce the mean magnitudes obtained from fully-sample light curves with only a handful of observations. The $V$ and $I$ band data from OGLE are also shown.}
    \label{fig:eg_validation}
\end{figure*}

\begin{figure*}
    \centering
    \includegraphics[width=0.9\textwidth]{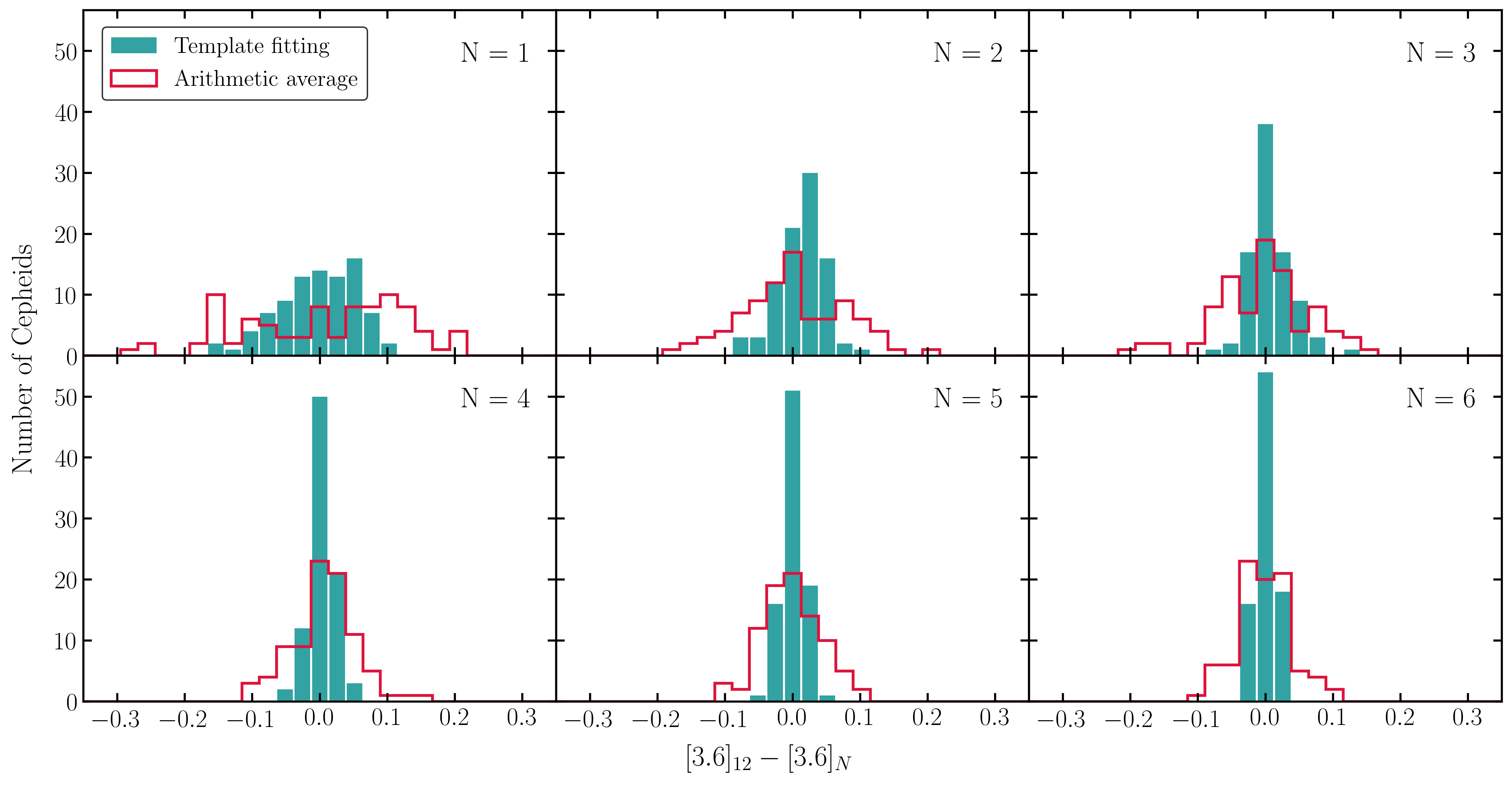}
    \caption{To validate the accuracy of our template fitting method, we created a mock dataset from the calibrating Cepheids that mimics the SAGE data, leaving between one and six observations to serve as our sparse dataset. In each plot, the blue, filled histograms show the mean magnitude differences between fitting these $N$ observations with our template light curves (blue) and those magnitudes obtained through GLOESS fitting to 12 fully-phased observations. Similarly, the red non-filled histogram in each panel shows the difference between GLOESS fitting to 12 fully-phased observations and those obtained from simply averaging the $N$ magnitudes. The plots show the distributions for $N = 1,2,3,4,5,6$ phase points. We only show the $[3.6]$ magnitudes for the SMC calibrating sample, but we find similar results for the SMC $[4.5]$ observations and the LMC sample as given in Table \ref{tab:mcmc_validation}.}
    \label{fig:template_test}
\end{figure*}

\begin{figure*}
    \centering
    \includegraphics[width=\textwidth]{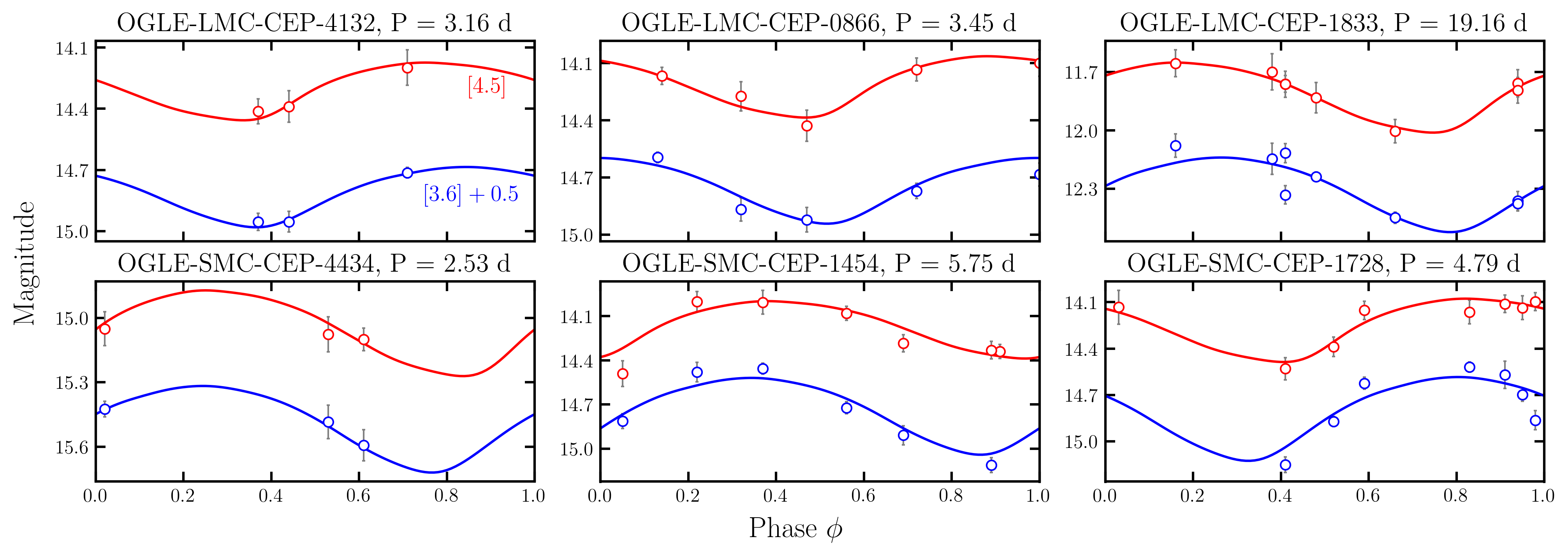}
    \caption{A selection of Cepheid light curves demonstrating the quality of the template fits to the photometric observations. The blue light curve and photometric points are for the $[3.6]$ band, while the red light curves and points are the $[4.5]$ band. These plots show the fits for various numbers of observations. The top panels show examples for LMC Cepheids, while the bottom panels show the results for the SMC.}
    \label{fig:temp_examples}
\end{figure*}

\begin{figure}
    \centering
    \includegraphics[width=0.4\textwidth]{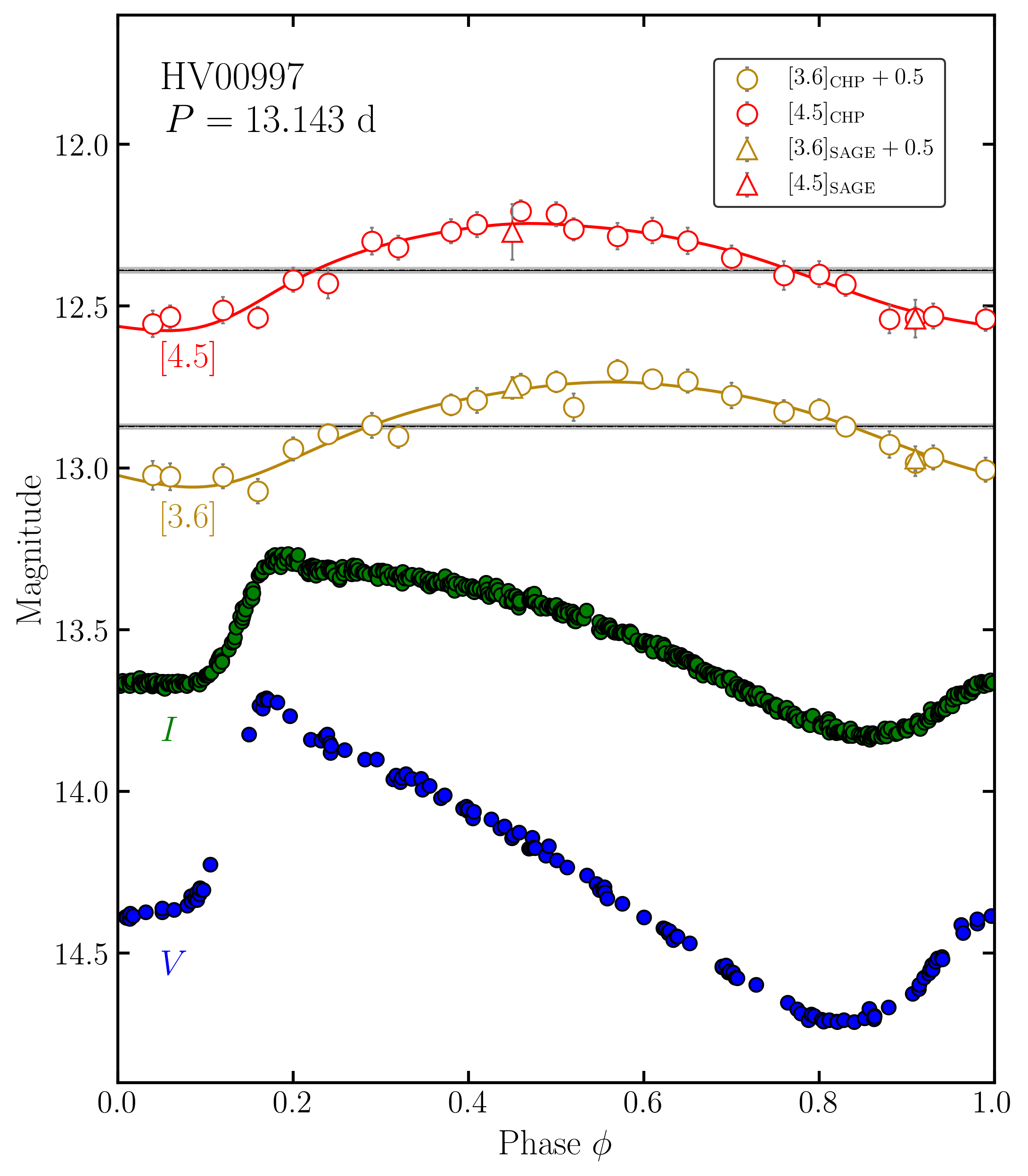}
    \caption{Example of the mid-IR light curves obtained from our template fitting procedure for the LMC calibrating Cepheid HV00997. Both the CHP (circles) and SAGE (triangles) observations were used in the template fitting process. For the $[3.6]$ and $[4.5]$ light curves, the solid black line is the mean magnitude obtained through GLOESS fitting while the black dashed-dot line shows the mean magnitude obtained through template fitting. The grey shaded region shows the uncertainty on the template-fitted mean magnitude. The $V$ and $I$ band light curves from OGLE are also plotted, given by the blue and green circles, respectively.} 
    \label{fig:chp_vs_sage_example}
\end{figure}

\subsection{Validation of template fitting method}
\label{sec:validation}

To assess the accuracy of our template fitting method, we used the observations of the calibrating sample to simulate a mock dataset that mimics typical SAGE data. As each Cepheid in the complete sample has a maximum of six observation epochs, we randomly selected between one and six observations for each of our calibrating Cepheids, which serves as their sparse set of observations. We allowed  observations to be selected multiple times, as it is likely that the SAGE observations will overlap in phase in some cases since the observations were not designed to be equally-spaced. While this test is not truly independent, as we are using the same data for both the validation of the method and the creation of the method itself, we see from the dispersion of our template fitting parameter relations in Tables \ref{tab:amp_ratios} and \ref{tab:phase_relations} that just using the mid-IR value inferred from a relation will not guarantee the optimal result. Therefore, this test does provide insight into the code's ability to find the best-fitting amplitude, phase and mean magnitude in each band.  

The template fitting code was run on this mock dataset for each calibrating Cepheid in the LMC and SMC and at both mid-IR wavelengths. Examples of the fitted mid-IR light curves are shown in Figure \ref{fig:eg_validation}. The observations used in the template fitting are given by the yellow and red points for the $[3.6]$ and $[4.5]$ band, respectively, while the grey points are the observations not used. The black solid lines are the ``true'' mean magnitudes obtained from GLOESS fitting to all available observations while the dashed-dot line shows the template-derived mean magnitude using the available observations, showing the excellent agreement between the two methods.

For each galaxy and wavelength, we computed the mean absolute difference between the template-derived magnitudes, $[3.6]_{N}$, for $N = 1,2,3,4,5,6$ observations and those obtained through GLOESS fitting to 12 or 24 equally-spaced observations. The results of this test are provided in Table \ref{tab:mcmc_validation}. An example of these results is shown by the blue filled distributions in Figure \ref{fig:template_test} for the SMC $[3.6]$ band. These results show that our template fitting method produces mean magnitudes that are in good agreement with the 
``true'' values obtained through 12 fully-phased observations and GLOESS fitting, with a precision that increases with the number of observations. For a single observation of an SMC Cepheid, the mean absolute difference was $\sim 0.05$ mag, reducing to $\sim 0.01$ mag for six observations. For three phase points, which is the most common number of observations for the SAGE data, the mean absolute difference was 0.02 mag.

It is also important to verify that our template fitting procedure provides more accurate mean magnitudes than just taking the arithmetic mean of the observations. The red, non-filled distributions in Figure \ref{fig:template_test} show the differences between GLOESS fitting to 12 fully-phased observations and those obtained from simply averaging the $N$ magnitudes. Comparing these distributions from the template fitting procedure clearly demonstrates the significant improvements provided by our technique. We observe that the template-fitted distributions are more concentrated around $\Delta\text{mag} = 0$, having significantly smaller dispersions than the arithemtic estimates. Thus, mean magnitudes derived from our template fitting method will be more precise than just taking the average of the available observations.

We conclude that our template fitting procedure is a suitable method for obtaining mid-IR mean magnitudes for Cepheids with few, sparse observations. We find that the fitting is more precise if the chosen observations provide good coverage of the entire pulsation cycle. However, we lose precision if the available observations are too close in phase, due to the fact that some observations become redundant. These redundant observations do not provide the fit with any additional information about the light curve as they are situated in the same region of the pulsation cycle as the other observations. These results populate the tails of the template-fitted distributions of Figure \ref{fig:template_test}.

\begin{table*}
  \caption{\label{tab:sage_catalogue} Final mid-IR catalogue for 2352 LMC and 2672 SMC fundamental mode Classical Cepheids in the Magellanic Clouds.}
  \centering
  \begin{tabular}{ccccccc}
\hline 
Galaxy & OGLE ID & Period (d)$^a$ & $[3.6]$ $^b$  & $[4.5]$ $^b$ & $E(B-V)$ $^c$ & Rejected?$^d$  \\ \hline \hline
LMC & OGLE-LMC-CEP-0002 & 3.118 & 14.833 $\pm$ 0.017 & 14.670 $\pm$  0.049 & 0.080 $\pm$ 0.017 & N \\
LMC & OGLE-LMC-CEP-0005 & 5.612 & 13.709 $\pm$ 0.031 & 13.624 $\pm$ 0.049 &  0.074 $\pm$ 0.015 & N\\
LMC & OGLE-LMC-CEP-0016 & 10.564 & 12.332 $\pm$ 0.032 & 12.314 $\pm$ 0.042 &  0.296 $\pm$ 0.004 & N\\
LMC & OGLE-LMC-CEP-0017 & 3.677 & 14.289 $\pm$ 0.054 & 14.199 $\pm$ 0.054 &  0.076 $\pm$ 0.012 & N\\
LMC & OGLE-LMC-CEP-0018 & 4.048 & 14.258 $\pm$ 0.026 & 14.173 $\pm$ 0.044 & 0.107 $\pm$ 0.024 & N\\

... & ... & ... & ... & ...  & ...  & ...                    \\
SMC & OGLE-SMC-CEP-4976 & 36.737 & 11.347 $\pm$ 0.011 & 11.361 $\pm$ 0.010 & 0.000 $\pm$ 0.000 & N\\
\hline
  \end{tabular}
\begin{itemize}
\small
\centering
\item[ ](This table is available in its entirety at \url{https://doi.org/10.15125/BATH-00915}.)
\begin{flushleft}
\item[\textbf{Notes:}]
\item[$^a$] Periods were taken from the OGLE-IV catalogue (\url{http://ogledb.astrouw.edu.pl/~ogle/OCVS/})
\item[$^b$] $[3.6]$ and $[4.5]$ are the intensity-averaged mean magnitudes in each mid-IR band and have not been corrected for reddening.
\item[$^c$] Colour excess values computed from the application of the \cite{Cardelli1989} and \cite{Indebetouw2005} reddening laws, assuming $R_{V} = 3.1$.
\item[$^d$] Flags whether a Cepheid is rejected from further analysis due to poor SAGE data.
\end{flushleft}
\end{itemize}
\end{table*}

\section{Results}
\label{sec:results}
Our template fitting method was performed on the SAGE photometric data for the complete sample of Cepheids. As the calibrating and complete samples overlap, we decided to incorporate the CHP observations for the calibrating sample into the template fitting for these Cepheids in order to utilise the vast number of observations available. Since the SAGE footprint does not cover as large an area as the OGLE footprint, a small percentage of the complete sample were observed in the optical bands but not in the mid-IR. We discarded from our sample the 125 LMC and 82 SMC Cepheids that did not have both $[3.6]$ and $[4.5]$ magnitudes. Our final complete sample contains 2352 LMC and 2672 SMC Cepheids.

\subsection{Mid-infrared Cepheid light curves}

In Figure \ref{fig:temp_examples}, we show a selection of Cepheid light curves fit using our template fitting procedure for different numbers of observations. Each plot shows both the $[3.6]$ (blue) and $[4.5]$ (red) fitted light curves, along with the available photometric observations used to fit the template. The top panels show examples from the LMC, while the bottom panels show light curves for a selection of SMC Cepheids. As can be seen from Figure \ref{fig:temp_examples}, our templates fit the available observations well. 

In Figure \ref{fig:chp_vs_sage_example}, we show the mid-IR and optical light curves for the LMC calibrating Cepheid HV00997. This Cepheid had mid-IR observations from both CHP and SAGE, and we incorporated all available observations into the template fitting procedure. Figure \ref{fig:chp_vs_sage_example} demonstrates the good agreement between the two photometry pipelines, as the additional SAGE observations (triangles) are consistent with the CHP observations (circles).

\begin{figure}
    \centering
    \includegraphics[width=0.4\textwidth]{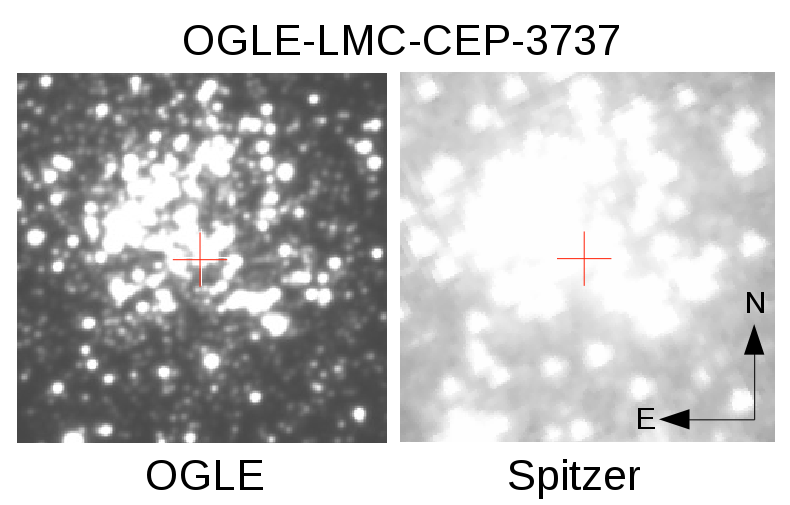}
    \caption{Optical and infrared images of the LMC Cepheid OGLE-LMC-CEP-3737. \textit{Left:} The OGLE-IV catalogue finding chart image, showing an area of $60\farcs0 \times 60\farcs$0. \textit{Right:} The \textit{Spitzer} mosaic on a log-scale for the same region containing the Cepheid. In both images, OGLE-LMC-CEP-3737 is at the centre of the image shown by the cross-hairs. We can see that accurate photometry for this Cepheid is not possible using these \textit{Spitzer} images.} 
    \label{fig:badData}
\end{figure}

\begin{figure*}
    \centering
    \includegraphics[width=\textwidth]{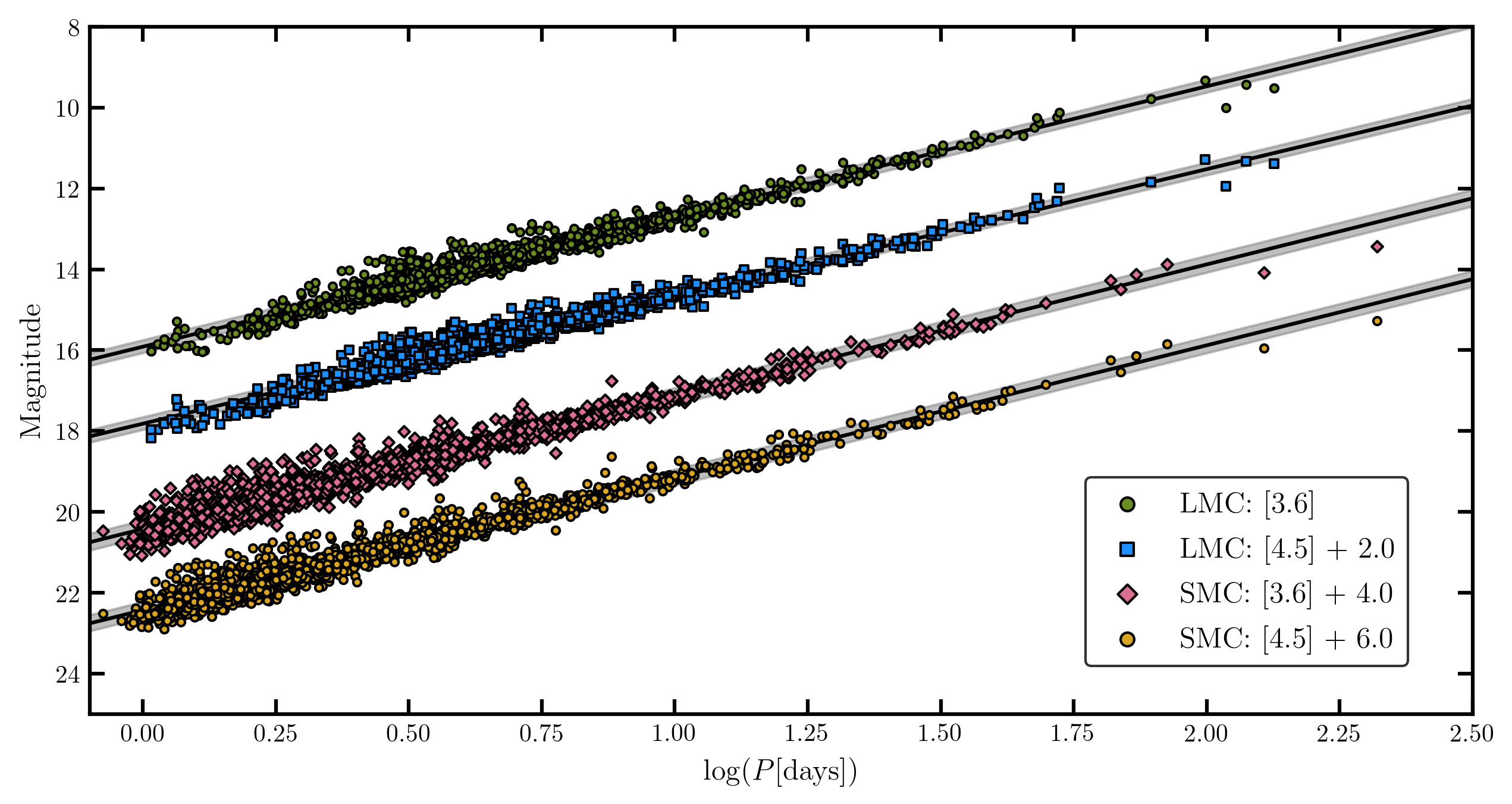}
    \caption{The mid-infrared Leavitt Law (LL) for the Magellanic Clouds incorporating all known fundamental mode Cepheids with \textit{Spitzer} observations. Cepheids with periods greater than $\log(P) > 1.90$ are classified as ultra-long period (ULP) Cepheids. The LMC $[3.6]$ and $[4.5]$ relations are shown by the green circles and blue squares, respectively. The SMC $[3.6]$ and $[4.5]$ relations are shown by the pink diamonds and yellow circles, respectively. Each LL is offset to improve clarity. A weighted least squares LL fit of the form given in Equation \ref{eq:LLform} was performed on each set of data points. The LL coefficients for these fits are given in Table \ref{tab:litPL}. The solid line shows the best-fitting relation for each dataset, with the filled region showing $\pm1\sigma$. Uncertainties associated with each individual Cepheid are smaller than the size of the points.}  
    \label{fig:LLfits}
\end{figure*}

\subsection{Mid-infrared mean magnitudes for Cepheids in the Magellanic Clouds}

Table \ref{tab:sage_catalogue} provides our mid-IR catalogue of fundamental mode Classical Cepheids. The colour excess $E(B-V)$ for each Cepheid was computed as described in Section \ref{sec:reddening} and is given in the final column. As described in Section \ref{sec:bad_data}, we removed a subset of our sample due to their poor mid-IR data. In doing so, our catalogue provides the mid-IR mean magnitudes for 2311 LMC and 2637 SMC Cepheids. The mean $[3.6]$ and $[4.5]$ magnitudes and associated uncertainties were computed from the application of our MCMC template fitting procedure described in Section \ref{sec:temp_fitting_method}. The magnitudes provided in Table \ref{tab:sage_catalogue} have not been corrected for reddening but can be corrected using $E(B-V)$ and Equations \ref{eq:reddening1} and \ref{eq:reddening2}. The average uncertainty on the mean magnitude for the LMC Cepheids is $0.032$ and $0.044$ mag in the $[3.6]$ and $[4.5]$, respectively. For the SMC, the average $[3.6]$ and $[4.5]$ uncertainties are $0.035$ and $0.041$ mag, respectively. The periods given here are taken from the OGLE-IV catalogue and have uncertainties on the order of $10^{-5}$s.

\subsubsection{Reddening correction}
\label{sec:reddening}

Although we are working at mid-IR wavelengths, where extinction is greatly reduced compared to optical wavelengths, we must still apply a reddening correction to our mid-IR magnitudes. While these corrections are small, they are a necessary step for ensuring precise mean magnitudes. To compute the size of the reddening correction for each Cepheid, we use one of two methods: application of the reddening law to multi-band photometry, or a nearest neighbours approach. The method used depends on whether OGLE-IV $V$ and $I$ band photometry is available for the Cepheid in question. 

For the vast majority of Cepheids ($\sim 95\%$ and $\sim 96\%$ for the LMC and SMC, respectively), we computed the reddening value by fitting appropriate reddening laws to multi-band photometry. We did not use the $[4.5]$ magnitude in this fit, as this band is known to be affected by the CO band-head \citep{Scowcroft2016b}. We fit the reddening laws of \cite{Cardelli1989} and \cite{Indebetouw2005} to the $V$, $I$ and $[3.6]$ magnitudes of each Cepheid by minimising the dispersion of the distance moduli around the values predicted by the reddening law. The distance modulus in each of the three bands was computed by adopting literature LLs as fiducial. For the $V$ and $I$ band, we used the relations given in \cite{Fouque2007}, while for the $[3.6]$ band we used the relation given in \cite{Monson2012}.

For the remaining Cepheids that did not have $V$, $I$ and $[3.6]$ magnitudes available, we computed the reddening as the average reddening value of the star's nearest neighbours. We set the search radius to be $\sqrt{1.2}/2 \sim 0.55$ deg and $\sqrt{0.22}/2 \sim 0.23$ deg for the LMC and SMC, respectively. These values were chosen as, when creating reddening maps for the LMC and SMC using OGLE-IV Cepheids, \cite{Joshi2019} divided each galaxy into segments with a mean spatial resolution of $1.2$ deg$^2$ and $0.22$ deg$^2$, respectively. The SMC was divided into smaller segments than the LMC as it contains a higher density of Cepheids in a smaller region of the sky.

The colour excess $E(B-V)$ for each Cepheid is given in the final column of Table \ref{tab:sage_catalogue}. The mean magnitudes provided in Table \ref{tab:sage_catalogue} have not been corrected for reddening. Reddening-corrected mid-IR mean magnitudes are computed using the extinction equations given in \cite{Monson2012},  

\begin{align} 
\hspace{2.4cm} A_{[3.6]} = 0.203 \times E(B-V) \label{eq:reddening1} \\ 
\hspace{2.5cm} A_{[4.5]} = 0.156 \times E(B-V),
\label{eq:reddening2}
\end{align}

\noindent where $A_{[3.6]}$ and $A_{[4.5]}$ are the extinction values in the $[3.6]$ and $[4.5]$ band, respectively. These equations are based on the reddening laws of \cite{Cardelli1989} and \cite{Indebetouw2005}, and assume $R_{V} = 3.1$.

\subsubsection{Bad data rejection}
\label{sec:bad_data}

To clean our data, we performed a preliminary unweighted least-squares fit of the LL to the dereddened magnitudes 
\begin{equation}
\hspace{2.5cm}
m = \alpha(\log(P) - 1.0) + \beta,
\label{eq:LLform}
\end{equation} 
where the $(\log(P) - 1.0)$ term is preferred over just a $\log(P)$ term as it reduces the correlation between the slope and the intercept. This fit was carried out for both galaxies and at both mid-IR wavelengths. We performed a visual inspection of the original images for stars which lay more than $3.5\sigma$ from the LL to check for potential contamination. In all cases, these Cepheids were either blended with a nearby companion or were part of a large unresolved `blob' of stars causing erroneously bright magnitude measurements. Figure \ref{fig:badData} shows an example of this effect for the LMC Cepheid OGLE-LMC-CEP-3737. As a result, we removed all Cepheids with such an effect in either of the two bands. In total, 41 LMC and 35 SMC Cepheids were discarded, resulting in a cleaned sample of 2311 LMC and 2637 SMC Cepheids.

\subsection{The mid-infrared Leavitt Law for the Magellanic Clouds}

Using our mid-IR catalogue in Table \ref{tab:sage_catalogue}, we show the dereddened  $[3.6]$ and $[4.5]$ LLs for the LMC and SMC in Figure \ref{fig:LLfits}. We perform a standard inverse-variance weighted least squares fit of the form given in Equation \ref{eq:LLform} for both bands and both galaxies. Table \ref{tab:litPL} gives the slope $(\alpha)$ and intercept $(\beta)$ values, their associated uncertainties, and the standard deviation of the fit. We stress that the slope and zero point values given here are not our final values. We devote the entirety of Section \ref{sec:LLdependence} to investigating the dependence of the LL coefficients on various characteristics of the sample. Our final adopted fits and their zero point calibration are given in Section \ref{sec:finalLL}. 

\section{Discussion}

\subsection{Dependence of the Leavitt Law on sample characteristics}
\label{sec:LLdependence}

\begin{table*}
\caption{\label{tab:litPL} The mid-IR LL for the Magellanic Clouds, employing various period cuts. A weighted least squares fitting of the form $m = \alpha(\log(P)-1.0) + \beta$ was performed with standard inverse variance weightings.}
\centering
\begin{tabular}{ccccccc}
\hline \hline
Galaxy & Wavelength & Period Range          & $\alpha$           & $\beta$            & $\sigma$ & $N$                       \\ \hline
LMC  & $[3.6]$    & No cut$^a$ & $-3.220 \pm 0.008$   & $12.694 \pm 0.003$   & 0.156    & 2311                   \\  
 &  & $0.8 < \log(P) < 1.8$ & $-3.241 \pm 0.024$   & $12.684 \pm 0.007$   & 0.137    & 420     \\
&  & $0.8 < \log(P_{\mathrm{cal}}) < 1.8$ $^b$ & $-3.285 \pm 0.060$   & $12.710 \pm 0.020$   & 0.107    & 80    \\
&  & $\log(P) > 0.4$ & $-3.213 \pm 0.009$   & $12.693 \pm 0.003$   & 0.155    & 2099      \\
&  & $\log(P) < 1.5$ & $-3.245 \pm 0.009$   & $12.687 \pm 0.004$   & 0.156    & 2290      \\
 \cline{2-7}
& $[4.5]$    & No cut$^a$ & $-3.147 \pm 0.008$   & $12.670 \pm 0.003$   & 0.158    & 2310                   \\  
&  & $0.8 < \log(P) < 1.8$ & $-3.140 \pm 0.026$   & $12.660 \pm 0.008$   & 0.136    & 420       \\
&  & $0.8 < \log(P_{\mathrm{cal}}) < 1.8$ $^b$ & $-3.219 \pm 0.065$   & $12.693 \pm 0.022$   & 0.116    & 80     \\
&  & $\log(P) > 0.4$ & $-3.134 \pm 0.008$   & $12.668 \pm 0.003$   & 0.154    & 2099      \\
&  & $\log(P) < 1.5$ & $-3.158 \pm 0.009$   & $12.669 \pm 0.003$   & 0.157    & 2290   \\
 \hline
SMC    & $[3.6]$    &   No cut$^a$  & $-3.272 \pm 0.009$ & $13.150 \pm 0.005$ &    0.198      & 2637        \\  
& &   $0.8 < \log(P) < 1.8$  & $-3.222 \pm 0.039$ & $13.147 \pm 0.010$ &    0.146      & 267               \\ 
& &   $0.8 < \log(P_{\mathrm{cal}}) < 1.8$ $^b$  & $-3.228 \pm 0.083$ & $13.150 \pm 0.023$ &    0.147     & 85                   \\ 
& &   $\log(P) < 0.4$  & $-3.313 \pm 0.057$ & $13.137 \pm 0.045$ &    0.221      & 1644                   \\  
 & &  $\log(P) > 0.4$   & $-3.197 \pm 0.014$ & $13.144 \pm 0.006$ &    0.174      & 993                     \\ 
  \cline{2-7}
 & $[4.5]$    &   No cut$^a$  & $-3.275 \pm 0.008$ & $13.154 \pm 0.005$ &    0.207      & 2637     \\  
& &   $0.8 < \log(P) < 1.8$  & $-3.189 \pm 0.039$ & $13.142 \pm 0.011$ &    0.150      & 267                     \\ 
& &   $0.8 < \log(P_{\mathrm{cal}}) < 1.8$ $^b$  & $-3.167 \pm 0.085$ & $13.135 \pm 0.025$ &    0.150     & 85                     \\ 
& &   $\log(P) < 0.4$  & $-3.381 \pm 0.055$ & $13.094 \pm 0.043$ &    0.220      & 1644                    \\  
 & &  $\log(P) > 0.4$   & $-3.192 \pm 0.014$ & $13.144 \pm 0.005$ &    0.171      & 993                   \\
\hline \hline
\end{tabular}
\begin{flushleft}
\begin{itemize}
\small
\item[ ]\textbf{Notes:}
\item[ ]$^a$ These fits are given by the solid lines in Figure \ref{fig:LLfits}.
\item[ ]$^b$ These fits used only the calibrating sample which have fully-phased observations of their light curve.
\end{itemize}
\end{flushleft}
\end{table*}

The large number of Cepheids in our sample allows for the exploration of selection effects on the LL coefficients. Rather than showing here the analysis for both the LMC and SMC at both $[3.6]$ and $[4.5]$ bands, we choose to focus our attention on the LMC $[3.6]$ relation, as it is the relation that is primarily used for distance measurements. We investigate the dependence of the LL coefficients on the number of observations for each Cepheid, the mean magnitude uncertainties for each Cepheid, the period distribution of the sample, and whether or not the relation is more appropriately described by a non-linear relation.

\subsubsection{Slope and zero point dependence on the number of observations}

\begin{figure}
    \centering
    \includegraphics[width=0.45\textwidth]{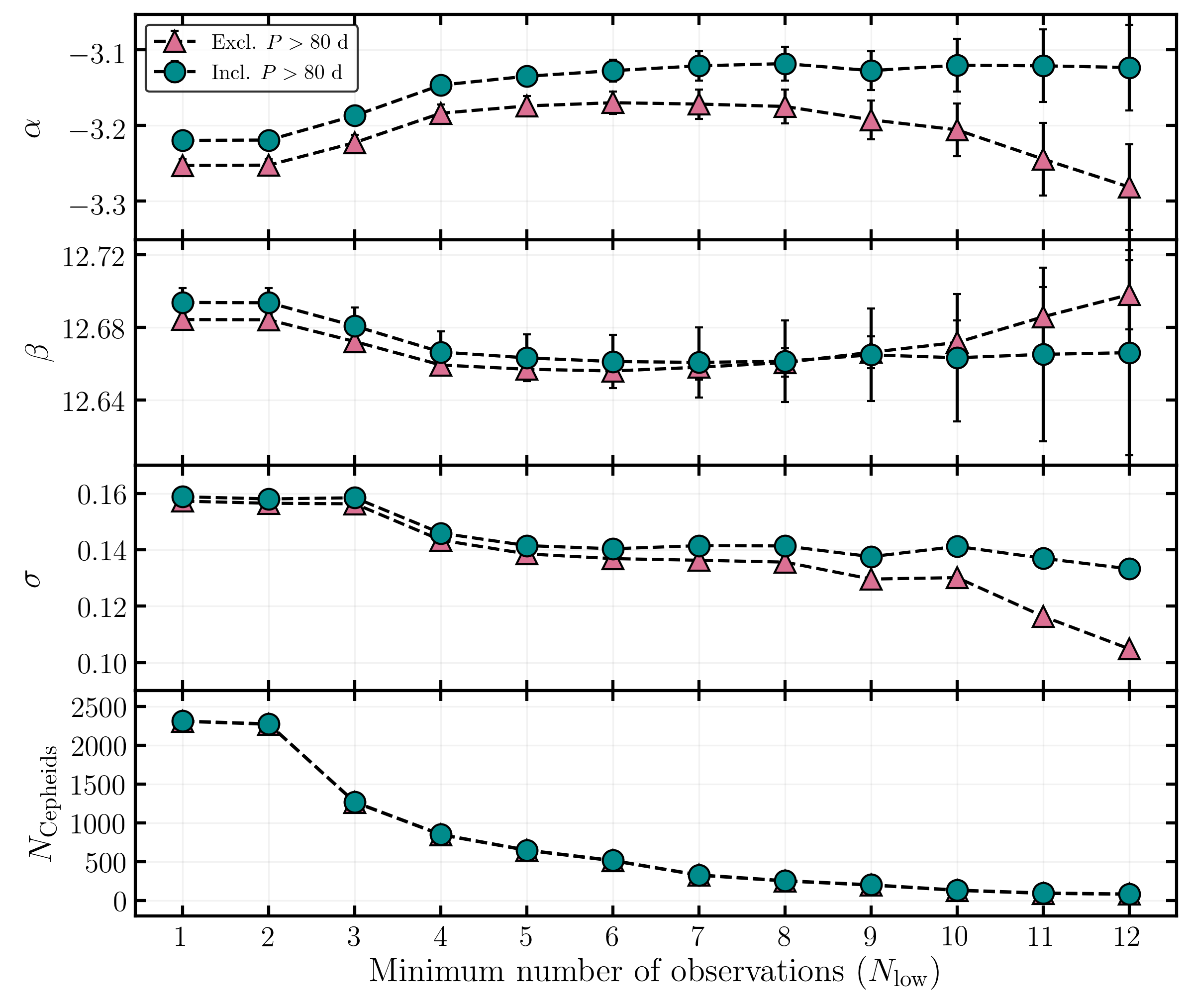}
    \caption{The dependence of the slope $(\alpha)$, zero point $(\beta)$, dispersion $(\sigma)$ and sample size $(N_{\text{Cepheids}})$ of the LMC $[3.6]$  LL on the number of observations available for each Cepheid. The analysis is performed on the entire sample of Cepheids (blue circles) and also when the four $P > 80$ days Cepheids have been removed (pink triangles). Both samples are shown in the $N_{\text{Cepheids}}$ panel (bottom). However, the two samples only differ by the four ULP Cepheids, causing the data to overlap.}
    \label{fig:nObs_effect}
\end{figure}

\begin{figure}
    \centering
    \includegraphics[width=0.45\textwidth]{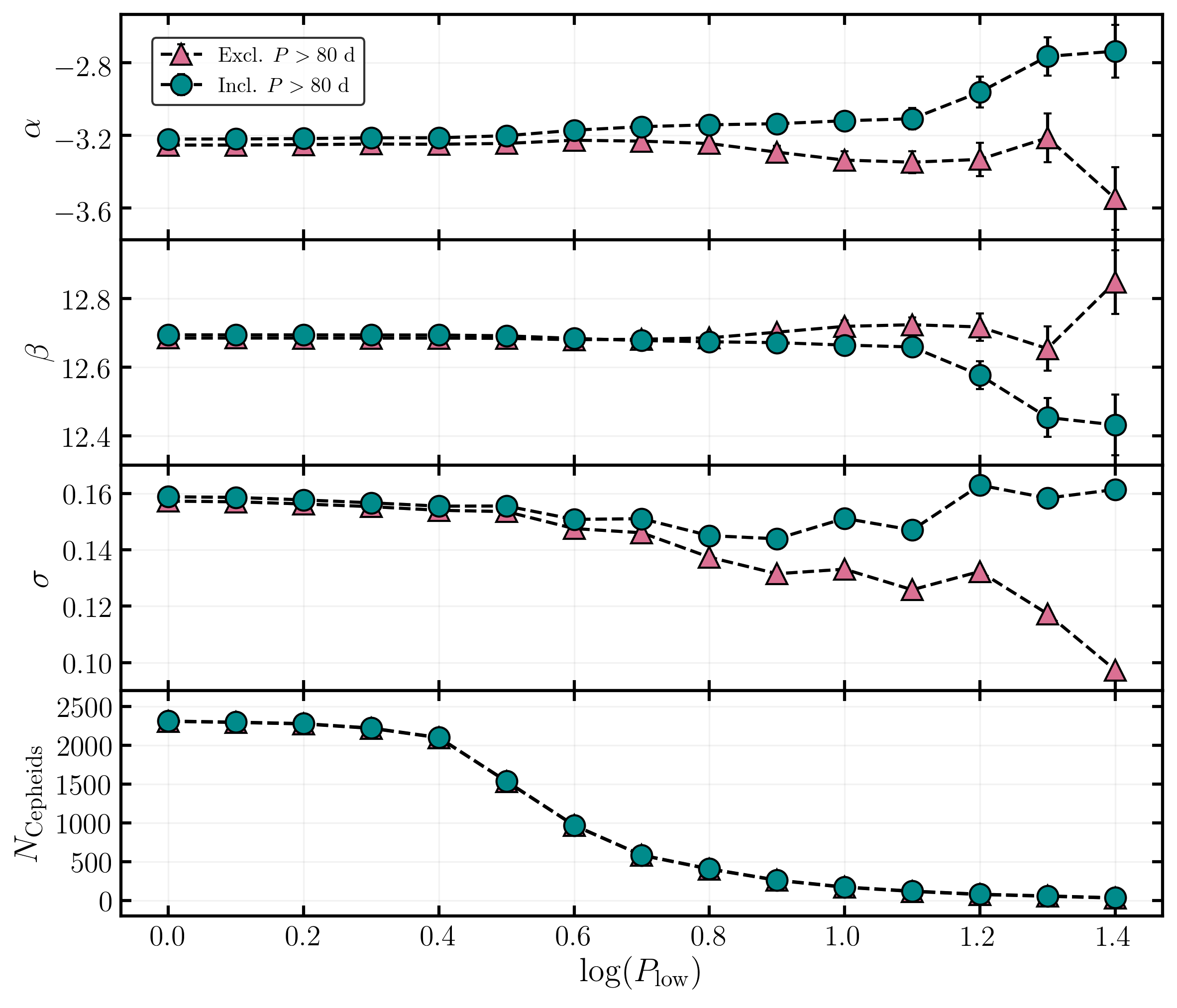}
    \caption{Same as Fig. \ref{fig:nObs_effect} but with cuts on pulsation period rather than the number of observations. Only Cepheids with periods greater than $P_{\text{low}}$ are included in the LL fit.}
    \label{fig:pCut_effect}
\end{figure}

Cepheids with a larger number of observations $(N_{\text{obs}})$ carry smaller uncertainties on their mean magnitude, such that $\sigma \propto 1/\sqrt{N_{\text{obs}}}$. Therefore, we tested what effect requiring a minimum number of observations has on the coefficients of the LL. As observed in Figure \ref{fig:LLfits}, there are four ultra-long period (ULP) Cepheids in the LMC sample, all of which are in our calibrating sample and have fully-phased observations. ULP Cepheids have periods greater than 80 days $(\log(P) = 1.90)$ \citep{Bird2009}. Cepheids with such long periods are often excluded from analysis of the LL as there is evidence that they obey their own LL (\citealt{Bird2009, Fiorentino2012}). Therefore, we performed our analysis twice: once including the ULP Cepheids and once excluding them, shown by the blue and pink points in Figure \ref{fig:nObs_effect}, respectively. We applied a minimum number of observations criteria for our sample, from $1 \leq N_{\text{low}} \leq 12$ observations. Cepheids that do not meet this criteria were discarded and a weighted linear least squares fit of the LL was performed on the remaining Cepheids.

The slope $(\alpha)$ of the LLs when including and excluding the ULP Cepheids are qualitatively similar up to $N_{\text{low}} = 10$, as they follow the same shape but with systematically steeper slopes when excluding the ULP Cepheids. For $N_{\text{low}} \geq 10$, there are substantially fewer Cepheids remaining in the fit, resulting in the ULP Cepheids having more weighting in these fits and causing the slope to remain shallower. All of the ULP Cepheids were observed by the CHP and therefore have $N_{\text{obs}} > 10$. Therefore, these Cepheids will never be knocked out for having too few observations and will have more weighting as $N_{\text{low}}$ increases and more Cepheids are removed. Without these Cepheids, we find that the slope becomes steeper as the minimum number of observations increases above 10. In both cases, the slope values are consistent for $4 \leq N_{\text{low}} \leq 10$, possibly suggesting that imposing a requirement of at least four observations is optimal. The same result is found for the zero point $(\beta)$, with consistent values for $4 \leq N_{\text{low}} \leq 10$.

For increasing $N_{\text{low}}$, the dispersion of the LL reduces, particularly when the ULP Cepheids are excluded. However, this improved dispersion must be counterbalanced by the fact that imposing this minimum observation criteria vastly reduces the number of Cepheids being used in the fit. For example, while the dispersion reduces from 0.157 to 0.143 mag for $N_{\text{low}} = 4$, the number of Cepheids reduces to only $37\%$ of the original sample ($N_{\text{Cepheids}} = 851$). We conclude that this improvement is not significant enough to warrant excluding the majority of our sample from the LL fit. While we do find significant improvements for $N_{\text{low}} > 10$, only $4\%$ of the Cepheids remain, resulting in a LL that is not representative of the entire Cepheid population.

\subsubsection{Slope and zero point dependence on the period range}
\label{sec:period_cuts}

When determining the Cepheid LL, period cuts are often applied to remove potentially contaminating Cepheids. For example, \cite{Ngeow2008} obtained the LMC mid-IR LLs by removing all Cepheids from their sample with $P < 2.5$ days in order to avoid potential contamination by overtone Cepheids. S11 applied a similar short-period cut for their LMC LL by excluding Cepheids with $P < 6$ days, in addition to a long-period cut at $P = 60$ days as these Cepheids are known to deviate significantly from the LL. We now focus on how the LL coefficients are affected by employing these period cuts.

For each low cut-off value $(P_{\text{low}})$, we discard all Cepheids with $P < P_{\text{low}}$ and fit a weighted least squares LL of the form given in Equation \ref{eq:LLform}. We test period cut values from $0 \leq \log(P_{\text{low}}) \leq 1.4$ in $\log(P_{\text{low}}) = 0.1$ steps. Similarly to the previous sections, we perform two fits for each value of $\log(P_{\text{low}})$ to include and exclude the ULP Cepheids. In Figure \ref{fig:pCut_effect}, we present the LL coefficients as function of $\log(P_{\text{low}})$. The top two panels show the slope $(\alpha)$ and zero point $(\beta)$, while the bottom two panels show the corresponding dispersion $(\sigma)$ and number of Cepheids remaining in the sample $(N_{\text{Cepheids}})$. In each panel, the blue points show the results when including the ULP Cepheids and the pink points show the results when excluding them. 

Regardless of whether we include the long period Cepheids, we observe consistent slope and zero point values (within $1\sigma$) up to a low cut off value of $\log(P_{\text{low}}) = 0.5$, corresponding to $P = 3.16$ days. However, above this value we find that the inclusion of the ULP Cepheids results in a monotonic decrease in the slope and zero point, causing the values to be inconsistent. In contrast, excluding the ULP Cepheids results in a larger proportion of consistent values, with only $\log(P_{\text{low}}) = 1.0, 1.1,$ and 1.4 not being in agreement within $1\sigma$. As the low cut off value becomes larger, fewer Cepheids are included in the fit, resulting in the four ULP Cepheids having more leverage on the LL coefficients than when many more Cepheids are included. This is likely the reason why the coefficients are much more consistent when excluding these Cepheids. Moreover, for larger cut off values, the derived coefficients carry larger uncertainties due to the relatively smaller sample size.

In addition to applying a period cut at 2.5 days, \cite{Ngeow2008} tested the sensitivity of the LL coefficients by employing period cuts from $0.4 \leq \log(P_{\text{low}}) \leq 0.9$ to in $\log(P) = 0.05$ steps. Their sample only includes two of our four ULP Cepheids, one of which is excluded from their analysis for being further than $2.5\sigma$ from their fitted LL. Therefore, it is more appropriate to compare their sample to our ULP-excluded sample. Our results show that the LL coefficients are unaffected by all period cuts less than $\log(P) \sim 0.9$, whereas \cite{Ngeow2008} obtain a smaller value of $\log(P) \sim 0.65$. These differing results could be due to our use of template fitting to obtain mean magnitudes rather than the use of single-epoch measurements as used in \cite{Ngeow2008}. Alternatively, it could be due to the much larger sample sizes used in our work ($N_{\text{LMC}} = 2311$ compared to their 613), resulting in our sample more uniformly occupying the Cepheid instability strip.

As previously mentioned, S11 derived the LMC mid-IR LL only using Cepheids with periods between 6 and 60 days (i.e. between $\log(P) \sim 0.8$ and 1.8). Imposing a short period cut off value of $\log(P) = 0.8$ and excluding the ULP Cepheids, we find that the slope and zero point from this sample are consistent within $1\sigma$ to the LL from the full sample with no period restrictions imposed. Therefore, we conclude that the LL with coefficients determined from only $6 < P < 60$ day Cepheids is still representative of an entire Cepheid population.

\subsubsection{The non-linearity of the Leavitt Law}
\label{sec:nonlinearity}

\begin{figure}
    \centering
    \includegraphics[width=0.45\textwidth]{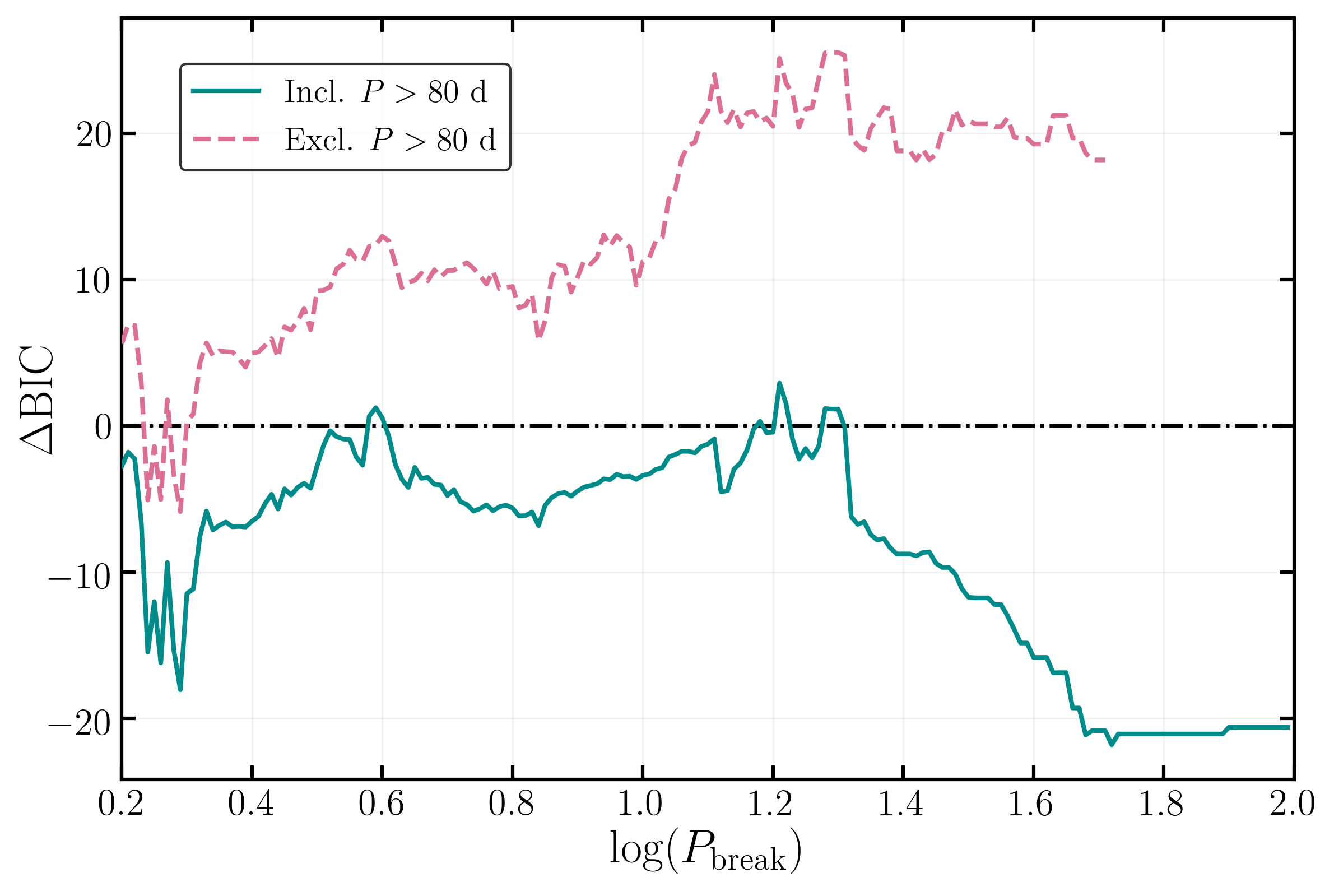}
    \caption{We test various values for a break in the LMC $[3.6]$ LL. We show how the BIC value differs from that of a simple linear LL fit, depending on the employed break value. Results below the dashed-dot line, such that $\Delta \text{BIC} < 0$, suggests that a non-linear relation with a break at $\log P_{\text{break}}$ is more representative of the data than a single linear fit, whereas a positive value indicates that the single LL fit is better. In each panel, we show the results when including the four long period Cepheids (blue solid line) and when excluding them (pink dashed line).}
    \label{fig:nonlinearity}
\end{figure}

There has been much debate within the literature surrounding the non-linearity of the LL, with many authors claiming that there is a break in the LL at around $P \sim 10$ days (\citealt{Tammann2003, Ngeow2005, Ngeow2008b, Kodric2015}). Using our large sample of Cepheids, we are able to investigate this claim at mid-IR wavelengths. 

The statistical test we use is the Bayesian Information Criterion (BIC) \citep{Schwarz1978}. While many statistical tests focus on which relation provides the smallest dispersion, BIC takes into account the number of parameters that are used in the model and penalises high-dimensional models. This is particularly useful when choosing the best fitting model of the LL, as it is necessarily the case that the higher dimension model will always maximise the likelihood. Therefore, BIC counteracts this by introducing a penalty term to penalise higher dimension models, so that there needs to be a significant improvement in the fit to the data for it to be considered the best model. 
We compute the BIC value of a model using the residual sum of squares (RSS) as 

\begin{equation}
\hspace{2.5cm}
\text{BIC} = n \ln\Big(\frac{RSS}{n}\Big) + k \ln(n),
\label{eq:BIC}
\end{equation} 
\noindent where $k$ is the number of model parameters and $n$ is the number of Cepheids in the sample. Using this formulation of the BIC equation is valid as our model errors, that is the residuals of the LL fit, are approximately Gaussian. As the model with the lowest BIC value is preferred, we see that the $k\ln(n)$ term in Equation \ref{eq:BIC} penalises those models with a larger number of parameters. 

\begin{figure}
    \centering
    \includegraphics[width=0.45\textwidth]{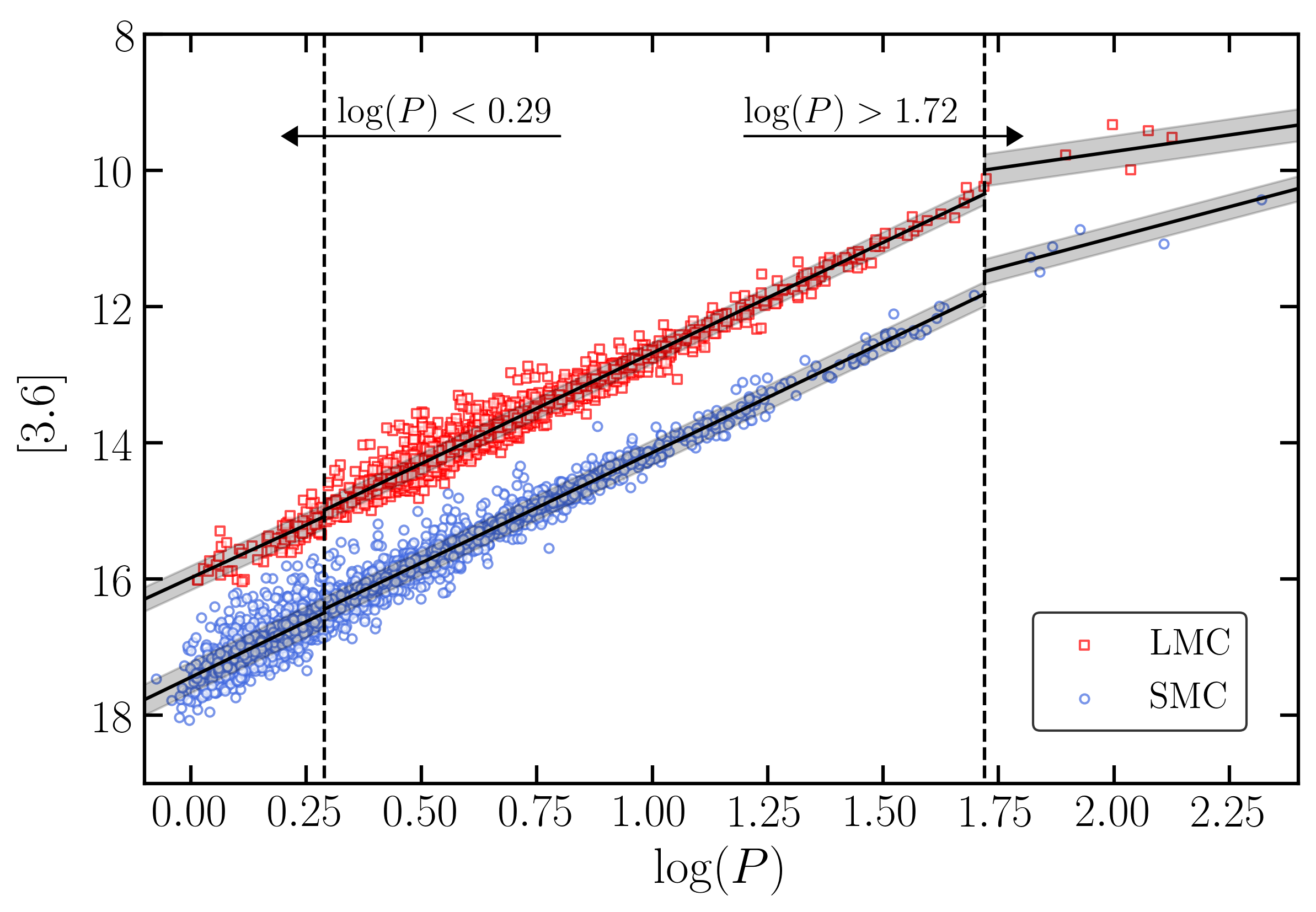}
    \caption{The LMC and SMC $[3.6]$ LLs applying breaks in the relation at $\log(P) = 0.29$ and $\log(P) = 1.72$. The red squares show the LMC Cepheids and the blue circles show the SMC Cepheids. Mean magnitudes have been dereddened using the process described in Section \ref{sec:reddening}. The vertical dashed lines show the breaks at $\log(P) = 0.29$ and $\log(P) = 1.72$. The best-fitting LL for each period range is shown by the black solid line. The shaded regions show $\pm1\sigma$ for each of the relations.}
    \label{fig:nonlinearLL}
\end{figure}

Our model for a single linear LL with $k = 2$ parameters is given by Equation \ref{eq:LLform}. For the non-linear relation with $k=4$ parameters, where we set a break value of $P_{\text{break}}$, we fit a short-period LL for Cepheids with $P < P_{\text{break}}$ and a long-period LL for Cepheids with $P \geq P_{\text{break}}$, both of the form given in Equation \ref{eq:LLform}. Fitting the non-linear relation in this way can introduce a small discontinuity at the $P = P_{\text{break}}$ boundary.

To determine whether the non-linear model with a certain break value is better than the linear model, we define $\Delta \text{BIC}$ as 
\begin{ceqn}
\begin{align}
\Delta \text{BIC} = \text{BIC}' - \text{BIC}_{\text{linear}},
\end{align}
\label{eq:deltaBIC}
\end{ceqn} 
where $\text{BIC}_{\text{linear}}$ is the BIC value for the linear LL and $\text{BIC}'$ is the value by employing a break and fitting a short-period and long-period LL. As the lowest BIC value is preferred, if $\Delta \text{BIC} > 0$ then the linear relation is the optimal fit, whereas if $\Delta \text{BIC} < 0$ then a non-linear model describes the data better, despite being penalised for the additional model parameters. How strong the evidence is for a certain model being preferred over another depends on the magnitude of $\Delta \text{BIC}$. For $|\Delta \text{BIC}| < 2$, there is no significant evidence for one model to be preferred. The larger the $|\Delta \text{BIC}|$, the stronger the evidence is for the model with the lower BIC value, with $|\Delta \text{BIC}| > 10$ providing very strong evidence that the model with the lowest BIC value is preferred \citep{Fabozzi2014}. 

We compute the $\Delta \text{BIC}$ values for period breaks  $0.20~\leq~\log(P_{\text{break}})~\leq 2.00$ in $\log(P_{\text{break}}) = 0.01$ steps. These results are shown in Figure \ref{fig:nonlinearity} for the LMC Cepheids in the $[3.6]$ band. As with previous analysis, we have performed each fit twice to include (blue solid line) and exclude (pink dashed line) the ULP Cepheids. We find very different results depending on whether or not the four ULP Cepheids are included. Excluding these Cepheids, we find the vast majority of break values do not result in a significantly improved fit compared to a linear LL as $\Delta \text{BIC} > 0$. However, looking at the results for the ULP-included sample, we observe just how much the four ULP Cepheids influence the $[3.6]$ LL. Apart from three small spikes at $\log(P_{\text{break}}) \sim 0.6, 1.2$ and 1.3, the ULP-included results are negative across the entire period range, suggesting a non-linear relation is optimal. The minimum $\Delta \text{BIC}$ occurs at $\log(P_{\text{break}}) = 1.72$ with $|\Delta \text{BIC}| = 21.80$, with $\Delta \text{BIC}$ values falling around this value for $1.68~\leq~\log(P_{\text{break}})~\leq 2.00$. $|\Delta \text{BIC}|$ values of this magnitude provide extremely strong evidence for a break at these long periods. For a break at $\log(P_{\text{break}}) = 1.72$, corresponding to 52.48 days, there are only 6 Cepheids in the long-period LL and 2305 in the short period LL. From this analysis, we conclude that there is sufficient evidence that suggests the ULP Cepheids do not follow the main LL and should be excluded from our final determination of the LL.

Aside from the large negative $\Delta \text{BIC}$ values that arise for $\log(P_{\text{break}}) \geq 1.68$ when the ULP Cepheids are included, we also observe a large decline in $\Delta \text{BIC}$ for  $0.22~\leq~\log(P_{\text{break}})~\leq 0.30$, with a minimum occurring at $\log(P_{\text{break}}) = 0.29$. This is the case regardless of whether the ULP Cepheids are included or not. Figure \ref{fig:nonlinearLL} shows the LMC and SMC Cepheid LLs based on period breaks at $\log(P_{\text{break}}) = 0.29$ and $1.72$, defining a short period LMC LL for $\log(P) < 0.29$ given by 

\begin{equation}
\hspace{0.9cm}[3.6] = -3.11(\log(P) - 1.0) + 12.90 \hspace{0.4cm} \sigma = 0.176,
\label{eq:shortLL}
\end{equation} 

\noindent a main LMC LL for $0.29~\leq~\log(P)~\leq1.72$ given by

\begin{equation}
\hspace{0.9cm}[3.6] = -3.25(\log(P) - 1.0) + 12.68,\hspace{0.4cm} \sigma = 0.155, 
\end{equation} 

\noindent and a long period LMC LL for $\log(P) > 1.72$ given by

\begin{equation}
\hspace{0.9cm}[3.6] = -0.97(\log(P) - 1.0) + 10.70 \hspace{0.4cm} \sigma = 0.235.
\label{eq:longLL}
\end{equation}

\noindent We now compare the use of these three LMC LLs, as shown in Figure \ref{fig:nonlinearLL}, to the single linear relation shown in Figure \ref{fig:LLfits}. We find that the short period Cepheids with $\log(P) < 0.29$ are more evenly distributed around their LL in Figure \ref{fig:nonlinearLL} compared to Figure \ref{fig:LLfits}, where we observe a larger proportion of Cepheids falling below the relation. At the long-period end of the relation, Figure \ref{fig:nonlinearLL} provides evidence that Cepheids with $\log(P) > 1.72$ should be excluded from the main LL fit as they appear to follow a separate, much shallower relation entirely. Despite being penalised for using triple the number of parameters used by the standard linear relation, the three relations shown in Figure \ref{fig:nonlinearLL} and given by Equations \ref{eq:shortLL} - \ref{eq:longLL} have $\Delta \text{BIC} = -32.44$, which is much more negative than the $\Delta \text{BIC}$ values for any of the relations with a single break point or no break point at all. Thus, we conclude that while we do not find strong evidence for a break midway through the LL, we do find that applying a small clipping at either end of the relation results in the optimal fit.

Figure \ref{fig:nonlinearLL} also shows the resulting SMC LLs when imposing breaks at $\log(P) = 0.29$ and $\log(P) = 1.72$. We find that the SMC slopes are consistent with the equivalent LMC values over the same period range. However, the consistency in the $\log(P) < 0.29$ and $\log(P) >~1.72$ relations is only due to their slopes carrying large uncertainties. On the other hand, the main LLs for $0.29~\leq~\log(P)~\leq1.72$ are consistent while also possessing small uncertainties. Over this period range, the LMC slope is $-3.246 \pm 0.008$ and the SMC slope is $-3.235 \pm 0.013$. This suggests that the LL slopes are not affected by metallicity across this period range.

Lastly, we tested whether adding a quadratic $\log(P)$ term, such that 

\begin{ceqn}
\begin{align}
m = \alpha\log(P)^2 + \beta\log(P) + \gamma,
\end{align}
\label{eq:quadratic}
\end{ceqn} 
\noindent where $\alpha, \beta, \gamma$ are the coefficients of the quadratic, fits the data better. We found that $\Delta \text{BIC} > 28,000$, suggesting that a quadratic model can be confidently ruled out, and that a simple linear model is much more appropriate to describe the data.  

\subsection{Comparison of the Leavitt Law to the literature}

\begin{figure}
    \centering
    \includegraphics[width=0.45\textwidth]{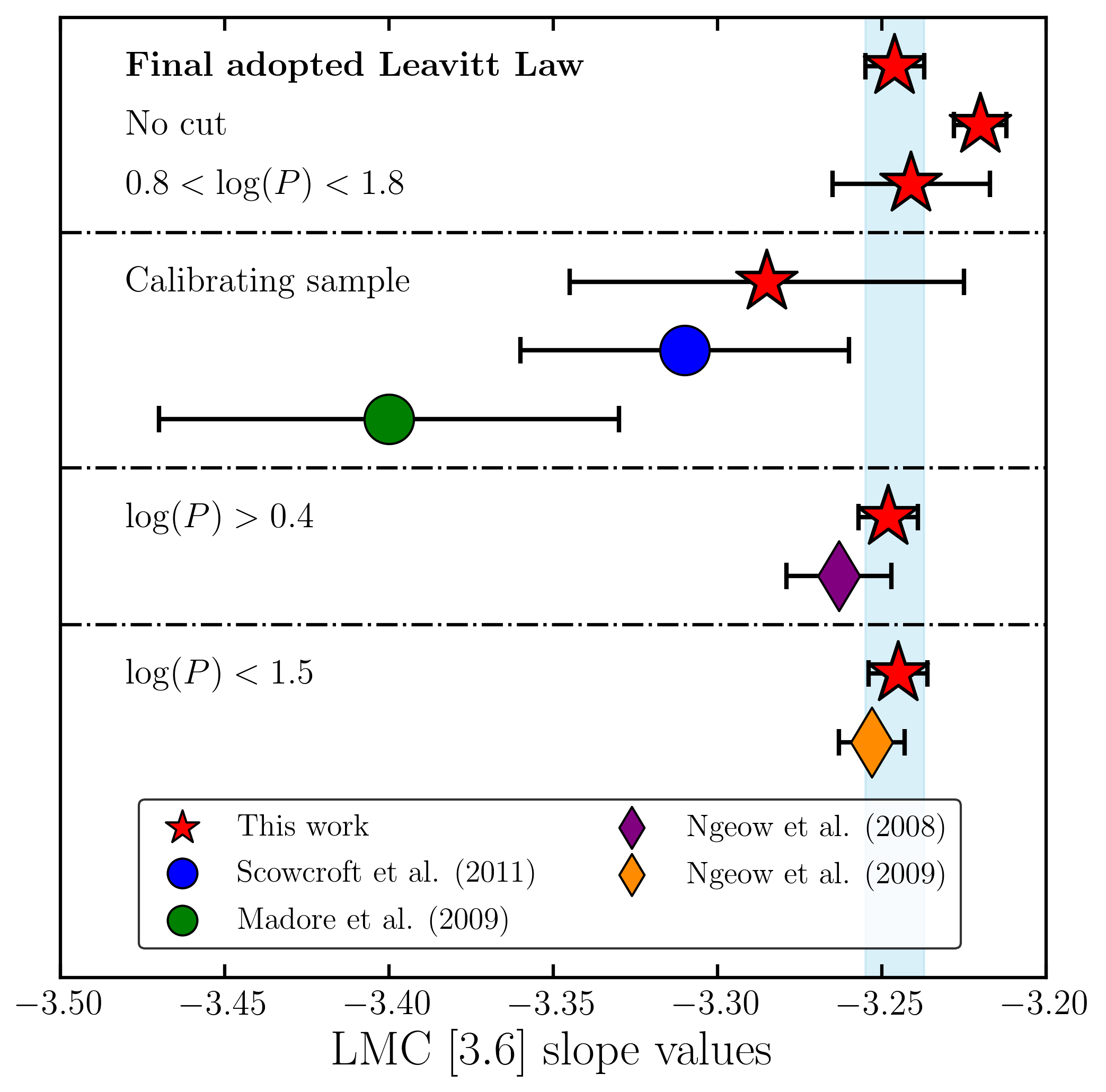}
    \caption{Our LMC $[3.6]$ slope values of the LL compared to those published in the literature. Identical period cuts have been applied to our data to match each literature value being compared. The blue shaded region shows our final adopted LL given in Section \ref{sec:finalLL}.}
    \label{fig:LMCslopes}
\end{figure}

\begin{table*}
\caption{\label{tab:finalLL} Final LL relations for the Magellanic Clouds. A weighted least squares fitting of the form $m_{\lambda} = \alpha(\log(P)-1.0) + \beta$ was performed with standard inverse variance weightings. Only Cepheids with $0.29 \leq \log(P) \leq 1.72$ were included in the fit.}
\centering
\begin{tabular}{ccccccc}
\hline
Galaxy & $\lambda$ & $\alpha$ & $\beta$ & $\beta_{cal}$ & $\sigma$ & $N$ \\ \hline
LMC    & $[3.6]$    &  $-3.246 \pm 0.008$     & $12.684\pm0.003$     &   $-5.784\pm0.030$         &  0.155 & 2221        \\
      & $[4.5]$    &  $-3.162 \pm 0.008$     &  $12.663 \pm 0.003$    &    $-5.751 \pm 0.030$        &  0.151 & 2221       \\
SMC    & $[3.6]$    &  $-3.246 \pm 0.008$     &   $13.143 \pm 0.005$   &     $-5.784\pm0.030$        & 0.180   & 1331        \\
      & $[4.5]$    &   $-3.162 \pm 0.008$    &   $13.160 \pm 0.005$   &   $-5.751 \pm 0.030$         &   0.177  & 1331    \\ \hline \hline
\end{tabular}
\end{table*}

\begin{table}
\caption{\label{tab:MW_parallaxes} The Milky Way sample used to calibrate the zero point of the LL. The period range for these Cepheids is $3 - 36$ days.}
    \centering
    \begin{tabular}{ccccc}
    \hline
Cepheid & $\varpi^a$ & $m_{[3.6]}$ $^{b}$ & $m_{[4.5]}$ $^b$ \\
\hline
$\delta$ Cep$^c$ & 3.364 $\pm$ 0.049             & $2.221 \pm 0.003$ & $2.217 \pm 0.003$ \\
CV Mon       & $0.523\pm 0.010$             &   $6.375 \pm 0.003$ &  $6.360 \pm 0.003 $                     \\
S Nor        & $1.025    \pm 0.004$             &   $3.652 \pm 0.003$ & $3.661 \pm 0.003$                                \\
U Sgr        & $1.514    \pm 0.003$             &  $3.824 \pm 0.003$ & $3.822 \pm 0.003$                  \\
V367 Sct     & $0.467    \pm 0.004$             &  $6.437 \pm 0.001$ & $6.410 \pm 0.002$   \\
V Cen        & $1.288\pm0.003$             &  $4.405 \pm 0.003$ & $4.400 \pm 0.003$ \\
DL Cas       & $0.511\pm0.002$             &  $5.783 \pm 0.003$  & $5.791 \pm 0.003$  \\
V340 Nor     & $0.443\pm0.002$             &  $5.453 \pm 0.002$ & $5.480 \pm 0.002$ \\
CF Cas       & $0.269\pm0.004$             &   $7.809 \pm 0.002$ & $7.783 \pm 0.002$ \\
TW Nor       & $0.383\pm0.006$             &  $6.152 \pm 0.003$ & $6.160 \pm 0.004$  \\ \hline
$\ell$ Car   & $2.01\pm0.20$              &   $0.925 \pm 0.004$ & $1.047 \pm 0.004$ \\
$\zeta$ Gem  & $2.78\pm0.18$              & $2.025 \pm 0.002$ & $2.037 \pm 0.003$ \\
$\beta$ Dor  & $3.14\pm0.16$              &  $1.858 \pm 0.003$ & $1.871 \pm 0.003$ \\
W Sgr        & $2.28\pm0.20$              &   $2.721 \pm 0.003$ & $2.719 \pm 0.004$  \\
X Sgr        & $3.00\pm0.18$              & $2.423 \pm 0.003$ & $2.409 \pm 0.003$ \\
Y Sgr        & $2.13\pm0.29$              &  $3.486 \pm 0.003$ & $3.483 \pm 0.003$ \\
FF Aql       & $2.81\pm0.18$              &  $3.378 \pm 0.001$ & $3.353 \pm 0.001$ \\
T Vul        & $1.90\pm0.23$              &  $4.114 \pm 0.003$ & $4.111 \pm 0.003$ \\
RT Aur       & $2.40\pm0.19$              &  $3.853 \pm 0.003$ & $3.849 \pm 0.003$ \\ \hline \hline
\end{tabular}
\begin{flushleft}
\begin{itemize}
\small
\item[ ] \textbf{Notes:}
\item[ ] $^a$ Parallaxes in the top section are from the parallax of the Cepheid's companion star or its host open cluster (\citealt{Cantat-Gaudin2018, Breuval2020}). These Cepheids were observed with \textit{Gaia} and the values stated here have not had a zero-point offset applied. The parallaxes in the bottom section are from the \textit{HST} Fine Guidance Sensors \citep{Benedict2007}. All values are given in milliarcseconds.
\item[ ] $^b$ $[3.6]$ and $[4.5]$ magnitudes are from M12.
\item[ ] $^c$ $\delta$ Cep was observed by both \textit{HST} and \textit{Gaia}. We decided to use the parallax value given by its companion star as it carried a smaller uncertainty.
\end{itemize}
\end{flushleft}
\end{table}

Observing campaigns have been performed that both serendipitously (e.g. SAGE) and purposefully (e.g. CHP) obtained mid-infrared observations of Cepheids in the Magellanic Clouds. We now compare our LL coefficients to those published in the literature. Our various relations with different period cuts are provided in Table \ref{tab:litPL}. The slope values of the LMC $[3.6]$ LL are also shown in Figure \ref{fig:LMCslopes}. 

Using the single-epoch measurements available at the time in the SAGE catalogue, \cite{Ngeow2008} obtained the mid-IR LL in the LMC for all four \textit{Spitzer} bands. The authors enforced a period cut of $\log(P) > 0.4$ to remove possible contamination by overtone Cepheids. They then used iterative $2.5\sigma$-clipping to obtain their LL coefficients. Comparing our $[3.6]$ LL for the LMC with no period cut applied, we find that our slope disagrees by just over $2\sigma$. However, if we apply the same period cut and also only retain the one Cepheid with $P > 60$ days that is common within both works,  we obtain coefficients of $\alpha = -3.248 \pm 0.009$ and $\beta = 12.684 \pm 0.003$, which are now both consistent with the values from \cite{Ngeow2008}, within the stated uncertainties. Regardless of whether we apply a period cut or not, our $[4.5]$ LL is inconsistent with these works by at least $3\sigma$. These discrepancies could be caused by a number of factors, including our sample being over three times larger, the use of template fitting rather than using single-epoch measurements, or the different pipelines used to obtain the photometry. In particular, the use of single-epoch measurements in the $[4.5]$ band are not appropriate due to the CO effect \citep{Scowcroft2016b}.

Building on their initial work, \cite{Ngeow2009} included an additional epoch of data and determined Cepheid mean magnitudes by averaging the two epochs to obtain the mid-IR LL using SAGE data. By cross-matching this data with the OGLE-III catalogue rather than OGLE-II, they were able to obtain the LL for a much larger sample size than their previous work. This time, the longest period Cepheids with $\log(P) \geq 1.5$ were discarded. Similarly to the LL based off of single-epoch measurements, we find a consistent $[3.6]$ slope value when also employing the same period cut ($\alpha = -3.245 \pm 0.009$), but inconsistent values in the $[4.5]$ band.

Both \cite{Madore2009b} and S11 use the sample of LMC Cepheids given in \cite{Persson2004} to obtain the mid-IR LL for 70 and 82 Cepheids, respectively. This is the same sample as our calibrating sample, with fully-phased observations and periods between 6 and 60 days. We see that these smaller samples give rise to steeper slopes than when many more Cepheids across the full period range are included. If we also restrict our sample to only include these longer period Cepheids, we obtain slopes of $-3.241 \pm 0.024$ and $-3.140 \pm 0.026$ for the $[3.6]$ and $[4.5]$, respectively, which are consistent with S11 but not \cite{Madore2009b}. However, if we further reduce our sample to only include the calibrating Cepheids, which are the identical set of Cepheids used by S11, we obtain slope values of $-3.283 \pm 0.058$ and $-3.217 \pm 0.062$ for the $[3.6]$ and $[4.5]$, respectively. These values are very similar to those of S11, possibly suggesting that discrepancies in the LL are caused by the Cepheid sample selection, with bright, isolated, long period Cepheids being preferential targets. As Figure \ref{fig:period_distribution} shows, the majority of Cepheids in the Magellanic Clouds have short periods. Therefore, such criteria may not necessarily represent the entire Cepheid population. 

Before our work, the only available SMC mid-IR relations that do not fix the slope of the relation to be that of the LMC are given by \cite{Ngeow2010}. Using only the first epoch of archival data taken as part of the SAGE-SMC program and applying iterative $2.5\sigma$-clipping, they obtained the SMC LL coefficients for all \textit{Spitzer} bands. Motivated by a potential slope change as different period cuts are employed (see their figure 4), they impose a break at $\log(P) = 0.4$ and obtain the coefficients for both the short-period and long-period relations. We find consistent slope values for all relations except the short-period $[3.6]$ band, where it differs from the equivalent \cite{Ngeow2010} relation by $\sim 2.2\sigma$.

Overall, we find that our LL coefficients are in generally good agreement with those found in the literature, particularly when we apply the same period cuts to our sample. 

\subsection{Final calibrated Leavitt Law from Magellanic Cloud Cepheids}
\label{sec:finalLL}

\subsubsection{Final adopted Leavitt Law}

For our final adopted LLs, we decided to only include Cepheids with periods $0.29 \leq \log(P) \leq 1.72$, using 2221 LMC and 1331 SMC Cepheids in the fit. We excluded the Cepheids with $\log(P) > 1.72$, which includes the ULP Cepheids, as Section \ref{sec:nonlinearity} showed that these Cepheids deviate from the fitted LL significantly and cause the LL to appear non-linear when they are included in the sample. Moreover, in Section \ref{sec:period_cuts}, these ULP Cepheis were shown to affect the LL coefficients substantially. Therefore, it is best to remove these Cepheids, particularly as there is growing evidence that they follow their own independent LL (\citealt{Bird2009, Fiorentino2012}). We also excluded Cepheids with $\log(P) < 0.29$, as Section \ref{sec:nonlinearity} provided strong evidence for a break in the LL at $\log(P) = 0.29$, with the shorter period Cepheids obeying a slightly shallower LL. As we showed in Section \ref{sec:period_cuts} that the LL coefficients are consistent up to $\log(P) \sim 0.9$, a period cut at $\log(P) = 0.29$ does not affect the LL coefficients significantly.

Our adopted relations are given in Table \ref{tab:finalLL}. For the SMC relations, we fixed the slope to be that of the LMC and computed the SMC zero point. We chose to do this because the LMC slope is defined for a larger number of Cepheids and to also prevent biases that would be present due to the geometry of the SMC. This method of adopting the LMC slopes as fiducial is appropriate as the $[3.6]$ LMC and SMC slopes are consistent within $1\sigma$ and the $[4.5]$ slopes within $3\sigma$ for Cepheids with $0.29 \leq \log(P) \leq 1.72$. The larger discrepancy in $[4.5]$ slopes is likely due to the effect of CO absorption on the $[4.5]$ magnitudes, while the $[3.6]$ magnitudes remain unaffected \citep{Scowcroft2016b}.

\subsubsection{Zero point calibration}

To calibrate the zero point, we combined the 10 MW Cepheids with geometric parallaxes determined by the Fine Guidance Sensors on-board the \textit{Hubble Space Telescope}, as given in \cite{Benedict2007}, with 10 MW Cepheids that have \textit{Gaia DR2} parallaxes determined by either their companion star or their host open cluster (\citealt{Cantat-Gaudin2018, Breuval2020}). For the \textit{Gaia} parallaxes, we applied a parallax zero-point offset value of -46 $\pm$ 13 $\mu$as, as determined by \cite{Riess2018}. The size of the parallax zero-point offset may depend on the object's colour and magnitude \citep{Lindegren2018}. Therefore, we use the value from \cite{Riess2018} as it was derived from MW Cepheids. Our sample of MW stars were chosen as they all have mid-IR mean magnitudes from M12. The parallaxes and magnitudes for these Cepheids are provided in Table \ref{tab:MW_parallaxes}. For $\delta$ Cep, both \textit{HST} and \textit{Gaia} parallaxes were available. We decided to use the Gaia parallax, as it carried a smaller uncertainty.

The calibrated zero point was computed using the Astrometric Based Luminosity (ABL) (\citealt{Feast1997, Arenou1999}), defined as 

\begin{ceqn}
\begin{align}
ABL = 10^{0.2(\alpha(\log(P)-1.0)+\beta)} = \varpi 10^{0.2m-2},
\end{align}
\label{eq:ABL}
\end{ceqn} 

\noindent where $\alpha$ is fixed from the apparent magnitude LL, $\varpi$ is the parallax in milliarcseconds and $m$ is the dereddened apparent magnitude. We fit the relation using \verb|curve_fit| from the \verb|scipy| library in Python to determine the value of the zero point, $\beta$. We find calibrated zero points of $-5.784\pm0.030$ and $-5.751\pm0.030$ for the $[3.6]$ and $[4.5]$ band, respectively.

Using the $[3.6]$ calibrated LL and the zero points for each galaxy, we calculated the apparent distance modulus of the LMC to be $18.47 \pm 0.03$ mag and the SMC to be $18.93 \pm 0.03$ mag. The differential distance modulus is $0.46 \pm 0.04$ mag. This result is in agreement with several other differential distance measurements from Cepheids including \cite{Groenewegen2000} ($0.50 \pm 0.10$ mag), \cite{Bono2010} ($0.44 \pm 0.12$ mag), \cite{Inno2013} ($0.48 \pm 0.03$ mag) and S16 ($0.48 \pm 0.01$ mag). Our result is also consistent with differential distance measurements from other methods such as eclipsing binaries \citep[][$0.472 \pm 0.026$ mag]{Graczyk2014} and RR Lyrae stars \citep[][$0.39 \pm 0.04$ mag]{Szewczyk2009}.

\section{Summary}

Using mid-IR observations of $\sim 5000$ fundamental mode Classical Cepheids in the Magellanic System, we have constructed the $[3.6]$ and $[4.5]$ LL for the Magellanic Clouds. Through our MCMC template light curve fitting procedure, where the templates were constructed from a subsample of Cepheids with well-sampled light curves from the CHP, we have obtained mid-IR mean magnitudes for all Cepheids within the SAGE footprint, amounting to 2311 LMC and 2637 SMC Cepheids. Our catalogue containing $[3.6]$ and $[4.5]$ mean magnitudes for these Cepheids is given in Table \ref{tab:sage_catalogue}, with the entire catalogue available for online download. This catalogue will be used in a companion paper (Chown et al. 2020, in prep) to study the 3D structure of the Magellanic System.

The LMC $[3.6]$ LL is the primary relation used in distance determination. We investigated the dependence of the coefficients of this LL on various period cuts. When excluding the ultra-long period Cepheids (ULP) (those with periods > 80 days), we find that our relation is unaffected by period cuts below $\log(P) \sim 0.9$, as the slope and zero point remain consistent with the relation that uses the entire sample. 

In addition, we investigated the non-linearity of the LMC $[3.6]$ LL. We used the Bayesian Information Criterion (BIC) to quantify our results. Our results varied greatly depending on whether the ULP Cepheids were included, with strong evidence for a linear relation when excluding them and strong evidence for a non-linear relation when including them. We found that the most likely periods for a break to occur were at $\log(P) = 0.29$ and $\log(P) = 1.72$.

Our final adopted relations make use of these results and only include Cepheids with $0.29 \leq \log(P) \leq 1.72$, resulting in 2221 LMC and 1331 SMC Cepheids being used in the LL fit. We showed in Section \ref{sec:period_cuts} that a period cut at $\log(P) = 0.29$ does not affect the slope and zero point significantly. The break value at $\log(P) = 1.72$ was used to exclude the ULP Cepheids, as these Cepheids were shown in Section \ref{sec:nonlinearity} to follow a much shallower LL.  Using parallaxes of Cepheids in the Milky Way from both the \textit{Hubble Space Telescope} and \textit{Gaia}, we calibrated the zero point of our LL using the Astrometric Based Luminosity. Our final relations are given by 
\begin{align} \label{eq:CalibratedLL}
M_{[3.6]} &= -3.246 (\pm 0.008)(\log(P)-1.0) -5.784(\pm0.030), \\
M_{[4.5]} &= -3.162 (\pm 0.008)(\log(P)-1.0) -5.751(\pm0.030).
\end{align}

\section*{Acknowledgements}

We thank the anonymous referee for their valuable comments which have significantly improved our paper. The authors would like to thank Raoul and Catherine Hughes for awarding the RCH Alan Hunter Scholarship to fund this research. This work is based on observations made with the \textit{Spitzer Space Telescope}, which is operated by the Jet Propulsion Laboratory, California Institute of Technology under a contract with NASA. This work makes use of the Optical Gravitational Lensing Experiment (OGLE) Collection of Variable Stars database. This research made use of Astropy,\footnote{\url{http://www.astropy.org}} a community-developed core Python package for Astronomy \citep{astropy:2013, astropy:2018}.

\section*{Data availability}

The data underlying this article are from \cite{ChownData}. The code used for this article are available at \url{https://github.com/abichown/DAOPHOT-Scripts}.




\bibliographystyle{mnras}
\bibliography{library} 

\begin{thebibliography}{}
\makeatletter
\relax
\def\mn@urlcharsother{\let\do\@makeother \do\$\do\&\do\#\do\^\do\_\do\%\do\~}
\def\mn@doi{\begingroup\mn@urlcharsother \@ifnextchar [ {\mn@doi@}
  {\mn@doi@[]}}
\def\mn@doi@[#1]#2{\def\@tempa{#1}\ifx\@tempa\@empty \href
  {http://dx.doi.org/#2} {doi:#2}\else \href {http://dx.doi.org/#2} {#1}\fi
  \endgroup}
\def\mn@eprint#1#2{\mn@eprint@#1:#2::\@nil}
\def\mn@eprint@arXiv#1{\href {http://arxiv.org/abs/#1} {{\tt arXiv:#1}}}
\def\mn@eprint@dblp#1{\href {http://dblp.uni-trier.de/rec/bibtex/#1.xml}
  {dblp:#1}}
\def\mn@eprint@#1:#2:#3:#4\@nil{\def\@tempa {#1}\def\@tempb {#2}\def\@tempc
  {#3}\ifx \@tempc \@empty \let \@tempc \@tempb \let \@tempb \@tempa \fi \ifx
  \@tempb \@empty \def\@tempb {arXiv}\fi \@ifundefined
  {mn@eprint@\@tempb}{\@tempb:\@tempc}{\expandafter \expandafter \csname
  mn@eprint@\@tempb\endcsname \expandafter{\@tempc}}}

\bibitem[\protect\citeauthoryear{{Arenou} \& {Luri}}{{Arenou} \&
  {Luri}}{1999}]{Arenou1999}
{Arenou} F.,  {Luri} X.,  1999, in {Egret} D.,  {Heck} A.,  eds,  Astronomical
  Society of the Pacific Conference Series Vol. 167, Harmonizing Cosmic
  Distance Scales in a Post-HIPPARCOS Era. pp 13--32 (\mn@eprint {arXiv}
  {astro-ph/9812094})

\bibitem[\protect\citeauthoryear{{Astropy Collaboration} et~al.,}{{Astropy
  Collaboration} et~al.}{2013}]{astropy:2013}
{Astropy Collaboration} et~al., 2013, \mn@doi [\aap]
  {10.1051/0004-6361/201322068}, \href
  {http://adsabs.harvard.edu/abs/2013A%26A...558A..33A} {558, A33}

\bibitem[\protect\citeauthoryear{Benedict et~al.,}{Benedict
  et~al.}{2007}]{Benedict2007}
Benedict G.~F.,  et~al., 2007, \mn@doi [The Astronomical Journal]
  {10.1086/511980}, 133, 1810

\bibitem[\protect\citeauthoryear{{Bird}, {Stanek}  \& {Prieto}}{{Bird}
  et~al.}{2009}]{Bird2009}
{Bird} J.~C.,  {Stanek} K.~Z.,   {Prieto} J.~L.,  2009, \mn@doi [\apj]
  {10.1088/0004-637X/695/2/874}, \href
  {https://ui.adsabs.harvard.edu/abs/2009ApJ...695..874B} {695, 874}

\bibitem[\protect\citeauthoryear{Bono, Caputo, Marconi  \& Musella}{Bono
  et~al.}{2010}]{Bono2010}
Bono G.,  Caputo F.,  Marconi M.,   Musella I.,  2010, \mn@doi [Astrophysical
  Journal] {10.1088/0004-637X/715/1/277}, 715, 277

\bibitem[\protect\citeauthoryear{{Breuval} et~al.,}{{Breuval}
  et~al.}{2020}]{Breuval2020}
{Breuval} L.,  et~al., 2020, arXiv e-prints, \href
  {https://ui.adsabs.harvard.edu/abs/2020arXiv200608763B} {p. arXiv:2006.08763}

\bibitem[\protect\citeauthoryear{{Cantat-Gaudin} et~al.,}{{Cantat-Gaudin}
  et~al.}{2018}]{Cantat-Gaudin2018}
{Cantat-Gaudin} T.,  et~al., 2018, \mn@doi [\aap]
  {10.1051/0004-6361/201833476}, \href
  {https://ui.adsabs.harvard.edu/abs/2018A&A...618A..93C} {618, A93}

\bibitem[\protect\citeauthoryear{Cardelli, Clayton  \& Mathis}{Cardelli
  et~al.}{1989}]{Cardelli1989}
Cardelli J.~A.,  Clayton G.~C.,   Mathis J.~S.,  1989, Astrophysical Journal,
  345, 245

\bibitem[\protect\citeauthoryear{Chown, Scowcroft  \& Wuyts}{Chown
  et~al.}{2020}]{ChownData}
Chown A.,  Scowcroft V.,   Wuyts S.,  2020, ``The mid-infrared Leavitt Law for
  Classical Cepheids in the Magellanic Clouds'' by Chown et al. (2020). Bath:
  University of Bath Research Data Archive, \url
  {https://doi.org/10.15125/BATH-00915}

\bibitem[\protect\citeauthoryear{{Cioni} et~al.,}{{Cioni}
  et~al.}{2011}]{Cioni2011}
{Cioni} M. R.~L.,  et~al., 2011, \mn@doi [\aap] {10.1051/0004-6361/201016137},
  \href {https://ui.adsabs.harvard.edu/abs/2011A&A...527A.116C} {527, A116}

\bibitem[\protect\citeauthoryear{{Fabozzi}, {Focardi}, {Rachev}  \&
  {Arshanapalli}}{{Fabozzi} et~al.}{2014}]{Fabozzi2014}
{Fabozzi} F.~J.,  {Focardi} S.~M.,  {Rachev} S.~T.,   {Arshanapalli} B.~G.,
  2014, {The Basics of Financial Econometrics: Tools, Concepts, and Asset
  Management Applications}.
John Wiley \& Sons, Inc., Hoboken, New Jersey

\bibitem[\protect\citeauthoryear{Fazio et~al.,}{Fazio et~al.}{2004}]{Fazio2004}
Fazio G.~G.,  et~al., 2004, \mn@doi [The Astrophysical Journal Supplement
  Series] {10.1086/422843}, 154, 10

\bibitem[\protect\citeauthoryear{{Feast} \& {Catchpole}}{{Feast} \&
  {Catchpole}}{1997}]{Feast1997}
{Feast} M.~W.,  {Catchpole} R.~M.,  1997, \mn@doi [\mnras]
  {10.1093/mnras/286.1.L1}, \href
  {https://ui.adsabs.harvard.edu/abs/1997MNRAS.286L...1F} {286, L1}

\bibitem[\protect\citeauthoryear{{Fiorentino} et~al.,}{{Fiorentino}
  et~al.}{2012}]{Fiorentino2012}
{Fiorentino} G.,  et~al., 2012, \mn@doi [\apss] {10.1007/s10509-012-1043-4},
  \href {https://ui.adsabs.harvard.edu/abs/2012Ap&SS.341..143F} {341, 143}

\bibitem[\protect\citeauthoryear{{Foreman-Mackey}, {Hogg}, {Lang}  \&
  {Goodman}}{{Foreman-Mackey} et~al.}{2013}]{Foreman-Mackey2013}
{Foreman-Mackey} D.,  {Hogg} D.~W.,  {Lang} D.,   {Goodman} J.,  2013, \mn@doi
  [\pasp] {10.1086/670067}, \href
  {https://ui.adsabs.harvard.edu/abs/2013PASP..125..306F} {125, 306}

\bibitem[\protect\citeauthoryear{Fouqu{\'{e}} et~al.,}{Fouqu{\'{e}}
  et~al.}{2007}]{Fouque2007}
Fouqu{\'{e}} P.,  et~al., 2007, \mn@doi [Astronomy {\&} Astrophysics]
  {10.1051/0004-6361:20078187}, 476, 73

\bibitem[\protect\citeauthoryear{Freedman et~al.,}{Freedman
  et~al.}{2011}]{Freedman2011a}
Freedman W.~L.,  et~al., 2011, \mn@doi [Astronomical Journal]
  {10.1088/0004-6256/142/6/192}, 142, 192

\bibitem[\protect\citeauthoryear{{Gieren}, {Storm}, {Barnes}, {Fouqu{\'e}},
  {Pietrzy{\'n}ski}  \& {Kienzle}}{{Gieren} et~al.}{2005}]{Gieren2005}
{Gieren} W.,  {Storm} J.,  {Barnes} Thomas~G. I.,  {Fouqu{\'e}} P.,
  {Pietrzy{\'n}ski} G.,   {Kienzle} F.,  2005, \mn@doi [\apj] {10.1086/430496},
  \href {https://ui.adsabs.harvard.edu/abs/2005ApJ...627..224G} {627, 224}

\bibitem[\protect\citeauthoryear{{Gordon} et~al.,}{{Gordon}
  et~al.}{2011}]{Gordon2011}
{Gordon} K.~D.,  et~al., 2011, \mn@doi [\aj] {10.1088/0004-6256/142/4/102},
  \href {https://ui.adsabs.harvard.edu/abs/2011AJ....142..102G} {142, 102}

\bibitem[\protect\citeauthoryear{{Graczyk} et~al.,}{{Graczyk}
  et~al.}{2014}]{Graczyk2014}
{Graczyk} D.,  et~al., 2014, \mn@doi [\apj] {10.1088/0004-637X/780/1/59}, \href
  {https://ui.adsabs.harvard.edu/abs/2014ApJ...780...59G} {780, 59}

\bibitem[\protect\citeauthoryear{{Groenewegen}}{{Groenewegen}}{2000}]{Groenewegen2000}
{Groenewegen} M.~A.~T.,  2000, \aap, \href
  {https://ui.adsabs.harvard.edu/abs/2000A&A...363..901G} {363, 901}

\bibitem[\protect\citeauthoryear{{Groenewegen}}{{Groenewegen}}{2018}]{Groenewegen2018}
{Groenewegen} M.~A.~T.,  2018, \mn@doi [\aap] {10.1051/0004-6361/201833478},
  \href {https://ui.adsabs.harvard.edu/abs/2018A&A...619A...8G} {619, A8}

\bibitem[\protect\citeauthoryear{{Hertzsprung}}{{Hertzsprung}}{1926}]{Hertzsprung1926}
{Hertzsprung} E.,  1926, \bain, \href
  {https://ui.adsabs.harvard.edu/abs/1926BAN.....3..115H} {3, 115}

\bibitem[\protect\citeauthoryear{Indebetouw et~al.,}{Indebetouw
  et~al.}{2005}]{Indebetouw2005}
Indebetouw R.,  et~al., 2005, \mn@doi [Astrophysical Journal] {10.1086/426679},
  619, 913

\bibitem[\protect\citeauthoryear{Inno et~al.,}{Inno et~al.}{2013}]{Inno2013}
Inno L.,  et~al., 2013, \mn@doi [Astrophysical Journal]
  {10.1088/0004-637X/764/1/84}, 764, 84

\bibitem[\protect\citeauthoryear{{Inno} et~al.,}{{Inno}
  et~al.}{2015}]{Inno2015}
{Inno} L.,  et~al., 2015, \mn@doi [\aap] {10.1051/0004-6361/201424396}, \href
  {https://ui.adsabs.harvard.edu/abs/2015A&A...576A..30I} {576, A30}

\bibitem[\protect\citeauthoryear{{Joshi} \& {Panchal}}{{Joshi} \&
  {Panchal}}{2019}]{Joshi2019}
{Joshi} Y.~C.,  {Panchal} A.,  2019, \mn@doi [\aap]
  {10.1051/0004-6361/201834574}, \href
  {https://ui.adsabs.harvard.edu/abs/2019A&A...628A..51J} {628, A51}

\bibitem[\protect\citeauthoryear{{Kodric} et~al.,}{{Kodric}
  et~al.}{2015}]{Kodric2015}
{Kodric} M.,  et~al., 2015, \mn@doi [\apj] {10.1088/0004-637X/799/2/144}, \href
  {https://ui.adsabs.harvard.edu/abs/2015ApJ...799..144K} {799, 144}

\bibitem[\protect\citeauthoryear{{Leavitt}}{{Leavitt}}{1908}]{Leavitt1908}
{Leavitt} H.~S.,  1908, Annals of Harvard College Observatory, \href
  {https://ui.adsabs.harvard.edu/abs/1908AnHar..60...87L} {60, 87}

\bibitem[\protect\citeauthoryear{Leavitt \& Pickering}{Leavitt \&
  Pickering}{1912}]{Leavitt1912}
Leavitt H.,  Pickering C.,  1912, Harvard College Observatory, 173, 1

\bibitem[\protect\citeauthoryear{{Lindegren} et~al.,}{{Lindegren}
  et~al.}{2018}]{Lindegren2018}
{Lindegren} L.,  et~al., 2018, \mn@doi [\aap] {10.1051/0004-6361/201832727},
  \href {https://ui.adsabs.harvard.edu/abs/2018A&A...616A...2L} {616, A2}

\bibitem[\protect\citeauthoryear{{Macri}, {Ngeow}, {Kanbur}, {Mahzooni}  \&
  {Smitka}}{{Macri} et~al.}{2015}]{Macri2015}
{Macri} L.~M.,  {Ngeow} C.-C.,  {Kanbur} S.~M.,  {Mahzooni} S.,   {Smitka}
  M.~T.,  2015, \mn@doi [\aj] {10.1088/0004-6256/149/4/117}, \href
  {https://ui.adsabs.harvard.edu/abs/2015AJ....149..117M} {149, 117}

\bibitem[\protect\citeauthoryear{{Madore}, {Rigby}, {Freedman}, {Persson},
  {Sturch}  \& {Mager}}{{Madore} et~al.}{2009a}]{Madore2009}
{Madore} B.~F.,  {Rigby} J.,  {Freedman} W.~L.,  {Persson} S.~E.,  {Sturch} L.,
    {Mager} V.,  2009a, \mn@doi [\apj] {10.1088/0004-637X/693/1/936}, \href
  {https://ui.adsabs.harvard.edu/abs/2009ApJ...693..936M} {693, 936}

\bibitem[\protect\citeauthoryear{Madore, Freedman, Rigby, Persson, Sturch  \&
  Mager}{Madore et~al.}{2009b}]{Madore2009b}
Madore B.~F.,  Freedman W.~L.,  Rigby J.,  Persson S.~E.,  Sturch L.,   Mager
  V.,  2009b, \mn@doi [The Astrophysical Journal]
  {10.1088/0004-637X/695/2/988}, 695, 988

\bibitem[\protect\citeauthoryear{{Makovoz} \& {Khan}}{{Makovoz} \&
  {Khan}}{2005}]{Makovoz2005}
{Makovoz} D.,  {Khan} I.,  2005, in {Shopbell} P.,  {Britton} M.,   {Ebert} R.,
   eds,  Astronomical Society of the Pacific Conference Series Vol. 347,
  Astronomical Data Analysis Software and Systems XIV. p.~81

\bibitem[\protect\citeauthoryear{Meixner et~al.,}{Meixner
  et~al.}{2006}]{Meixner2006}
Meixner M.,  et~al., 2006, \mn@doi [The Astronomical Journal] {10.1086/508185},
  132, 2268

\bibitem[\protect\citeauthoryear{{Mighell}, {Glaccum}  \& {Hoffmann}}{{Mighell}
  et~al.}{2008}]{Mighell2008}
{Mighell} K.~J.,  {Glaccum} W.,   {Hoffmann} W.,  2008, in Space Telescopes and
  Instrumentation 2008: Optical, Infrared, and Millimeter. p. 70102W,
  \mn@doi{10.1117/12.789801}

\bibitem[\protect\citeauthoryear{{Moffett}, {Gieren}, {Barnes}  \&
  {G{\'o}mez}}{{Moffett} et~al.}{1998}]{Moffett1998}
{Moffett} T.~J.,  {Gieren} W.~P.,  {Barnes} Thomas~G. I.,   {G{\'o}mez} M.,
  1998, \mn@doi [\apjs] {10.1086/313116}, \href
  {https://ui.adsabs.harvard.edu/abs/1998ApJS..117..135M} {117, 135}

\bibitem[\protect\citeauthoryear{Monson, Freedman, Madore, Persson, Scowcroft,
  Seibert  \& Rigby}{Monson et~al.}{2012}]{Monson2012}
Monson A.~J.,  Freedman W.~L.,  Madore B.~F.,  Persson S.~E.,  Scowcroft V.,
  Seibert M.,   Rigby J.~R.,  2012, \mn@doi [Astrophysical Journal]
  {10.1088/0004-637X/759/2/146}, 759, 146

\bibitem[\protect\citeauthoryear{{Ngeow} \& {Kanbur}}{{Ngeow} \&
  {Kanbur}}{2008}]{Ngeow2008}
{Ngeow} C.,  {Kanbur} S.~M.,  2008, \mn@doi [\apj] {10.1086/586704}, \href
  {https://ui.adsabs.harvard.edu/abs/2008ApJ...679...76N} {679, 76}

\bibitem[\protect\citeauthoryear{Ngeow \& Kanbur}{Ngeow \&
  Kanbur}{2010}]{Ngeow2010}
Ngeow C.~C.,  Kanbur S.~M.,  2010, \mn@doi [Astrophysical Journal]
  {10.1088/0004-637X/720/1/626}, 720, 626

\bibitem[\protect\citeauthoryear{Ngeow, Kanbur, Nikolaev, Buonaccorsi, Cook  \&
  Welch}{Ngeow et~al.}{2005}]{Ngeow2005}
Ngeow C.-c.,  Kanbur S.~M.,  Nikolaev S.,  Buonaccorsi J.,  Cook K.~H.,   Welch
  D.~L.,  2005, Monthly Notices of the Royal Astronomical Society, 363, 831

\bibitem[\protect\citeauthoryear{{Ngeow}, {Kanbur}  \& {Nanthakumar}}{{Ngeow}
  et~al.}{2008}]{Ngeow2008b}
{Ngeow} C.,  {Kanbur} S.~M.,   {Nanthakumar} A.,  2008, \mn@doi [\aap]
  {10.1051/0004-6361:20078812}, \href
  {https://ui.adsabs.harvard.edu/abs/2008A&A...477..621N} {477, 621}

\bibitem[\protect\citeauthoryear{{Ngeow}, {Kanbur}, {Neilson}, {Nanthakumar}
  \& {Buonaccorsi}}{{Ngeow} et~al.}{2009}]{Ngeow2009}
{Ngeow} C.-C.,  {Kanbur} S.~M.,  {Neilson} H.~R.,  {Nanthakumar} A.,
  {Buonaccorsi} J.,  2009, \mn@doi [\apj] {10.1088/0004-637X/693/1/691}, \href
  {https://ui.adsabs.harvard.edu/abs/2009ApJ...693..691N} {693, 691}

\bibitem[\protect\citeauthoryear{{Ngeow}, {Kanbur}, {Bhardwaj}  \&
  {Singh}}{{Ngeow} et~al.}{2015}]{Ngeow2015}
{Ngeow} C.-C.,  {Kanbur} S.~M.,  {Bhardwaj} A.,   {Singh} H.~P.,  2015, \mn@doi
  [\apj] {10.1088/0004-637X/808/1/67}, \href
  {https://ui.adsabs.harvard.edu/abs/2015ApJ...808...67N} {808, 67}

\bibitem[\protect\citeauthoryear{Persson, Madore, Krzemin, Freedman, Roth  \&
  Murphy}{Persson et~al.}{2004}]{Persson2004}
Persson S.~E.,  Madore B.~F.,  Krzemin W.,  Freedman W.~L.,  Roth M.,   Murphy
  D.~C.,  2004, \mn@doi [The Astronomical Journal] {10.1086/424934}, 128, 2239

\bibitem[\protect\citeauthoryear{{Price-Whelan} et~al.,}{{Price-Whelan}
  et~al.}{2018}]{astropy:2018}
{Price-Whelan} A.~M.,  et~al., 2018, \mn@doi [\aj] {10.3847/1538-3881/aabc4f},
  \href {https://ui.adsabs.harvard.edu/#abs/2018AJ....156..123T} {156, 123}

\bibitem[\protect\citeauthoryear{Reach et~al.,}{Reach et~al.}{2005}]{Reach2005}
Reach W.~T.,  et~al., 2005, \mn@doi [Publications of the Astronomical Society
  of the Pacific] {10.1086/432670}, 117, 978

\bibitem[\protect\citeauthoryear{Riebel et~al.,}{Riebel
  et~al.}{2015}]{Riebel2015}
Riebel D.,  et~al., 2015, \mn@doi [Astrophysical Journal]
  {10.1088/0004-637X/807/1/1}, 807, 1

\bibitem[\protect\citeauthoryear{{Riess}}{{Riess}}{2019}]{Riess2019}
{Riess} A.~G.,  2019, \mn@doi [Nature Reviews Physics]
  {10.1038/s42254-019-0137-0}, \href
  {https://ui.adsabs.harvard.edu/abs/2019NatRP...2...10R} {2, 10}

\bibitem[\protect\citeauthoryear{{Riess} et~al.,}{{Riess}
  et~al.}{2018}]{Riess2018}
{Riess} A.~G.,  et~al., 2018, \mn@doi [\apj] {10.3847/1538-4357/aac82e}, \href
  {https://ui.adsabs.harvard.edu/abs/2018ApJ...861..126R} {861, 126}

\bibitem[\protect\citeauthoryear{Ripepi et~al.,}{Ripepi
  et~al.}{2016}]{Ripepi2016a}
Ripepi V.,  et~al., 2016, \mn@doi [The Astrophysical Journal Supplement Series]
  {10.3847/0067-0049/224/2/21}, 224, 21

\bibitem[\protect\citeauthoryear{{Sandage}, {Tammann}  \& {Reindl}}{{Sandage}
  et~al.}{2009}]{Sandage2009}
{Sandage} A.,  {Tammann} G.~A.,   {Reindl} B.,  2009, \mn@doi [\aap]
  {10.1051/0004-6361:200810550}, \href
  {https://ui.adsabs.harvard.edu/abs/2009A&A...493..471S} {493, 471}

\bibitem[\protect\citeauthoryear{{Schwarz}}{{Schwarz}}{1978}]{Schwarz1978}
{Schwarz} G.,  1978, Annals of Statistics, \href
  {https://projecteuclid.org/euclid.aos/1176344136#abstract} {6, 461}

\bibitem[\protect\citeauthoryear{{Scowcroft}, {Bersier}, {Mould}  \&
  {Wood}}{{Scowcroft} et~al.}{2009}]{Scowcroft2009}
{Scowcroft} V.,  {Bersier} D.,  {Mould} J.~R.,   {Wood} P.~R.,  2009, \mn@doi
  [\mnras] {10.1111/j.1365-2966.2009.14822.x}, \href
  {https://ui.adsabs.harvard.edu/abs/2009MNRAS.396.1287S} {396, 1287}

\bibitem[\protect\citeauthoryear{Scowcroft, Freedman, Madore, Monson, Persson,
  Seibert, Rigby  \& Sturch}{Scowcroft et~al.}{2011}]{Scowcroft2011}
Scowcroft V.,  Freedman W.,  Madore B.~F.,  Monson A.~J.,  Persson S.~E.,
  Seibert M.,  Rigby J.~R.,   Sturch L.,  2011, \mn@doi [Astrophysical Journal]
  {10.1088/0004-637X/743/1/76}, 743, 76

\bibitem[\protect\citeauthoryear{{Scowcroft}, {Seibert}, {Freedman}, {Beaton},
  {Madore}, {Monson}, {Rich}  \& {Rigby}}{{Scowcroft}
  et~al.}{2016a}]{Scowcroft2016b}
{Scowcroft} V.,  {Seibert} M.,  {Freedman} W.~L.,  {Beaton} R.~L.,  {Madore}
  B.~F.,  {Monson} A.~J.,  {Rich} J.~A.,   {Rigby} J.~R.,  2016a, \mn@doi
  [\mnras] {10.1093/mnras/stw628}, \href
  {https://ui.adsabs.harvard.edu/abs/2016MNRAS.459.1170S} {459, 1170}

\bibitem[\protect\citeauthoryear{{Scowcroft}, {Freedman}, {Madore}, {Monson},
  {Persson}, {Rich}, {Seibert}  \& {Rigby}}{{Scowcroft}
  et~al.}{2016b}]{Scowcroft2016a}
{Scowcroft} V.,  {Freedman} W.~L.,  {Madore} B.~F.,  {Monson} A.,  {Persson}
  S.~E.,  {Rich} J.,  {Seibert} M.,   {Rigby} J.~R.,  2016b, \mn@doi [\apj]
  {10.3847/0004-637X/816/2/49}, \href
  {https://ui.adsabs.harvard.edu/abs/2016ApJ...816...49S} {816, 49}

\bibitem[\protect\citeauthoryear{Soszynski, Gieren  \& Pietrzynski}{Soszynski
  et~al.}{2005}]{Soszynski2005}
Soszynski I.,  Gieren W.,   Pietrzynski G.,  2005, \mn@doi [Publications of the
  Astronomical Society of the Pacific] {10.1086/431434}, 117, 823

\bibitem[\protect\citeauthoryear{Soszy{\'{n}}ski et~al.,}{Soszy{\'{n}}ski
  et~al.}{2015}]{Soszynski2015}
Soszy{\'{n}}ski I.,  et~al., 2015, Acta Astronomica, 65, 297

\bibitem[\protect\citeauthoryear{{Soszy{\'n}ski} et~al.,}{{Soszy{\'n}ski}
  et~al.}{2017}]{Soszynski2017}
{Soszy{\'n}ski} I.,  et~al., 2017, \mn@doi [\actaa]
  {10.32023/0001-5237/67.2.1}, \href
  {https://ui.adsabs.harvard.edu/abs/2017AcA....67..103S} {67, 103}

\bibitem[\protect\citeauthoryear{Stetson}{Stetson}{1987}]{Stetson1987}
Stetson P.~B.,  1987, \mn@doi [Publications of the Astronomical Society of the
  Pacific] {10.15713/ins.mmj.3}, 99, 191

\bibitem[\protect\citeauthoryear{Stetson}{Stetson}{1994}]{Stetson1994}
Stetson P.~B.,  1994, \mn@doi [Publications of the Astronomical Society of the
  Pacific] {10.1086/133378}, 106, 250

\bibitem[\protect\citeauthoryear{Stetson \& Harris}{Stetson \&
  Harris}{1988}]{Stetson1988}
Stetson P.~B.,  Harris W.~E.,  1988, Astronomical Journal, 96, 3

\bibitem[\protect\citeauthoryear{{Szewczyk}, {Pietrzy{\'n}ski}, {Gieren},
  {Ciechanowska}, {Bresolin}  \& {Kudritzki}}{{Szewczyk}
  et~al.}{2009}]{Szewczyk2009}
{Szewczyk} O.,  {Pietrzy{\'n}ski} G.,  {Gieren} W.,  {Ciechanowska} A.,
  {Bresolin} F.,   {Kudritzki} R.-P.,  2009, \mn@doi [\aj]
  {10.1088/0004-6256/138/6/1661}, \href
  {https://ui.adsabs.harvard.edu/abs/2009AJ....138.1661S} {138, 1661}

\bibitem[\protect\citeauthoryear{Tammann, Sandage  \& Reindl}{Tammann
  et~al.}{2003}]{Tammann2003}
Tammann G.~A.,  Sandage A.,   Reindl B.,  2003, \mn@doi [Astronomy {\&}
  Astrophysics] {10.1051/0004-6361}, 448, 423

\bibitem[\protect\citeauthoryear{Udalski, Szyma{\'{n}}ski, Kubiak,
  Pietrzy{\'{n}}ski, Soszy{\'{n}}ski, Wozniak  \& Zebrun}{Udalski
  et~al.}{1999}]{Udalski1999}
Udalski A.,  Szyma{\'{n}}ski M.~K.,  Kubiak M.,  Pietrzy{\'{n}}ski G.,
  Soszy{\'{n}}ski I.,  Wozniak P.,   Zebrun K.,  1999, \mn@doi [Acta
  Astronomica] {10.15713/ins.mmj.3}, 49, 201

\bibitem[\protect\citeauthoryear{{Udalski}, {Szymanski}, {Soszynski}  \&
  {Poleski}}{{Udalski} et~al.}{2008}]{Udalski2008}
{Udalski} A.,  {Szymanski} M.~K.,  {Soszynski} I.,   {Poleski} R.,  2008,
  \actaa, \href {https://ui.adsabs.harvard.edu/abs/2008AcA....58...69U} {58,
  69}

\makeatother
\end{thebibliography}




\appendix

\section{Mid-infrared light curves for the calibrating Cepheids}
\label{sec:ap1}

\begin{figure}
    \centering
    \includegraphics[width=0.45\textwidth]{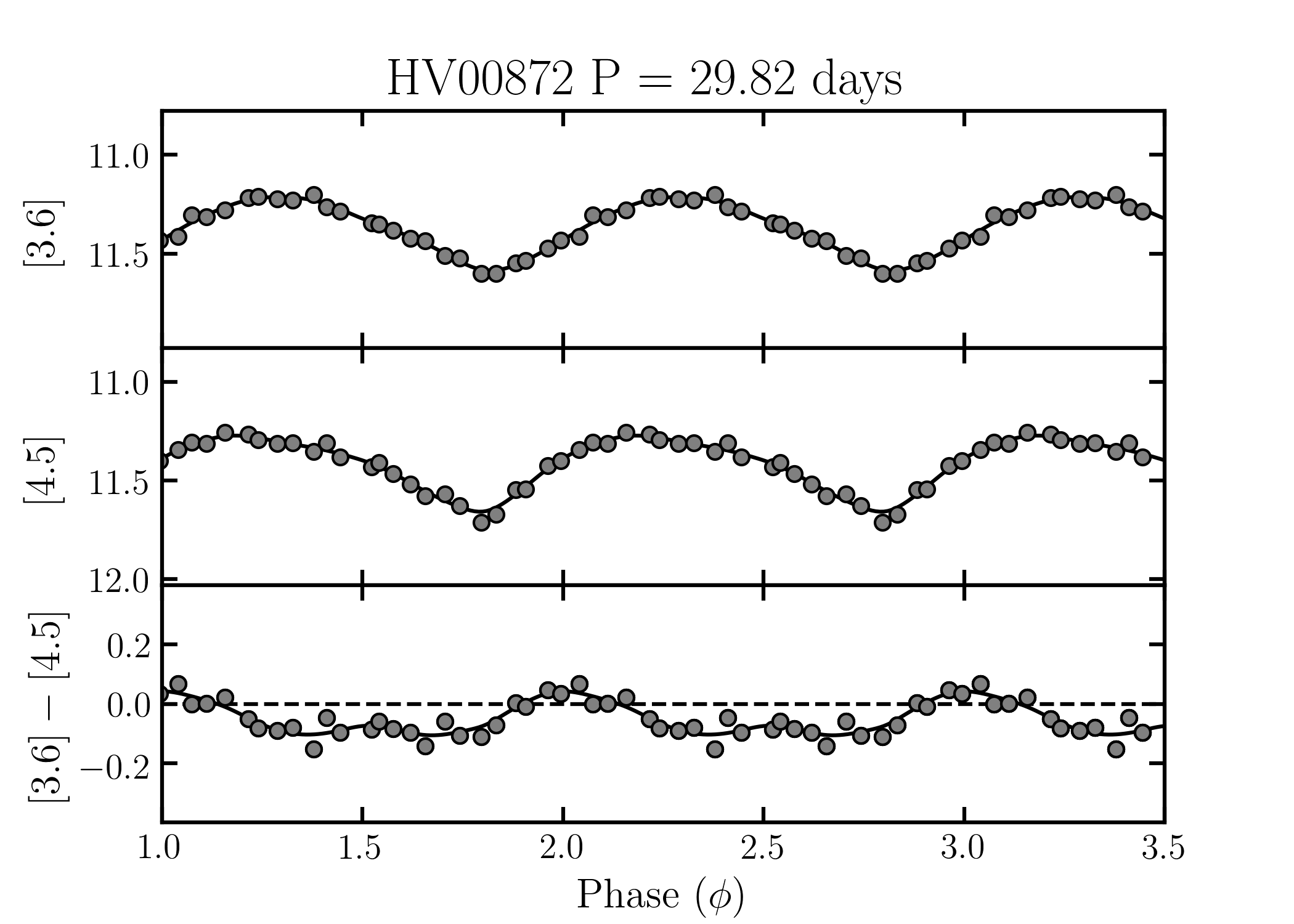}
    \caption{Mid-IR light and colour curves for Cepheids in the calibrating sample. The full collection of light curves are given in the online version of the journal.}
    \label{fig:allLCs}
\end{figure}

We present the full sample of calibrating sample Cepheids in Figure \ref{fig:allLCs} of the online version.

\section{Choice of priors}
\label{sec:ap2}

\begin{figure}
    \centering
    \includegraphics[width=0.4\textwidth]{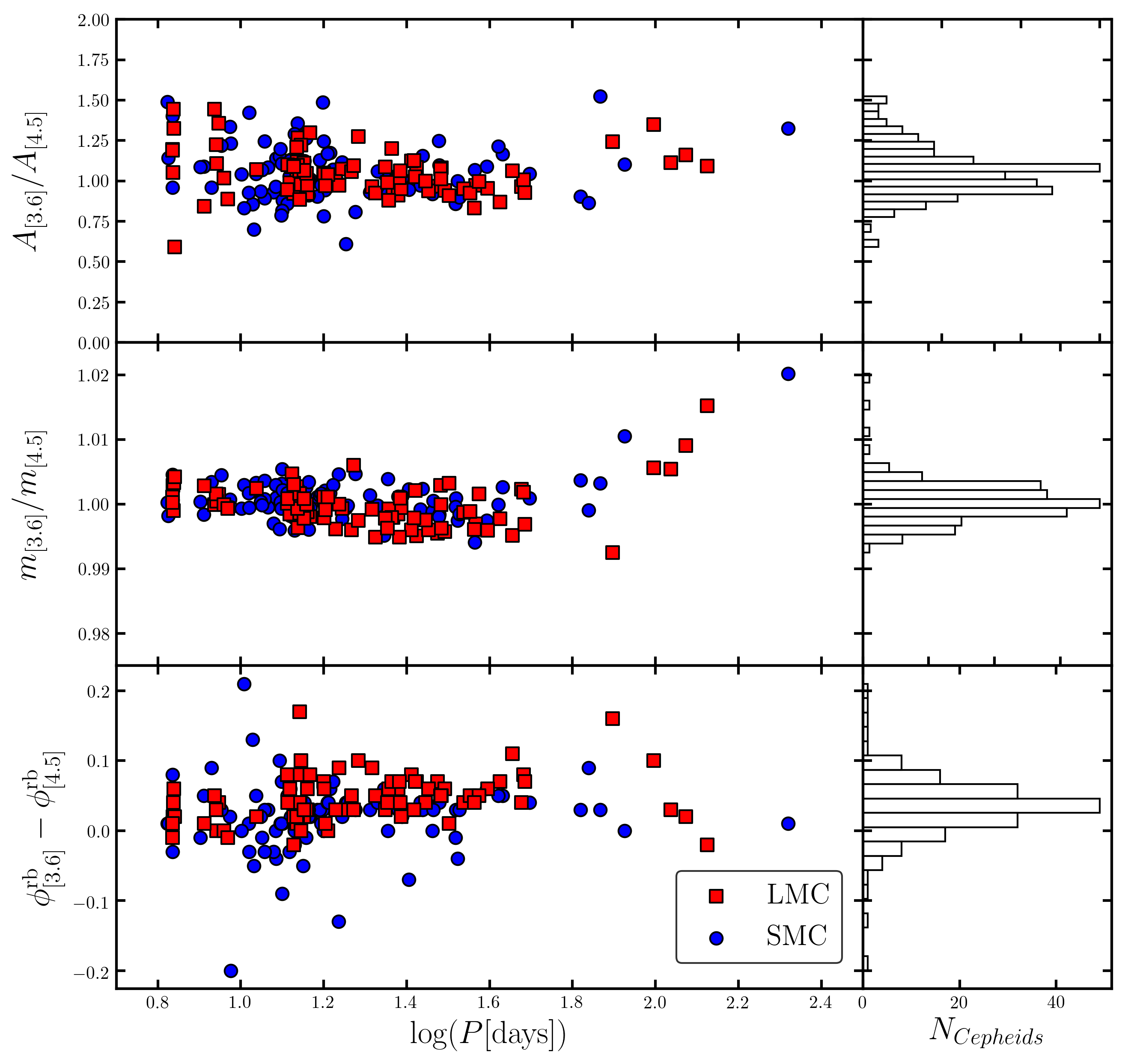}
    \caption{Mid-infrared parameter correlations that were used as priors for the MCMC code. \textit{Top:} mid-IR amplitude ratio ($A_{[3.6]}/A_{[4.5]}$). \textit{Middle:} mid-IR magnitude ratio ($m_{[3.6]}/m_{[4.5]}$). \textit{Bottom:} mid-IR phase lag ($\phi_{[3.6]}^{\text{rb}} - \phi_{[4.5]}^{\text{rb}}$). The red squares and blue circles show the LMC and SMC calibrating Cepheids, respectively.  }
    \label{fig:ir_priors}
\end{figure}

In Section \ref{sec:temp_fitting_method}, we discussed several uniform priors that were used in our MCMC template fitting method. These priors were imposed on the $[3.6]$ and $[4.5]$ mean magnitudes, amplitudes and phases and were based on Figure \ref{fig:ir_priors}. A uniform prior was placed on the mid-IR amplitude ratios $A_{[3.6]}/A_{[4.5]}$ (top panel of Figure \ref{fig:ir_priors}), the mid-IR magnitude ratio $m_{[3.6]}/m_{[4.5]}$ (middle panel of Figure \ref{fig:ir_priors}), and mid-IR phase lag $\phi_{[3.6]}^{\text{rb}} - \phi_{[4.5]}^{\text{rb}}$ (bottom panel of Figure \ref{fig:ir_priors}). In each of the panels, the distribution appears Gaussian. However, we opted for a uniform prior as we found that when using a Gaussian prior resulted in the MCMC walkers being unable to explore the parameter space fully.

In the middle panel of Figure \ref{fig:ir_priors}, there appears to be a negative correlation for $\log(P) < 1.8$, with an inversion for $\log(P) \geq 1.8$. This shape mimics the period-colour relation shown in \cite{Scowcroft2016b} (their figure 5).

\bsp	
\label{lastpage}
\end{document}